\documentclass[aps,manuscript,nofootinbib,superscriptaddress]{revtex4}
\usepackage{rotfloat}
\usepackage{amsmath}
\usepackage{graphicx}
\usepackage{amssymb}

\begin{document}

\global\long\def\oo{\mathcal{O}}
\global\long\def\met{\not\!\! E_{T}}
\preprint{ANL-HEP-PR-09-107,~EFI-09-31,~IPMU09-0162}

\title{Effective Dark Matter Model:\\
Relic density, CDMS II, Fermi LAT and LHC}

\author{Hao Zhang}
\email{haozhang.pku@pku.edu.cn}
\affiliation{Department of Physics and State Key Laboratory of Nuclear Physics
and Technology, Peking University, Beijing 100871, China}
\affiliation{Enrico Fermi Institute, University of Chicago, Chicago, Illinois
60637, U.S.A.}

\author{Qing-Hong Cao}
\email{caoq@hep.anl.gov}
\affiliation{Department of Physics and State Key Laboratory of Nuclear Physics
and Technology, Peking University, Beijing 100871, China}
\affiliation{Enrico Fermi Institute, University of Chicago, Chicago, Illinois
60637, U.S.A.}
\affiliation{HEP Division, Argonne National Laboratory, Argonne, Illinois 60439,
U.S.A.}

\author{Chuan-Ren Chen}
\email{chuan-ren.chen@ipmu.jp}
\affiliation{Institute for the Physics and Mathematics of the Universe, The University
of Tokyo, Chiba 277 8568, Japan }

\author{Chong Sheng Li}
\email{csli@pku.edu.cn}
\affiliation{Department of Physics and State Key Laboratory of Nuclear Physics
and Technology, Peking University, Beijing 100871, China}

\begin{abstract}
The Cryogenic Dark Matter Search recently announced the observation
of two signal events with a $77\%$ confidence level. Although statistically
inconclusive, it is nevertheless suggestive. 
In this work we present
a model-independent analysis on the implication of direct and indirect searches
for the dark matter using effective operator approach. 
Assuming that the interactions between
(scalar, fermion or vector) dark matter and the standard model are mediated
by unknown new physics at the scale $\Lambda$, we examine various
dimension-6 tree-level induced operators and constrain them using
the current experimental data,
including the WMAP data of the relic abundance, CDMS II direct detection of the spin-independent scattering, and indirect
detection data (Fermi LAT cosmic gamma-ray). Finally, the LHC search is also explored.

\end{abstract}
\maketitle
\tableofcontents{}

\newpage{}

\section{Introduction}

The observational data that have accumulated for 
decades
point to the existence of a significant amount of 
dark matter (DM), and the nature 
of this substance has become one
of the key problems at the interface between particle physics, astrophysics
and cosmology. Since the Standard Model (SM) of particle physics does
not possess any candidate for such 
dark matter, this
problem constitutes a major indication for new physics beyond the SM. Our current
knowledge of the proprieties of DM are inferred solely from astrophysical
and cosmological observations, which provide no information about
its most basic characteristics such as the spin and mass.
Recently, many anomalies
in various cosmic-ray observations, for example, INTEGRAL~\cite{Strong:2005zx}, ATIC~\cite{:2008zzr},
PAMELA~\cite{Adriani:2008zr} and Fermi LAT~\cite{Abdo:2009zk}, were found and attracted many investigators' 
attention.
Due to our poor understanding of astrophysical 
backgrounds, it is not affirmative to say those excesses are indeed
caused by dark matter. On the other hand, the direct detection is
considered more robust.
The Cryogenic Dark Matter Search (CDMS II) recently announced the
observation of two signal events with a $77\%$ confidence level~\cite{Ahmed:2009zw}
and additional two events just outside the signal region.
While
statistically such an observation is not significant, it is nevertheless
suggestive and stimulates a lot of studies of its implication~\cite{Kadastik:2009ca,Kadastik:2009gx,
Bottino:2009km,Feldman:2009pn,Ibe:2009pq,Kopp:2009qt,Allahverdi:2009sb,
Endo:2009uj,Cao:2009uv,Holmes:2009uu}.

\begin{figure}
\includegraphics[scale=0.4]{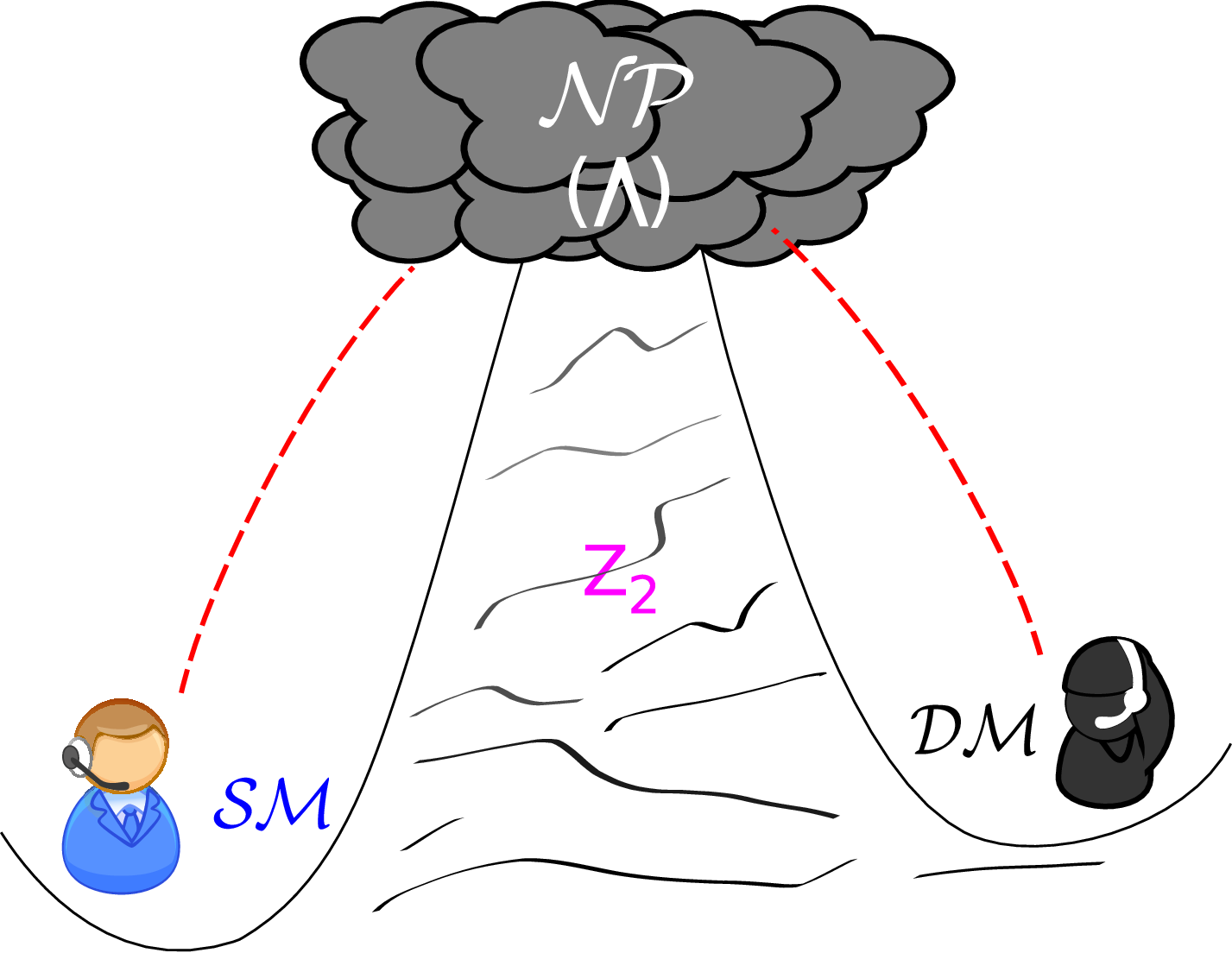}
\caption{Pictorial illustration of our EFT model, in which the SM particles
interact with the (unknown) DM particles through the new physics which
appears at the scale $\Lambda$.\ \label{fig:model}}
\end{figure}

There are many proposals for the dark matter in the literature:
the lightest supersymmetric particle
in the supersymmetric models with conserved $R$-parity~\cite{Jungman:1995df,Bertone:2004pz},
the lightest KK excitation 
in the universal extra dimension models
with conserved KK-parity\ \cite{Servant:2002aq,Cheng:2002ej,Servant:2002hb},
the lightest T-odd particle in the little Higgs models
with conserved T-parity\ \cite{Cheng:2003ju,Cheng:2004yc,Birkedal:2006fz}, 
and dark scalar
models\ \cite{Silveira:1985rk,McDonald:1993ex,Barger:2006dh,Barbieri:2006dq,
LopezHonorez:2006gr,Gustafsson:2007pc,Barger:2007im,Cao:2007rm,
Cao:2009yy,Kadastik:2009dj,Kadastik:2009cu,Zhang:2009dd}, etc. Beside the DM candidate, all
the new physics models mentioned above also provide many novel predictions
that can (and will) be probed at the Large Hadron Collider (LHC);
observing any such new effect will shed light on the theory underlying
the SM. But it is also possible that all new physics effects will
be hidden from direct observation at the LHC\ \cite{Wilczek:2007gsa}.
This is the case, for example, when the DM belongs to a ``hidden
sector'' that interacts very weakly with the SM particles  (e.g. if the
mediating particles are very heavy), see Fig.~\ref{fig:model} for illustration.
Communication between the hidden sector and the SM can then be
described by an effective field theory (EFT). 
In this paper, we investigate the dimension six, tree-level generated and
$SU(3)_c\times SU(2)_L\times U(1)_Y$ invariant effective operators, 
which describe the interactions of DM and SM particles.
Similar studies in the effective Lagrangian approach have been carried out
in the literature~\cite{Barger:2008qd,Beltran:2008xg,Shepherd:2009sa,
Goodman:2010ku,Goodman:2010qn,Goodman:2010yf,Agrawal:2010ax,Agrawal:2010fh,Bai:2010hh,Fan:2010gt}, 
which focus on different operators. 

We add one new field $D$ (the DM candidate) and define
the effective Lagrangian as 
\begin{equation}
\mathcal{L}_{{\rm eff}}=\mathcal{L}_{SM}+
\sum\frac{f_{i}^{(5)}}{\Lambda}\oo_{i}^{(5)}
+\sum\frac{f_{i}^{(6)}}{\Lambda^{2}}\oo_{i}^{(6)}+...\,,
\label{eq:eft}
\end{equation}
where $\oo_{i}^{(n)}$ is  a dimension-$n$ operator which consists of a gauge-invariant
combination of the SM and $D$ fields\ \cite{Buchmuller:1985jz,Arzt:1994gp};
$\Lambda$ then denotes the new physics scale (we have assumed that
the new physics decouples in the limit $\Lambda\to\infty$). 
Since the loop-induced operator suffers from $1/(16\pi^{2})$ loop-integration factor, 
we focus our attention on the tree-level induced operators throughout this work.

The DM candidate field $D$ is assumed to carry an odd quantum
number under a discrete $Z_{2}$ symmetry\footnote{This eliminates possible contributions to observables that are strongly
constrained by existing data. For example, if $D$ is a vector field,
the $Z_{2}$ symmetry forbids contributions to 4-Fermi current-current
interactions whose coefficients are tightly bound~\cite{Amsler:2008zzb}.
} under which all SM fields are even. Such a $Z_{2}$ symmetry could
be an unbroken remnant of some underlying $U(1)$ gauge symmetry~\cite{Krauss:1988zc}.
When $D$ is a scalar or fermion, $\mathcal{L}_{eff}$ is then required
to be invariant under $SU(3)_{c}\times SU(2)_{L}\times U(1)_{Y}\times Z_{2}$.
When $D$ is a vector, it will be assumed to be the gauge boson of
an additional $U(1)_{X}$ symmetry and to carry no SM quantum numbers;
all SM particles are assumed to be $U(1)_{X}$ singlets and $\mathcal{L}_{eff}$
will be invariant under $SU(3)_{c}\times SU(2)_{L}\times U(1)_{Y}\times U(1)_{X}$.
The main object of this work is to examine the constraints imposed
by current DM searches on the coefficients $f_{i}^{(n)}$. We also
explore the collider signatures at the LHC and probe these SM-DM interactions in future DM search
experiments. 

The rest of this paper is organized as follows. In Sec.\ \ref{sec:cosmo} we briefly
review some general aspects of dark matter. In Sec.\ \ref{sec:fermion},\ \ref{sec:Scalar}
and \ref{sec:Vector} we present our studies of fermionic dark matter,
scalar dark matter and vector dark matter, respectively. Finally,
in Sec.\ \ref{sec:Conclusion}, we review the results of our study
and present our conclusions.

\section{General aspects of cold dark matter\ \label{sec:cosmo}}

In order to survive until the present epoch, any 
dark matter particles must either be stable or have a lifetime 
longer than
 the present age of the universe. Furthermore, if the dark matter
particles have electromagnetic or strong interactions, they would
bind to nucleons and form anomalous heavy isotopes. Such isotopes
have been sought but not found~\cite{Smith:1979rz,Smith:1982qu,Verkerk:1991jf}.
Thus, the dark matter particles can, at best, participate in weak
(and gravitational) interaction, or, at worst, only participate in gravitational
interaction. One obvious possibility satisfying the foregoing constraints
is that dark matter consists of neutrinos. However, the present data
on neutrino masses show that, although neutrinos might barely account
for the inferred DM mass density, they cannot generate the observed
structure: simulations of galaxy and cluster formation require the \emph{cold}
dark matter.
That is, the weakly
interacting massive particles (WIMPs) are favored as  DM candidates.
 
The contribution to the energy density from a DM, $\Omega_{DM}$, 
of mass $m_{D}$ is~\cite{KolbTurner}
\begin{equation}
\Omega_{DM}h^{2}\approx\frac{1.04\times10^{9}}{M_{Pl}}
\frac{x_{F}}{\sqrt{g_{*}}}\frac{1}{(a+3b/x_{F})},
\label{eq:omega}
\end{equation}
where $h=0.73\pm0.04$ is the scaled Hubble parameter,
 $x_{F}\equiv m_{D}/T_{F}$ with $T_{F}$ being the freeze-out
temperature, $M_{Pl}=G_{N}^{-1/2}=1.22\times10^{19}\,{\rm GeV}$,
and $g_{*}$ counts the number of relativistic degrees of freedom
at the freeze-out temperature $T_{F}$. As the freeze-out temperature
is below the DM mass roughly by a factor of 20-25, the dark matter
freeze-out mechanism is independent of  the uncertain early thermal
history of the universe and possible new interactions at high energy scales.
The parameters $a$ and $b$ should be derived from $\left\langle \sigma v_{rel}\right\rangle $,
the thermally averaged annihilation cross section of DM ($\sigma$) times
the relative velocity ($v_{rel}$), cf. Eq.\ (\ref{eq:ab}). The relic
abundance of DM in the universe has been precisely measured: combining
the results from the WMAP Collaboration with those from the Sloan
Digital Sky Survey gives\ \cite{Spergel:2006hy} 
\begin{equation}
\Omega_{DM}h^{2}=0.111_{-0.015}^{+0.011}\,\,\,\,(2\sigma\,{\rm level}).
\label{eq:omega_WMAP}
\end{equation}
As to be shown below, the accuracy of about $10\%$
in the expected DM
abundance imposes an extremely strong bound on 
the effective operators, and leads to 
a strong correlation between the dark matter mass and
the new physics scale $\Lambda$, e.g., the heavier the dark matter the larger
the scale $\Lambda$. We focus on heavy dark matter,
say $m_{D}\gtrsim100\,{\rm GeV}$, all through this paper and will
present the study of the light dark matter elsewhere. 

Besides cosmological bounds on dark matter abundance, there are other
constraints on the properties of dark matter coming from direct detection
searches of WIMPs in the halo of the Milky Way. The idea is that if
WIMPs constitute the dominant component of the halo of our galaxy,
it is expected that some may cross the Earth at a reasonable rate
and can be detected by measuring the energy deposited in a low background
detector through the scattering of a dark matter particle with nuclei
of the detector. Several experiments have obtained upper bounds on
scattering cross sections as a function of WIMP mass. The elastic
scattering of a WIMP with nuclei can be separated into spin-independent
(SI) and spin-dependent (SD) contributions. SI scattering can take
place coherently with all of the nucleons in a nucleus, leading to
a cross section proportional to the square of the nuclear mass. As
a result, the current constraints on spin-independent
scattering are considerably stronger than that on the spin-dependent component.
As to be shown, the direct search of dark matter has a significant
impact on the effective operators of both fermionic and scalar dark
matters. 

Additional constraints on relic DM can be derived from the expectation
that DM will collect and become gravitationally bound to the center
of the galaxy, the center of the Sun and the center of the Earth.
If this happens, then a variety of \emph{indirect} dark matter detection
opportunities arise.
Detecting the annihilation products of dark matter particles in the
form of gamma-rays, antimatters and neutrinos are collectively known
as indirect search.
 Among these cosmic-ray observations,  gamma-ray
is thought to be more robust. 
Therefore, we focus our attention
on the detection of the anomalous cosmic gamma-rays and examine the
impact of current measurements and future projected sensitivities
on the effective operators. Studies of antimatter and neutrino searches
are also interesting but are beyond the scope of this project. 

If the DM particles are light enough,
they will be pair-produced at Large Hadron Collider (LHC) and subsequently escape from the
detector, resulting in a spectacular collider signature characterized
by a large missing transverse momentum. 
These events
can
be tagged by requiring that they also contain specific SM particles,
such as hard photons, charged leptons, or jets. Extracting the DM 
signal from this type of events reduces essentially to
a counting experiment since a resonance in the invariant mass distribution
of the DM particle pair cannot be extracted. Thus, one needs to have
better control of the SM background to affirm the existence of DM
particles. Here we focus our attention only to the statistical uncertainty
of the background events, but needless to say, other uncertainties,
such as the systematic error and parton luminosity uncertainty, etc.,
must be included in order to provide realistic predictions.

Each experiment mentioned above will
have different sensitivities to each individual dark operator (i.e. the effective operators) and
thus can probe the parameter space of the dark operators independently~
\footnote{Systematic studies of the neutralino dark matter along this direction
have been carried out in Refs.\ \cite{Allanach:2004xn,Nojiri:2005ph,Baltz:2006fm}.}. Based on the time-line of the experimental programs, it can be achieved
in the following steps:
\begin{itemize}
\item Current relic abundance measurements severely constrain the dark operators
and induce a non-trivial relation between the dark matter mass and
the new physics scale.
\item Direct and indirect searches of dark matter, mostly the SI scattering
experiment, further constrain the parameter space of the dark operators.
\item Dark matter pair can be copiously produced at the incoming LHC: discovering
events with large missing energy can determine the possible mass of
the DM. 
\item Future direct and indirect detections of dark matter can probe more
parameter space lying beyond the reach of LHC. 
\item Consistently checking all experimental measurements may shed light
on the underlying theory.
\end{itemize}
The main goal of the remainder of this paper is to show the close
connection between dark matter searches and the experiments foreseen
at the LHC. We will address this question individually for
the fermion, scalar and vector dark matter below.

\section{Fermionic dark matter $\chi$\ \label{sec:fermion}}

We first analyze the case that the DM candidate is a fermion
$\chi$, which is assumed to be odd under a $Z_{2},$ and is a SM gauge
singlet; we  also assume that $\chi$ has no chiral interactions.
In this case the (tree-level generated) effective interactions with
the SM are of two types
\begin{itemize}
\item four-fermion: 
\begin{eqnarray}
 &  & \oo_{u\chi}=\frac{1}{2}\left(\bar{u}\gamma^{\mu}u\right)
 \left(\bar{\chi}\gamma_{\mu}\chi\right),\quad\oo_{d\chi}=
 \frac{1}{2}\left(\bar{d}\gamma^{\mu}d\right)
 \left(\bar{\chi}\gamma_{\mu}\chi\right),\nonumber \\
 &  & \oo_{e\chi}=\frac{1}{2}\left(\bar{e}\gamma^{\mu}e\right)\,
 \left(\bar{\chi}\gamma_{\mu}\chi\right),\quad\oo_{\ell\chi}=
 \left(\bar{\ell}\chi\right)\left(\bar{\chi}\ell\right),
 \quad\oo_{q\chi}=\left(\bar{q}\chi\right)\left(\bar{\chi}q\right),
 \label{eq:fermion:fermiononly}
\end{eqnarray}

\item Vectors, fermions and scalars~
\footnote{The dimension-5 operator $\bar{\chi}\chi\phi^{\dagger}\phi$ has been
studied in Refs.~\cite{Kim:2006af,Cao:2007fy}.
}\begin{equation}
\oo_{\phi\chi}=i\left(\phi^{+}D_{\mu}\phi\right)\left(\bar{\chi}\gamma_{\mu}\chi\right)+h.c.\,\,.\label{eq:fermon:fermion-vector}\end{equation}

\end{itemize}
Here, $q\left(\ell\right)$ denotes the left-handed doublets while
$u(d,e)$ denotes the right-handed singlet. In the unitary gauge the
Higgs doublet is given by\[
\phi=\frac{v+h}{\sqrt{2}}\left(\begin{array}{c}
0\\
1\end{array}\right),\]
where $h$ is the SM Higgs boson and $v=246\,{\rm GeV}$ is the vacuum
expectation value. All through this study we will choose $m_{h}=120\,{\rm GeV}$.
With the help of the Fierz identity, the scalar-scalar-current in eq~(\ref{eq:fermion:fermiononly}) can be written in the vector-vector-current form. 
In this work, we will focus on the four-fermion effective
Lagrangian for the $\bar{\chi}\chi\bar{f}f$ interaction in the form
of \begin{equation}
\mathcal{L}_{\chi f}^{(6)}=\frac{1}{\Lambda^{2}}\left(\bar{\chi}\gamma_{\mu}P_{R}\chi\right)\;\left(\bar{f}\gamma^{\mu}(g_{L}^{f}P_{L}+g_{R}^{f}P_{R})f\right),\label{eq:fermion:ffff}\end{equation}
where $f=u,d,e$ and \begin{eqnarray}
 &  & g_{L}^{u}=-\frac{1}{2}\alpha_{q\chi},\quad g_{R}^{u}=\frac{1}{2}\alpha_{u\chi},\quad g_{L}^{d}=-\frac{1}{2}\alpha_{q\chi},\quad g_{R}^{d}=\frac{1}{2}\alpha_{d\chi},\nonumber \\
 &  & g_{L}^{e}=-\frac{1}{2}\alpha_{\ell\chi},\quad g_{R}^{e}=\frac{1}{2}\alpha_{e\chi},\quad g_{L}^{\nu}=-\frac{1}{2}\alpha_{\ell\chi},\quad g_{R}^{\nu}=0.\label{eq:coeff:ffff}\end{eqnarray}
Note that the right-handed projector $P_R$ in eq~(\ref{eq:fermion:ffff}) originates 
from the fact that the fermionic DM particle considered in this work 
is a SM gauge singlet. We assume \[
\alpha_{q\chi}=\alpha_{\ell\chi}\equiv\alpha_{f\chi}^{L},\qquad\alpha_{u\chi}=\alpha_{d\chi}=\alpha_{e\chi}\equiv\alpha_{f\chi}^{R},\]
and these operators are also diagonal in the flavor space  for simplicity. The fermionic
DM can also annihilate into a vector-scalar boson pair via the following
vertex:\begin{equation}
\mathcal{L}_{\chi\phi}^{(6)}=\frac{m_{Z}}{\Lambda^{2}}\alpha_{\phi\chi}\left(\bar{\chi}\gamma^{\mu}\chi\right)Z_{\mu}h.\label{eq:fermion:ffss}\end{equation}
The operator $\oo_{\phi\chi}$ also gives rise to the vertex $Z\chi\chi$
as\begin{equation}
\mathcal{L}_{Z\chi\chi}^{(6)}=\frac{\alpha_{\phi\chi}\, v\, m_{Z}}{2\Lambda^{2}}Z_{\mu}\bar{\chi}\gamma^{\mu}\chi,\end{equation}
which contributes to the annihilation processes of $\chi\bar{\chi}\to Z\to f\bar{f},~Zh~{\rm or}~W^+W^-$.
For simplicity we consider the operators $\oo_{f\chi}^{L}$, $\oo_{f\chi}^{R}$
and $\oo_{\phi\chi}$ and use the coefficients, $\alpha_{f\chi}^{L/R}$
and $\alpha_{\phi\chi}$, to denote the operators hereafter.

\begin{figure}
\includegraphics[clip,scale=0.4]{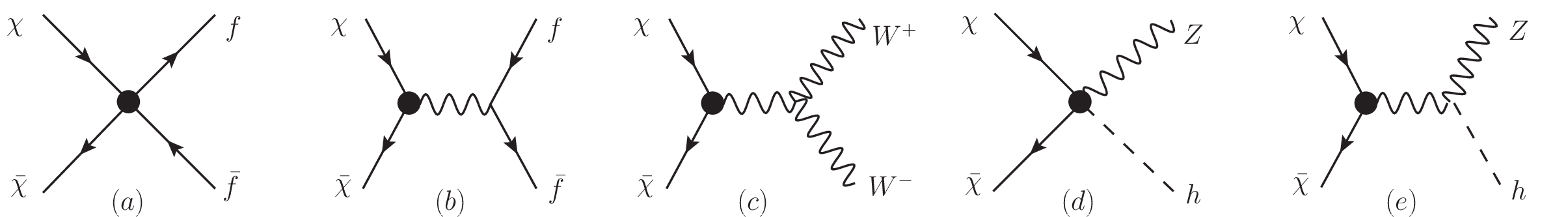}

\caption{Feynman diagrams for $\chi\bar{\chi}$ annihilations. Blobs denote
the effective vertices induced by the dim-6 operators. \label{fig:feyn-ff}}

\end{figure}

\begin{table}
\caption{Sensitivities of different experiments to the dark operators where
the round brackets denote the Feynman diagrams in Fig.\ \ref{fig:feyn-ff}
while the square brackets denotes the dark operators. The last three
columns denote the LHC collider signatures of those dark operators,
$\gamma+\met\,(q\bar{q}\to\gamma\chi\bar{\chi})$, $j+\met\,(q\bar{q}\to j\chi\bar{\chi})$
and $h+\met\,(q\bar{q}\to h\chi\bar{\chi})$; see Fig.~\ref{fig:fermion-lhc-feyn}
for details.\ \label{tab:sensitivities-fermion}}

\begin{tabular}{clllllllllll}
\hline 
 & $\Omega_{DM}h^{2}$ & $\qquad$ & $\chi$-nucleon & $\qquad$ & cosmic $\gamma$-${\rm ray}^{\,\star}$ & $\qquad$ & $\gamma+\met$ & $\qquad$ & $j+\met$ & $\qquad$ & $h+\met$\tabularnewline
\hline 
 & (a) {[}$\alpha_{f\chi}^{L/R}${]} &  & (a) {[}$\alpha_{f\chi}^{L/R}${]} &  & (a) {[}$\alpha_{f\chi}^{L/R}${]} &  & (a) {[}$\alpha_{f\chi}^{L/R}${]} &  & (a) {[}$\alpha_{f\chi}^{L/R}${]} &  & (d) {[}$\alpha_{\phi\chi}${]}\tabularnewline
 & (d) {[}$\alpha_{\phi\chi}${]} &  & (b) {[}$\alpha_{\phi\chi}${]} &  & (d) {[}$\alpha_{\phi\chi}${]} &  & (b) {[}$\alpha_{\phi\chi}${]} &  & (b) {[}$\alpha_{\phi\chi}${]} &  & (e) {[}$\alpha_{\phi\chi}${]}\tabularnewline
\hline 
 & \multicolumn{11}{l}{$\star$: also sensitive to the sign of the coefficients. }\tabularnewline
\end{tabular}
\end{table}

The fermionic DM annihilates into the SM fermions, 
vector bosons, or
into a pair of Higgs boson and $Z$-boson. The Feynman diagrams of
the annihilations are shown in Fig.\ \ref{fig:feyn-ff}, where the
diagram (a) is related to $\alpha_{f\chi}^{L(R)}$ and the diagrams
(b-e) are related to $\alpha_{\phi\chi}$. These five diagrams have
different effects on various detection experiments; see Table\ \ref{tab:sensitivities-fermion}.
For the annihilation of a heavy $\chi$, the contributions from diagrams
(b), (c) and (e), being $s$-channel processes, are highly suppressed by
$1/(4m_{\chi}^{2}$) from the $s$-channel propagator. Therefore,
only diagrams (a) and (d) need to be considered for a heavy $\chi$
annihilation. It is worthy mentioning that if $\chi$ is very light,
say $\sim m_{Z}/2$, then the diagram (b) receives a large enhancement
from the threshold and becomes dominant. Detailed study of such light
dark matter will be presented in a future work. For elastic scattering
from a nucleus, only diagrams (a) and (b) contribute, whereas for
cosmic gamma-ray detection, only diagrams (a) and (d) contribute sizably.
Moreover, diagrams (a,\ b) can be probed at the LHC using the signature
of mono-photon plus missing energy while diagrams (d,\ e) can be
probed using the signature of single Higgs scalar plus missing energy, 
and diagram (c) can be probed using the signal of two forward jets with missing energy.
All these issues will be explored in details in the rest of this section.

\subsection{Relic abundance}

 One can calculate the cross
sections of $\chi\bar{\chi}$ annihilation and obtain the leading
terms ($a$ and $b$) in the non-relativistic expansion. The 
$x_{F}$ can be determined using Eq.\ (\ref{eq:xf}). After
substituting $x_{F}$ and $(a,\, b)$ into Eq.\ (\ref{eq:omega}),
one can evaluate the relic abundance of the DM. Detailed calculations
are given in Appendix\ \ref{sec:Dark-matter-annihilation}.

\begin{figure}
\includegraphics[clip,scale=0.5]{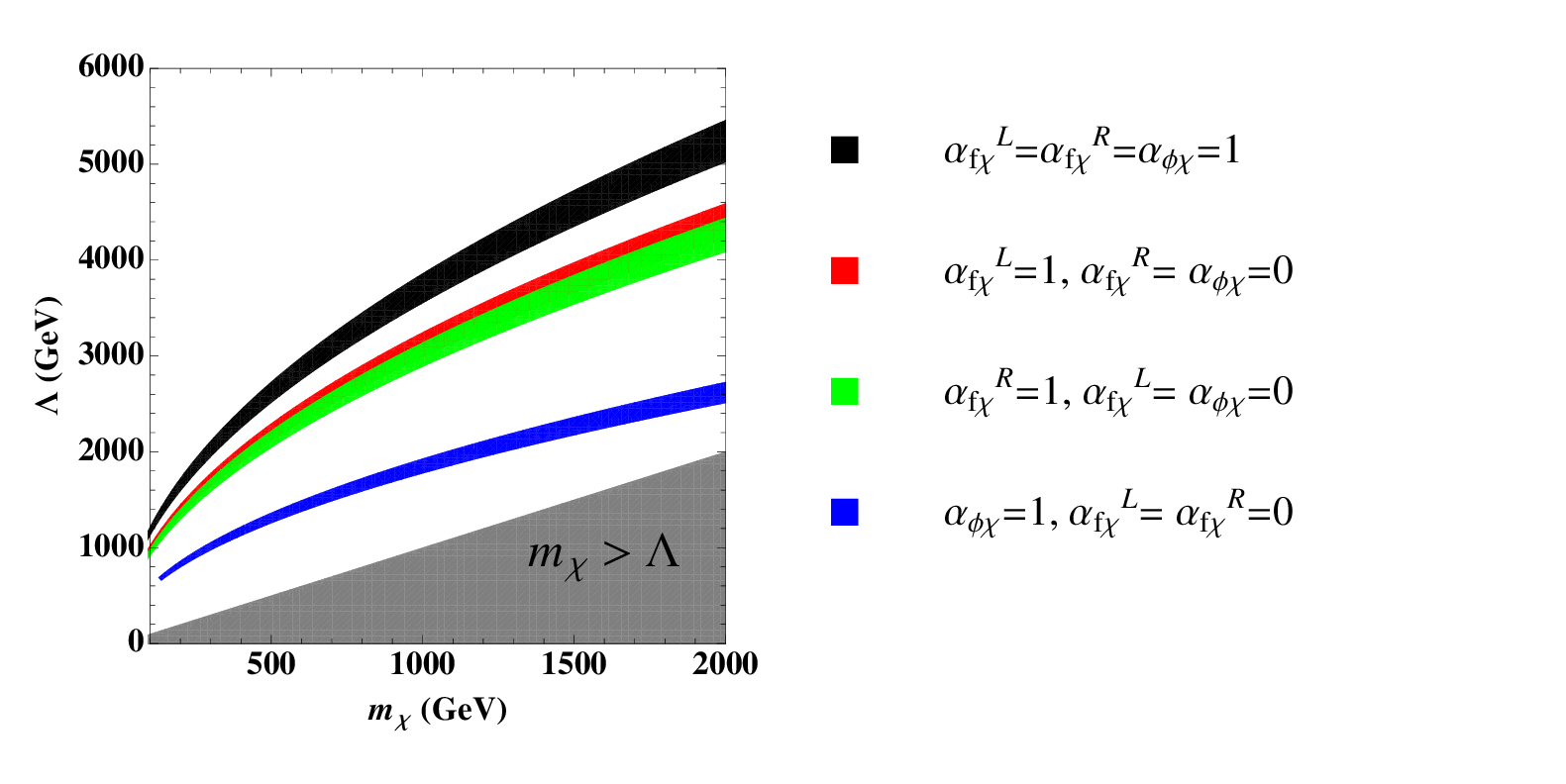}

\caption{Allowed parameter set $\left(m_{\chi},\Lambda\right)$ for a universal
coupling constant $\alpha$ when DM is a fermion. The upper (lower)
boundary of each band corresponds to the upper (lower) limit of $\Omega_{DM}h^{2}$.
\label{fig:fermion}}

\end{figure}

The allowed parameter set $(m_{\chi},\Lambda)$ with respect to WMAP
data is shown in Fig.\ \ref{fig:fermion}. For a 
heavy $\chi$,
say $m_{\chi}\gtrsim300\,{\rm GeV}$, all the SM particles can be
treated as massless. One can then obtain the leading terms $a$ and $b$
as follows: 
\begin{eqnarray}
a & = & \frac{m_{\chi}^{2}}{\Lambda^{4}}\left(0.24\alpha_{f\chi}^{L\,2}+0.21\alpha_{f\chi}^{R\,2}+0.03\alpha_{\phi\chi}^{2}\right),\\
b & = & \frac{m_{\chi}^{2}}{\Lambda^{4}}\left(0.11\alpha_{f\chi}^{L\,2}+0.10\alpha_{f\chi}^{R\,2}+0.009\alpha_{\phi\chi}^{2}\right).
\end{eqnarray}
We note that $x_{F}$ hardly varies, $x_{F}\simeq25$\ 
\footnote{In the equation that determines $T_{F}$, $x$ appears in a Boltzmann
factor $e^{-x}$. Taking the logarithm, one can show $x_{F}\approx25$
for a wide range of values of the annihilation cross section.
}, in the entire allowed parameter space, but the freeze-out temperature
$T_{F}$ varies from $20-80\,{\rm GeV}$ for $m_{\chi}\sim500-2000\,{\rm GeV}$.
Choosing $x_{F}=25$, we obtain the following relation between $m_{\chi}$
and $\Lambda$,\begin{equation}
\frac{2.22\times10^{-2}}{\left(\Omega_{DM}h^{2}\right)\left(0.252\alpha_{f\chi}^{L\,2}+0.221\alpha_{f\chi}^{R\,2}+0.031\alpha_{\phi\chi}^{2}\right)}=\left(\frac{m_{\chi}}{100\,{\rm GeV}}\right)^{2}\left(\frac{{\rm TeV}}{\Lambda}\right)^{4}.\label{eq:fermion-relation}\end{equation}
We have verified that this relation agrees very well with the numerical
results shown in Fig.\ \ref{fig:fermion}. 

For illustrations, we adopt the value of $\alpha$ to be 1 in the following discussions,
but it is  straightforward
to include its effects by rescaling $m$. When $m_{\chi}=500\,{\rm GeV}$,
the second lightest new particle is expected to have a mass of about
$2\,{\rm TeV}$ if only $\alpha_{f\chi}^{L}$ ($\alpha_{f\chi}^{R}$)
is non-zero, but it should appear around $1\,{\rm TeV}$ if $\alpha_{\phi\chi}$
is present only, see respectively the red (green) and blue bands.
 The bands of $\alpha_{f\chi}^{L}$ (red) and $\alpha_{f\chi}^{R}$
(green) overlap while the $\alpha_{f\chi}^{L}$ band is slightly higher
because the coefficient of $\alpha_{f\chi}^{L}$ is slightly larger
than the one of $\alpha_{f\chi}^{R}$, see Eq.\ (\ref{eq:fermion-relation}).
This small difference comes from three neutrino annihilation channels
which only contribute to $\alpha_{f\chi}^{L}$ but not to $\alpha_{f\chi}^{R}$.
Furthermore, the coefficients of $\alpha_{f\chi}^{L}$ and $\alpha_{f\chi}^{R}$
are both larger than the coefficients of $\alpha_{\phi\chi}$ by an
order of magnitude, i.e., the annihilation into fermions is dominant
in the annihilation cross sections. 
Eq.\ (\ref{eq:omega}) implies
that, for a given $m$, the scale $\Lambda$ has to be small
in order to compensate the smaller coefficient of $\alpha_{\phi\chi}$,
leading to a much lower band. Finally, when these three couplings
all are present, the allowed new physics scale $\Lambda$ becomes
larger, e.g. $\Lambda\sim2.6-5.5\,{\rm TeV}$ for $m_{\chi}\sim500-2000\,{\rm GeV}$. 

It is worthy mentioning that for each operator the region below the
band is also allowed even though less relic abundance is produced
and additional annihilation channels are needed to explain the current
relic abundance.

\subsection{Direct detection of $\chi$}

The elastic scattering of a fermion dark matter $\chi$ from a nucleus
is induced by effective four fermion operators:\begin{eqnarray}
\mathcal{L}_{\chi\chi qq} & = & \frac{1}{4\Lambda^{2}}\left(\alpha_{q\chi}^{R}-\alpha_{q\chi}^{L}+\alpha_{\phi\chi}\frac{g}{2\cos\theta_{W}}\frac{v}{m_{Z}}C_{V}^{q}\right)\bar{\chi}\gamma_{\mu}\chi\bar{q}\gamma^{\mu}q\nonumber \\
 & + & \frac{1}{4\Lambda^{2}}\left(\alpha_{q\chi}^{R}+\alpha_{q\chi}^{L}+\alpha_{\phi\chi}\frac{g}{2\cos\theta_{W}}\frac{v}{m_{Z}}C_{A}^{q}\right)\bar{\chi}\gamma_{\mu}\gamma_{5}\chi\bar{q}\gamma^{\mu}\gamma_{5}q,\label{eq:FDM-nucleon-interaction}\end{eqnarray}
with $C_{V}^{q}=T_{3}^{q}-2Q_{q}\sin^{2}\theta_{W}$ and $C_{A}^{q}=T_{3}^{q}$.
Here, $T_{3}$ and $Q$ are the weak isospin and electric charge of
the quark $q$, respectively, and $\theta_{W}$ is the Weinberg angle.
The first term results in a spin-independent scattering from a nucleus
while the second term leads to a spin-dependent scattering from a
nucleus. It is convenient to consider the cross section with a single
nucleon for comparing with experiments. 

The DM-nucleus cross section in this case is\ 
\footnote{Note that Eq.\ (\ref{eq:fermion-xsec-SI}) does not correspond to the total DM-nucleus cross section, but to that at zero momentum transfer,
usually named as {}``standard cross section''.}
\begin{equation}
\sigma_{\chi N}^{SI}=\frac{m_{\chi}^{2}m_{N}^{2}b_{N}^{2}}{\pi\left(m_{\chi}+m_{N}\right)^{2}},\label{eq:fermion-xsec-SI}
\end{equation}
where $b_{N}$ is the effective DM-nucleus coupling. Through simple
algebra (see Appendix\ \ref{sub:Fermionic-DM-nucleon-interaction}
for details), we obtain the DM-proton and DM-neutron cross sections
as given below: 
\begin{eqnarray}
\sigma_{\chi p}^{SI} & \approx & \left(6.98\times10^{-5}\,{\rm pb}\right)\left(\frac{{\rm TeV}}{\Lambda}\right)^{4}\left(\alpha_{q\chi}^{R}-\alpha_{q\chi}^{L}+0.013\alpha_{\phi\chi}\right)^{2},\label{eq:fermion-proton-SI}\\
\sigma_{\chi n}^{SI} & \approx & \left(6.98\times10^{-5}\,{\rm pb}\right)\left(\frac{{\rm TeV}}{\Lambda}\right)^{4}\left(\alpha_{q\chi}^{R}-\alpha_{q\chi}^{L}-0.162\alpha_{\phi\chi}\right)^{2},\label{eq:fermion-neutron-SI}\end{eqnarray}
where $m_{\chi}\gg m_{p}\simeq m_{n}$ are used to derive the approximate
results. We note that $\sigma_{\chi p}^{SI}\simeq\sigma_{\chi n}^{SI}$
when either $\alpha_{q\chi}^{L}$ or $\alpha_{q\chi}^{R}$ is non-zero,
whereas $\sigma_{\chi n}^{SI}\gg\sigma_{\chi p}^{SI}$ when only $\alpha_{\phi\chi}$
presents. 

The spin-dependent DM-nuclei elastic scattering cross section can
be expressed as
\begin{equation}
\sigma_{\chi N}^{SD}\approx\frac{32m_{\chi}^{2}m_{N}^{2}}{\pi\left(m_{\chi}+m_{N}\right)^{2}}\left[\Lambda_{N}^{2}J\left(J+1\right)\right],\label{eq:fermion-xsec-SD}
\end{equation}
where $J$ is the total angular momentum of the nucleus and
 $\Lambda_{N}\left(\propto1/J\right)$
depends on the axial couplings of DM to the quarks. For odd-proton
nuclei the spin-dependent DM-nucleus cross section is mainly due
to the DM-proton interactions, whereas for odd-neutron nuclei it
is dominated by DM-neutron scattering. For even-even nuclei the
spin-dependent cross-section is highly suppressed.
For the proton/neutron as the target, Eq.\ (\ref{eq:fermion-xsec-SD}) is transformed
into the cross section from DM-proton/neutron interactions with
the proton/neutron spin, which are given as
\begin{eqnarray}
\sigma_{\chi p}^{SD} & \approx & \left(4.183\times10^{-6}\,{\rm pb}\right)\left(\frac{{\rm TeV}}{\Lambda}\right)^{4}\left(\alpha_{q\chi}^{R}+\alpha_{q\chi}^{L}-4.53\alpha_{\phi\chi}\right)^{2},\label{eq:fermion-proton-SD}\\
\sigma_{\chi n}^{SD} & \approx & \left(4.183\times10^{-6}\,{\rm pb}\right)\left(\frac{{\rm TeV}}{\Lambda}\right)^{4}\left(\alpha_{q\chi}^{R}+\alpha_{q\chi}^{L}+3.53\alpha_{\phi\chi}\right)^{2}.\label{eq:fermion-neutron-SD}
\end{eqnarray}
It indicates that for a given $\Lambda$ the contribution of $\oo_{\phi\chi}$
dominates over the other two operators.
Furthermore, we note that $\sigma_{\chi p}^{SD}\approx\sigma_{\chi n}^{SD}$ 

\begin{figure}
\includegraphics[clip,scale=0.5]{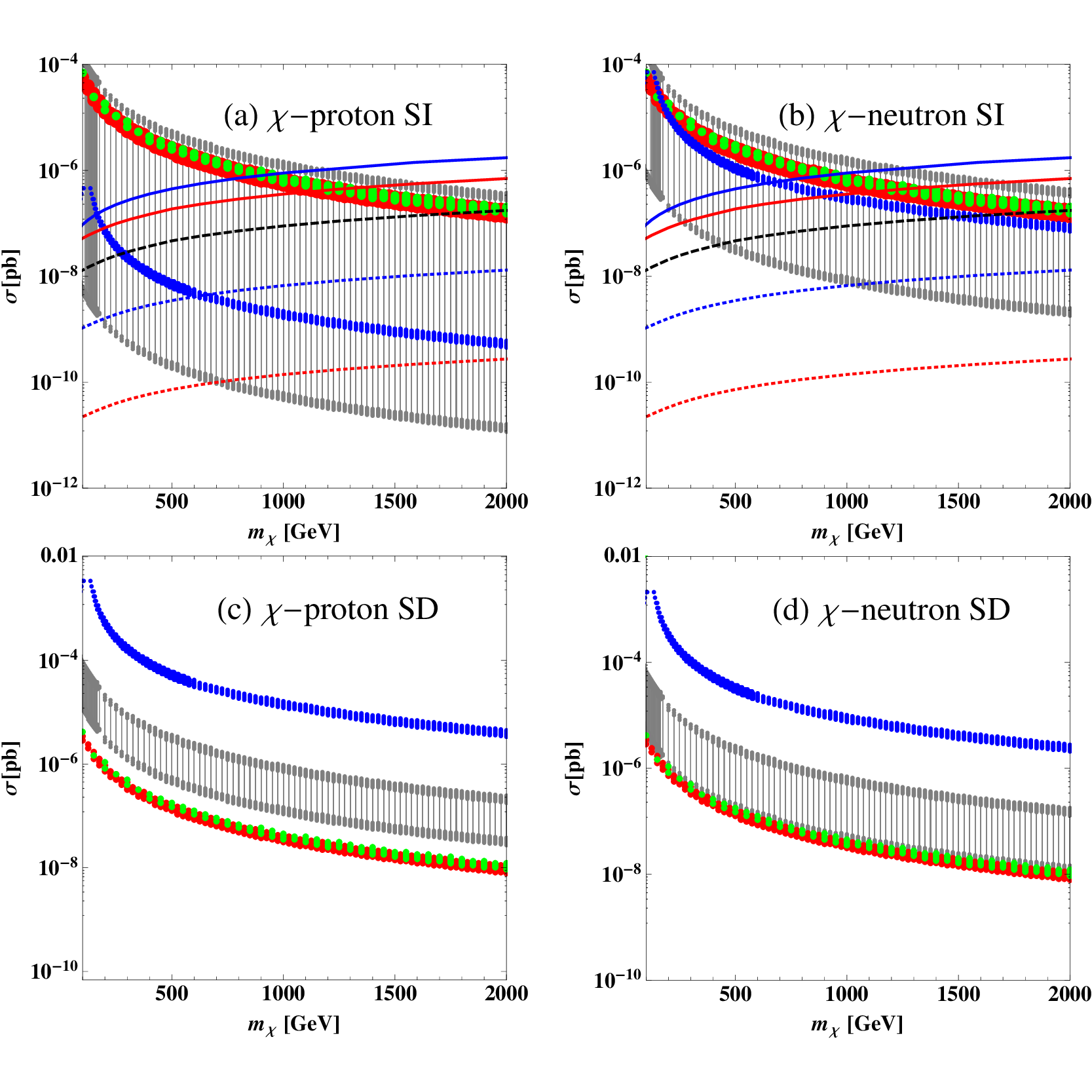}

\caption{
Prediction of the fermionic DM-nucleon cross sections with respect
to $m_{\chi}$ and $\Lambda$ for allowed parameter set given in Fig.\ \ref{fig:fermion};
(a, b) for the spin-independent DM search; (c, d) for the spin-dependent
DM search. The red (green, blue) band denotes the cross section for
$\alpha_{\chi q}^{L}=1$ ($\alpha_{\chi q}^{R}=1$, $\alpha_{\phi\chi}=1$),
respectively, when one operator is considered at a time. The gray
shaded region denotes the cross sections when all these three operators
contribute, which corresponds to the black band in Fig.\ \ref{fig:fermion};
see the text for details. In the upper panel, the blue-solid (red-solid)
line labels the upper limit ($90\%$ confidence level) on the SI DM-nucleon
scattering cross section from the current XENON10 (CDMS II 2009) direct
search, respectively. The black-dashed line denotes the near term
expected sensitivity from the CDMS II experiment, whereas the blue-dotted
(red-dotted) line represents the longer term projection for the Super-CDMS
Phase-A (Phase-C), respectively. \label{fig:fermion-SI-SD}}

\end{figure}

In Figs.\ \ref{fig:fermion-SI-SD}(a) and \ref{fig:fermion-SI-SD}(b),  we compare the spin-independent
elastic scattering cross section of the fermionic DM $\chi$ with the
current and projected sensitivities of direct detection experiments;
the blue-solid curve denotes the XENON10 exclusion limit~\cite{Angle:2007uj}
while the red curve shows the CDMS II 2009 exclusion limit\ \cite{Akerib:2005kh}.
In the near future, the CDMS II is projected to increase the sensitivities
of spin-independent scattering roughly by a factor of 4; see the black-dashed
curve. The projected sensitivities of SuperCDMS\ \cite{Akerib:2006rr},
whose first phase is proposed to start operating in 2011, are also
plotted; the blue-dashed line represents phase-A while the red-dashed
line phase-C. The red, green and blue bands represent the cross sections
for $\alpha_{f\chi}^{L}=1$, $\alpha_{f\chi}^{R}=1$ and $\alpha_{\phi\chi}=1$,
respectively, with the assumption of only one parameter being non-zero
at a time. We note that the bands of $\alpha_{f\chi}^{L}$ and $\alpha_{f\chi}^{R}$
overlap for the whole mass range we are interested, whereas the band
of $\alpha_{\phi\chi}$ is much higher than those two. It is consistent
with the relic abundance constraints shown in Fig.\ \ref{fig:fermion}.
The exclusion limit of $m_{\chi}$ (denoted by the symbol {}``\textgreater{}'')
and the potential reach of future experiments (denoted by {}``\textless{}'')
are summarized in Table\ \ref{tab:fermion}. If the recent CDMS II
observation is a hint that direct detection is {}``around the corner'',
the the dark matter mass should hide in between the CDMS II 2009 exclusion
limit and CDMS II projected sensitivity. Nevertheless, the entire
range of $m_{\chi}$ considered here can be covered by the Phase-C
of SuperCDMS.

\begin{table}
\caption{Exclusion limits of $m_{\chi}$ (GeV) from current SI direct search
(denoted by the symbol {}``\textgreater{}'') and potential reach
of future experiments (denoted by the symbol {}``\textless{}'')
when one parameter is considered at a time. The color of each operator
refers to Fig.~\ref{fig:fermion-SI-SD}.~\label{tab:fermion}}

\begin{tabular}{lcccccccccc}
\hline 
 & \ \ \  & XENON10 & \ \ \  & CDMS II 2009 & \ \ \  & CDMS II proj & \ \ \  & SuperCDMS-A & \ \ \  & SuperCDMS-C\tabularnewline
\hline 
$\alpha_{f\chi}^{L}$ (red) &  & $>750$ &  & $>1050$ &  & $<2000$ &  & $<2000$ &  & $<2000$\tabularnewline
$\alpha_{f\chi}^{R}$ (green) &  & $>800$ &  & $>1130$ &  & $<2000$ &  & $<2000$ &  & $<2000$\tabularnewline
$\alpha_{\phi\chi}$(blue) &  & $>600$ &  & $>870$ &  & $<1700$ &  & $<2000$ &  & $<2000$\tabularnewline
\hline
\end{tabular}
\end{table}

Considering now the case when all three operators contribute, for which
the scattering cross sections are shown as the broad gray bands. Although
the relic abundance imposes very tight constraints on $m_{\chi}$,
$\Lambda$ and the coefficients, see Eq.\ (\ref{eq:fermion-relation}),
the signs of the coefficients are not determined as the relic abundance
is only proportional to the square of the coefficients. On the contrary,
the scattering cross sections are sensitive to the signs of the coefficients;
see Eqs.\ (\ref{eq:fermion-proton-SI}-\ref{eq:fermion-neutron-SD}).
The upper limit of the $\chi$-proton SI scattering and the $\chi$-neutron
SI scattering corresponds to $\alpha_{f\chi}^{L}=-\alpha_{f\chi}^{R}=\alpha_{\phi\chi}=\pm1$
and $-\alpha_{f\chi}^{L}=\alpha_{f\chi}^{R}=\alpha_{\phi\chi}=\pm1$,
respectively. The lower limits of both scattering corresponds to $\alpha_{f\chi}^{L}=\alpha_{f\chi}^{R}$,
i.e., solely determined by the contribution of $\alpha_{\phi\chi}$.
The large difference between two lower limits, for which $\chi$-neutron
scattering is much larger than $\chi$-proton scattering, is attributed
to the the coefficients of $\alpha_{\phi\chi}$ in Eqs.\ (\ref{eq:fermion-proton-SI})\ and\ (\ref{eq:fermion-neutron-SI}).

For completeness, in Figs.\ \ref{fig:fermion-SI-SD}(c) and \ref{fig:fermion-SI-SD}(d) we also present
spin-dependent cross section which is obviously well below the present
and even future experimental reach. Picking out such a signal 
amongst the large range of noise sources is a considerable challenge.
See Ref.~\cite{Cao:2009uv} for a brief comment on the impact of spin-dependent
measurement on the PAMELA antiproton observation.

\subsection{Indirect gamma-ray detection of $\chi$}

Photons from dark matter annihilation in the center of the galaxy
also provide an indirect signal. Detection of cosmic gamma-rays is
carried out in both space-based and ground-based telescopes; the former
directly observe the cosmic gamma-rays and can cover  
the low energy regime (GeV), whereas the latter indirectly observe the gamma-rays
through the detection of secondary particles and the Cerenkov light
originating from their passage through the Earth's atmosphere and
will cover 
the high energy regime (100 GeV to TeV). 
 Thus, both types
of telescopes can be complementary to each other. The sensitivity
of present and future detectors in gamma-ray astrophysics 
has been studied
in Ref.\ \cite{Morselli:2002nw}. 

\begin{figure}
\includegraphics[scale=0.4]{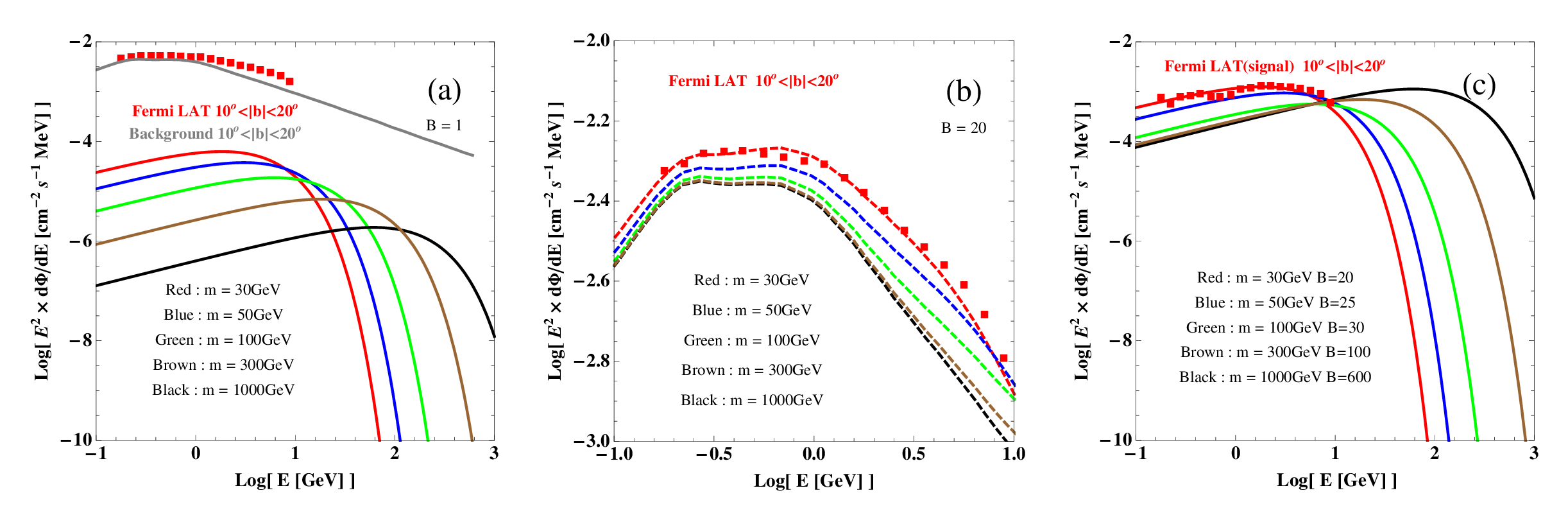}

\caption{(a) Predicted gamma-ray spectra for the annihilation of fermionic DM $\chi$ (solid
lines) with the NFW density profile. The Fermi LAT observation
(Galactic background) is also presented by the red-box (grey-solid line);
(b) Comparison between the DM signal plus background and the Fermi
LAT observation. Note that a common boost factor of 20 is applied to all the DM
signals; (c) The comparison between the DM signal with the difference of Fermi LAT observation and background.
The different boost factors are adopted for different DM masses 
so that the DM signal will not exceed the data.}
\label{fig:dflux-fermion}
\end{figure}

As the DM $\chi$ cannot annihilate into photon pairs directly, we
can only detect continuum photon signals. Using Eq.\ (\ref{eq:gamma-flux})
we obtain the differential flux of the gamma-rays observed on Earth from DM
annihilation as follows:
\begin{eqnarray}
\frac{d\Phi}{dE_{\gamma}} & \approx & \left(1.4\times10^{-11}{\rm s}^{-11}{\rm cm}^{-2}{\rm GeV}^{-1}\right)\bar{J}\left(\Delta\Omega\right)\Delta\Omega\nonumber \\
 & \times & \left(\frac{100\,{\rm GeV}}{m_{\chi}}\right)\left(\frac{{\rm TeV}}{\Lambda}\right)^{4}x^{-1.5}e^{-6.5x}\left(\alpha_{f\chi}^{L\,2}+\alpha_{f\chi}^{R\,2}+0.042e^{-2.26x}\alpha_{\phi\chi}^{2}\right),\label{eq:fermion-diff-flux}
\end{eqnarray}
where $x\equiv E_{\gamma}/m_{\chi}$ and $\bar{J}\left(\Delta\Omega\right)\Delta\Omega$
counts the dependence on the DM halo profile (see Appendix\ \ref{sec:Dark-matter-indirect}
for details). For simplicity, we assume a standard NFW density profile\ \cite{Navarro:1995iw,Navarro:1996gj}
for the DM in our galaxy, i.e. $\bar{J}\left(\Delta\Omega\right)\Delta\Omega\sim1$
for a $\Delta\Omega=10^{-3}\,{\rm sr}$ region around the direction
of the galactic center. The predicted differential gamma-ray fluxes from DM annihilation are plotted in Fig.\ \ref{fig:dflux-fermion}(a)
with the existing Fermi LAT observations (red box)\ \cite{FermiLAT1:2009,FermiLAT2:2009}
and the galactic background (gray-solid curve).
See Ref.~\cite{Barger:2009yt} for details of the galactic background and its uncertainties. 
For illustration, we choose $m_{\chi}$ as 30 (50, 100, 300, 1000) GeV with the corresponding
scale $\Lambda$  which is  derived for each mass from the relic 
abundance, cf. Fig.\ \ref{fig:fermion}. 
We only present the distribution of
$\alpha_{f\chi}^{L}=1$ in the figure as all three operators give rise
to almost the same distributions. It is because the shape of gamma-ray
spectrum is almost independent of the specific annihilation
process and the normalization is also fixed by the relic abundance.
Clearly, all flux distributions of the gamma-rays are consistent with
the current Fermi LAT observations 
since the dark matter signals are far below the observation and background.

%

However, the distribution of dark matter might be clumpy. Such small-scale structure would enhance 
DM signal by a \emph{boost factor} ($B$) defined in Eq.\ (\ref{eq:boost}). 
Due to large uncertainties of Galactic background estimations, even there is small discrepancy 
between signal and background, as shown in Fig.~\ref{fig:dflux-fermion}, one should not treat such small deviation too seriously. Instead,  one can extract upper bounds on the allowed boost factor for dark matter.
We vary the boost
factor such that the distributions of cosmic gamma-rays for the given
$m_{\chi}$ plus background satisfy the Fermi LAT measurement. When the coefficients of the dark operators are set to be 1, we found that a boost factor $B=20$  is allowed by the Fermi LAT data for the $30\,{\rm GeV}$ dark matter; see Figs.~\ref{fig:dflux-fermion}(b) and (c). However, it is in contradiction to the CDMS II observation when one interprets the two events as signal. 
For a heavier dark matter ($m_{\chi} > 30$ GeV), the allowed boost factor can be larger.
With a large boost factor, such a heavy dark matter would be detected by Fermi LAT with good sensitivity in higher energy region since the dark matter annihilation produces much harder energy spectrum of the cosmic gamma-ray; see Fig.~\ref{fig:dflux-fermion}(c).

\begin{figure}
\includegraphics[scale=0.7]{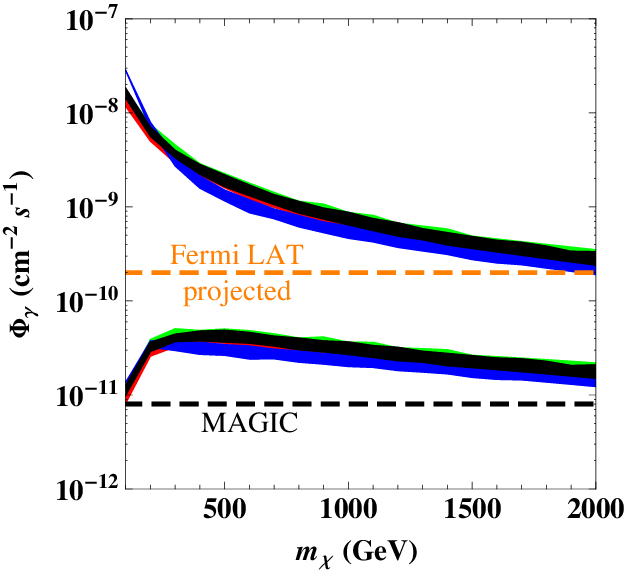}

\caption{Integrated photon flux as a function of $m_{\chi}$ for energy threshold
of $1\,{\rm GeV}$ (upper bands) and $50\,{\rm GeV}$  (lower bands). The notation of the color band
is the same as Fig.\ \ref{fig:fermion-SI-SD}). The plot assumes
$\bar{J}(\Psi,\Delta\Omega)\Delta\Omega=1$; all fluxes scale linearly
with this parameter, see the Appendix \ref{sec:Dark-matter-indirect} for details. \label{fig:integrateflux-fermion}}

\end{figure}

The integrated photon flux is also interesting. The integrated photon
flux above some photon energy threshold $E_{th}$ is given by Eq.\ (\ref{eq:integral-flux}).
We plot the integrated photon fluxes in Fig.\ \ref{fig:integrateflux-fermion}
as a function of $m_{\chi}$ for two representative $E_{th}$: $1\,{\rm GeV}$,
accessible to space-based detectors, and $50\,{\rm GeV}$, characteristic
of ground-based telescopes. Estimated sensitivities for two 
 promising experiments, Fermi LAT~\cite{Sadrozinski:2001wu}
and MAGIC II~\cite{Petry:1999fm,Baixeras:2004wa}, are also shown.
It is clear to see from the Fig.~\ref{fig:integrateflux-fermion} that the Fermi LAT and MAGIC II could probe a heavy $\chi$ if the background
is well understood.

\subsection{Collider search for $\chi$}

\begin{figure}
\includegraphics[scale=0.8]{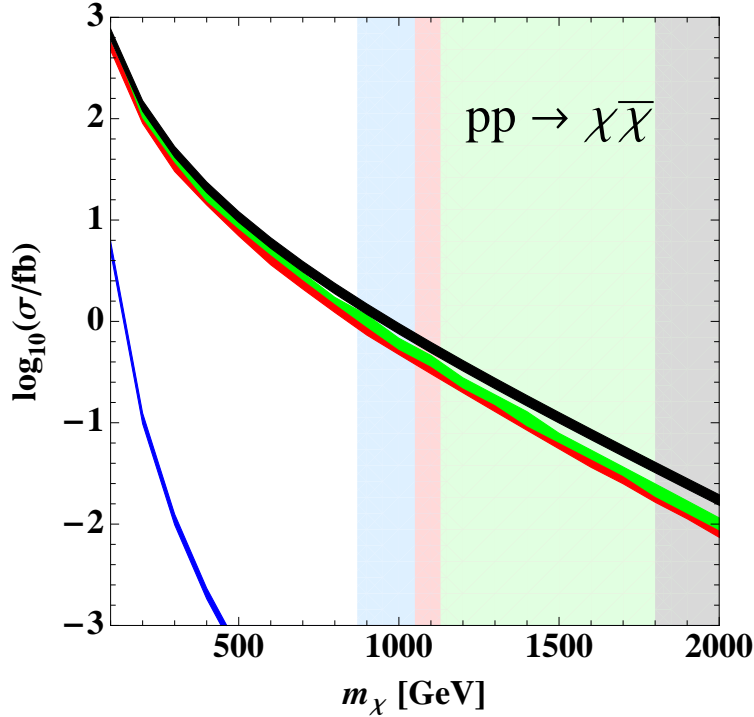}

\caption{Cross section of the $\chi\bar{\chi}$ pair production at the LHC.
The lower narrow band (blue) denotes the cross section for $\alpha_{\phi\chi}=1$
while the upper narrow bands (black, green and red) denote the cross
sections for $\alpha_{f\chi}^{L}=1$, $\alpha_{f\chi}^{R}=1$, or
$\alpha_{f\chi}^{L}=\alpha_{f\chi}^{R}=\alpha_{\phi\chi}=1$. The
shaded regions denote the preferred DM mass region by the CDMS II
positive signal; see Table~\ref{tab:fermion}. 
\label{fig:lhc-xsec-fermion0}}

\end{figure}

Since $x_{F}\approx25$ as pointed out above, one can substitute it
into the equation of the relic abundance Eq.\ (\ref{eq:omega}) and
obtain the proper thermally averaged annihilation cross section, which
gives rise to the correct relic abundance, as\ \cite{Peskin:2007nk}
\[
\left\langle \sigma_{ann}v_{rel}\right\rangle \approx1\,{\rm {\rm pb}}.\]
Inverting the annihilation process, we can produce the DM pair at
the LHC though the initial state only consists of light quarks.
The $\chi\bar{\chi}$ pair can be produced via the processes 
\begin{equation}
q\bar{q}\to\chi\bar{\chi},\qquad q\bar{q}\to Z\to\chi\bar{\chi},\qquad
\label{eq:lhc-qq2ff}
\end{equation}
Shown in Fig.\ \ref{fig:lhc-xsec-fermion0} is the cross section
of the $\chi\bar{\chi}$ pair production ($\sigma_{prod}$) as a function
of $m_{\chi}$. The upper narrow band denotes the cross section for
$\alpha_{f\chi}^{L}=1$ (red), or $\alpha_{f\chi}^{R}=1$ (green),
or $\alpha_{f\chi}^{L}=\alpha_{f\chi}^{R}=\alpha_{\phi\chi}=1$ (black),
whereas the blue narrow band below denotes the cross section for $\alpha_{\phi\chi}=1$.
The upper three narrow bands almost degenerate due to the relic abundance
constraints. We note that $\sigma_{prod}\sim1\,{\rm pb}$ for $m_{\chi}\sim100\,{\rm GeV}$,
but $\sigma_{prod}$ drops steeply with increasing $m_{\chi}$. 
The 
large suppression is due to two facts. The first is the decrease of coupling, since the $\Lambda$ gets larger.
The second reason is due to the parton distribution functions
(PDFs) which drop rapidly in the large Bjorken-$x$ region. Typically,
$\left\langle x\right\rangle \approx2m_{\chi}/\sqrt{s}$ with $\sqrt{s}=14\,{\rm TeV}$
for LHC and $\sqrt{s}=1.96\,{\rm TeV}$ for the Tevatron. Such a large
suppression restricts the DM search at the LHC to small $m_{\chi}$.
The $\alpha_{\phi\chi}$ operator suffers from a much severer suppression
as it contributes only via the $Z$-boson-exchange process, i.e. the
second term in Eq.\ (\ref{eq:lhc-qq2ff}). For a heavy DM pair, the
off-shell $Z$ boson in the propagator receives a large $1/s$ suppression
which makes the cross section drop much faster than other bands. The
shaded blue (green, red, gray) region denotes the preferred dark matter
region for $\alpha_{\phi\chi}$ ($\alpha_{f\chi}^{R}$, $\alpha_{f\chi}^{L}$,
$\alpha_{f\chi}^{L}=\alpha_{f\chi}^{R}=\alpha_{\phi\chi}=1$) by the
CDMS II positive signal.

\begin{figure}
\includegraphics[clip,scale=0.65]{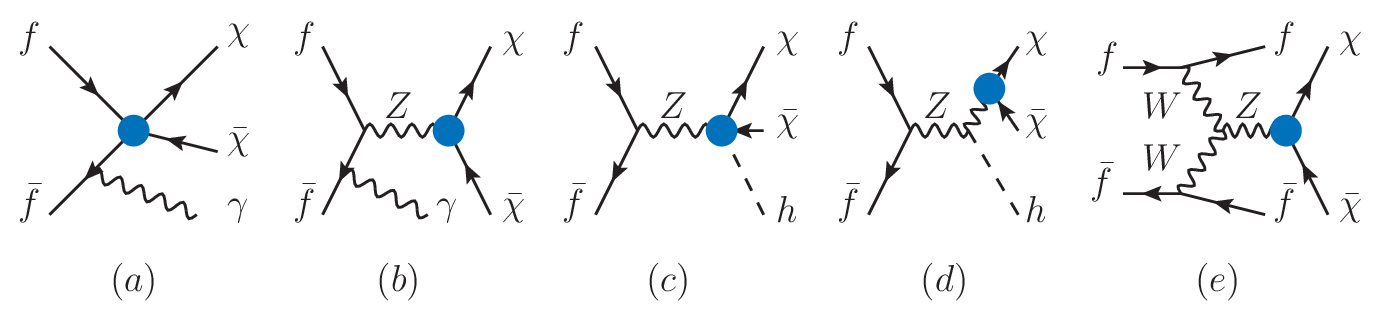}

\caption{Schematic diagrams of the searching channels of DM $\chi$ at the
colliders.
\label{fig:fermion-lhc-feyn}}

\end{figure}

In order to detect the DM signal, additional SM particles are needed.
A hard photon is a good probe which can be produced in association
with the DM pair via the following processes
\begin{equation}
q\bar{q}\to\gamma\chi\bar{\chi},\qquad q\bar{q}\to Z\gamma\to\chi\bar{\chi}\gamma,
\end{equation}
as shown in Fig.\ \ref{fig:fermion-lhc-feyn}(a)\ and\ (b), giving
rise to a collider signature of mono-photon plus large missing transverse
momentum ($\gamma+\met$). Similarly, one can also search for the
DM $\chi$ using the signature of mono-jet plus missing transverse
momentum ($j+\met$) which involves the processes as follows: 
\begin{eqnarray}
q\bar{q}\to\chi\bar{\chi}g, & \qquad & q\bar{q}\to Zg\to\chi\bar{\chi}g,\nonumber\\
qg\to\chi\bar{\chi}q, &  & qg\to Zq\to\chi\bar{\chi}q.\label{eq:lhc-fermion-mono-j}
\end{eqnarray}
Furthermore, the $\chi\bar{\chi}$ pair can also be produced associated
with a Higgs boson and through the W-boson fusion process, i.e. 
\begin{equation}
q\bar{q}\to Z\to\chi\bar{\chi}h,\qquad q\bar{q}\to Zh\to\chi\bar{\chi}h,\qquad
q q^{\prime} \to W^{+*}W^{-*} q q^{\prime}\to Z^{*} q q^{\prime} \to \chi \bar{\chi} q q^{\prime},
\label{eq:lhc-fermion-hZ}
\end{equation}
as shown in Fig\@.\ \ref{fig:fermion-lhc-feyn}(c), (d) and (e). 
Unfortunately, the cross sections of such processes Eq.\ (\ref{eq:lhc-fermion-hZ})
are too small to be detected. We then focus our attention on the $\gamma+\met$
and $j+\met$ signatures below. 

\begin{figure}
\includegraphics[clip,scale=0.6]{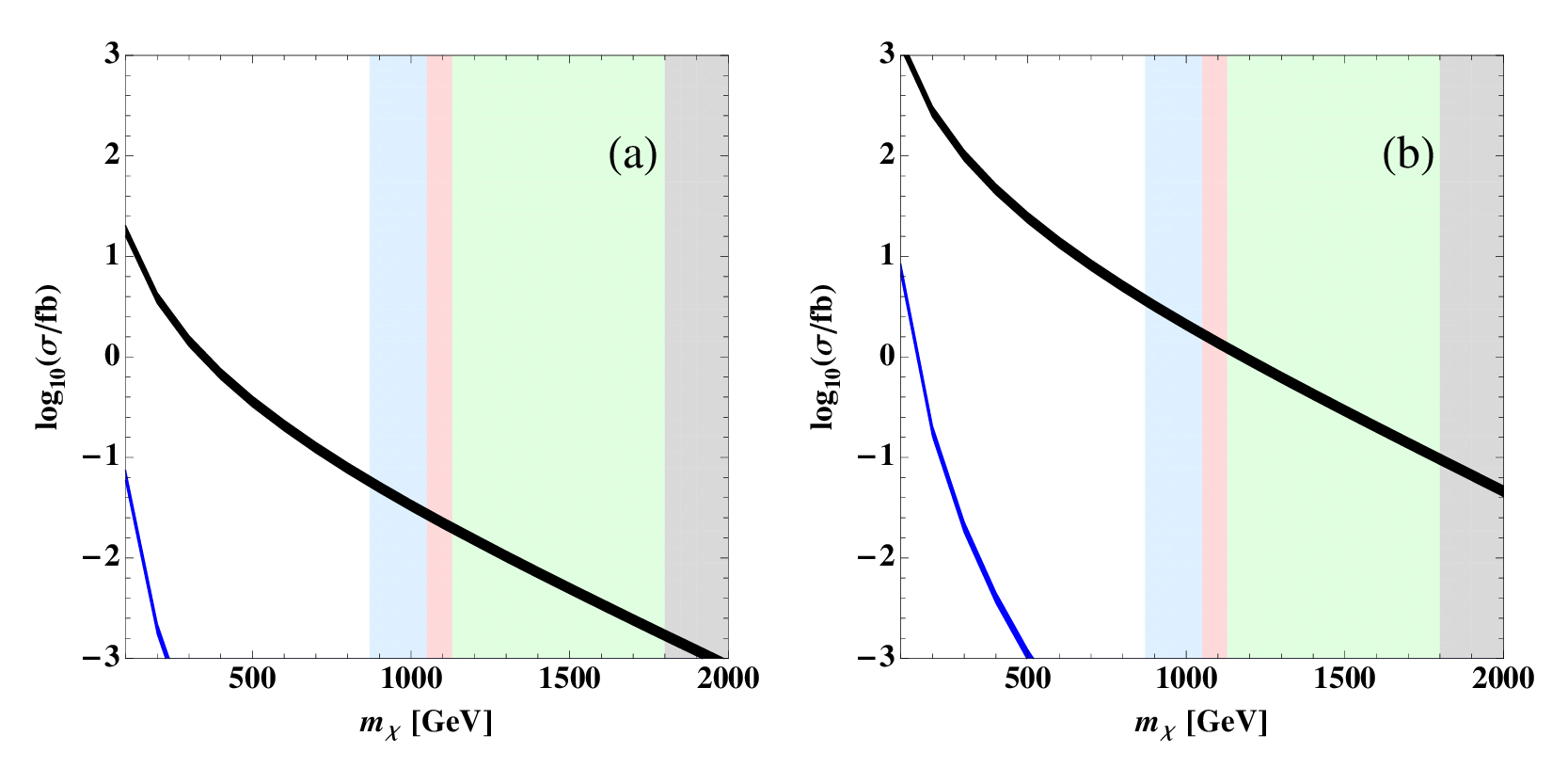}

\caption{Production cross sections (in the unit of fb) of the $\chi$ pair
together with a SM particle at the LHC: (a) mono-photon plus $\met$,
(b) mono-jet plus $\met$, where a kinematics cut $p_{T}^{\gamma}>5\,{\rm GeV}$
(or $p_{T}^{j}>5\,{\rm GeV}$) is imposed in order to avoid the collinear
singularities arising from the light quark propagators. The lower
band (blue) denotes the cross section for $\alpha_{\phi\chi}=1$ while
the upper bands (black, green and red) denote the cross sections for
$\alpha_{f\chi}^{L}=1$, or $\alpha_{f\chi}^{R}=1$, or $\alpha_{f\chi}^{L}=\alpha_{f\chi}^{R}=\alpha_{\phi\chi}=1$.
The shaded regions denote the preferred DM mass region by the CDMS
II positive signal; see Table~\ref{tab:fermion}. \label{fig:lhc-xsec-fermion}
}
\end{figure}

Search for new physics using the collider signatures of $\gamma+\met$ and $j+\met$ 
has been carried out at the Tevatron Run-I
($\sqrt{s}=1.8\,{\rm TeV}$) and Run II ($\sqrt{s}=1.96\,{\rm TeV}$ ), which set upper
bounds on the production cross sections of above processes~\cite{Abazov:2003gp,Aaltonen:2008hh}. For the
parameter space given in Fig.\ \ref{fig:fermion} the production
cross section is below the Tevatron direct search bound and is not
shown here. We plot the relevant cross sections at the LHC in Fig.\ \ref{fig:lhc-xsec-fermion}:
(a) $\gamma+\met$ and (b) $j+\met$. In Fig.\ \ref{fig:lhc-xsec-fermion}(b)
we have summed over the processes shown in Eq.\ (\ref{eq:lhc-fermion-mono-j}). 
Similar
to Fig.~\ref{fig:lhc-xsec-fermion0}, the lower blue narrow band
denotes the cross section for $\alpha_{\phi\chi}=1$ while the upper
narrow bands denote the cross sections for $\alpha_{f\chi}^{L}=1$
(red), $\alpha_{f\chi}^{R}=1$ (green) and $\alpha_{f\chi}^{L}=\alpha_{f\chi}^{R}=\alpha_{\phi\chi}=1$
(black), respectively. In order to avoid the collinear singularities
arising from the light quark propagator, we impose a kinematical cut
on the transverse momentum ($p_{T}$) of the final state photon or
jet as $p_{T}^{\gamma(j)}\geq5\,{\rm GeV}$. The shaded regions denote
the preferred DM mass region by the CDMS II positive signal (see Table~\ref{tab:fermion}).

Besides the PDF suppression, the production of the DM pair associated
with one single photon (or with one single jet) is suppressed by two
additional factors: (i) additional coupling ($\alpha_{EM}$ or $\alpha_{s}$)
and (ii) large volume of three-body phase space. We note that the
cross sections of the mono-photon production are much smaller than
the ones of the mono-jet production. It is because the former only
comes from the quark-quark initial state while the latter comes from
both the quark-quark initial state and the quark-gluon initial state.
Although having a smaller cross section, the $\gamma+\met$ process
has a relatively clean signature, i.e. hard photon, which could be
detected at the LHC. On the other hand, the mono-jet process suffers
from huge QCD backgrounds which makes its detection more challenging.
Furthermore, the PDFs receive large uncertainties in the large DM
mass region, i.e., the large $x$ region. Such large uncertainties
will further make the detection of the signal event intricate. 

The blue narrow band, i.e. the cross section for $\alpha_{\phi\chi}=1$,
is much lower than other three narrow bands because $\alpha_{\phi\chi}$
only contributes via the Feynman diagram (b) in Fig.\ \ref{fig:fermion-lhc-feyn}.
Bearing in mind that 
Higgs-associated production is also rare, we then conclude
that it is hard to observe the \emph{sole} $\alpha_{\phi\chi}$ contribution
at the LHC, i.e. one can probe $\alpha_{\phi\chi}$
only via the cosmology observations. For the other operators, the
positive CDMS II signal prefers the DM mass
$m_{\chi}\gtrsim  900$ GeV,
where the mono-photon and mono-jet production cross sections are of
order ${\cal O}(10^{-2})$ fb and ${\cal O}(1)$ fb, respectively.  It is plausible to cross-check 
the cosmological and collider measurements when the good knowledge of the SM backgrounds
and detector sensitivity at the LHC is achieved.
For example, see Refs.~\cite{Goodman:2010yf,Goodman:2010ku,Bai:2010hh} for a detailed study of bounds 
from existing Tevatron searches for monojets as well as expected LHC reaches for a discovery.

\section{Scalar DM $\varphi$\ \label{sec:Scalar}}

Let's now consider the scalar type dark matter. After adding a $Z_{2}$-odd
real scalar singlet $\varphi$ into the SM, we obtain the effective operators
as follows:
\begin{itemize}
\item scalar only\begin{eqnarray}
\oo_{\phi1}=\frac{1}{2}\left(\phi^{+}\phi\right)^{2}\left(\varphi\varphi\right), & \qquad & \oo_{\phi2}=\partial_{\mu}\left(\phi^{+}\phi\right)\partial^{\mu}\left(\varphi\varphi\right).\label{eq:scalar:scalaronly}\end{eqnarray}

\item scalar-vector\begin{equation}
\oo_{\phi3}=\left(\varphi\varphi\right)\left(D_{\mu}\phi^{+}D^{\mu}\phi\right)\label{eq:scalar:scalar-vector}\end{equation}

\item scalar fermion\begin{eqnarray}
\oo_{e\phi}=\left(\varphi\varphi\right)\left(\bar{\ell}e\phi\right), & \qquad & \oo_{u\phi}=\left(\varphi\varphi\right)\left(\bar{q}u\tilde{\phi}\right),\nonumber \\
\oo_{d\phi}=\left(\varphi\varphi\right)\left(\bar{q}d\phi\right).\label{eq:scalar:scalar-fermion}\end{eqnarray}

\end{itemize}
Besides the operators listed above, there exists dimension-4 operator
$\varphi\varphi\phi^{\dagger}\phi$ and dimension-5 operator $\varphi\varphi\bar{f}f$
which has been studied in Ref.~\cite{Cao:2007fy}. There are also
a few loop-induced operators which are suppressed by factors of $1/(16\pi^{2})$.
In this work we only consider the tree-level induced operator, but
it is worthy mentioning that those loop-induced operators, e.g. $\varphi\varphi B^{\mu\nu}B_{\mu\nu}$,
are also very interesting as they may provide a spectral line feature
observable in indirect detection experiments if their contributions
dominate in dark matter annihilation\ \cite{Jungman:1995df,LopezHonorez:2006gr,Bertone:2009cb,Jackson:2009kg}.

These operators induce the following DM annihilation processes, \[
\varphi\varphi\to hh,\,\,\varphi\varphi\to VV,\,\,{\rm and}\,\,\varphi\varphi\to f\bar{f},\]
where $h(V,f)$ denotes the SM Higgs boson (gauge boson, fermion).
The vertex of the DM annihilating into the Higgs scalars
is \begin{equation}
\mathcal{L}_{\varphi\varphi hh}^{(6)}=\frac{1}{\Lambda^{2}}g_{\varphi\varphi hh}\,\varphi\varphi hh,\label{eq:scalar:sshh}\end{equation}
where\begin{equation}
g_{\varphi\varphi hh}=\frac{3v^{2}}{4}\alpha_{\phi1}+\frac{s}{4}\left(4\alpha_{\phi2}+\alpha_{\phi3}\right)\label{eq:scalar:sshh:coupling}\end{equation}
with $s$ being the square of the system energy. The vertices of the
DM annihilating into the vector bosons are summarized as follows,\begin{equation}
\mathcal{L}_{\varphi\varphi VV}^{(6)}=\frac{1}{\Lambda^{2}}\left(\alpha_{\phi3}\, m_{W}^{2}\;\varphi\varphi W_{\mu}^{+}W^{-\mu}+\frac{1}{2}\alpha_{\phi3}\, m_{Z}^{2}\;\varphi\varphi Z_{\mu}Z^{\mu}\right).\label{eq:scalar:ssvv}\end{equation}
The vertices of the DM annihilating into the fermions are \begin{equation}
\mathcal{L}_{ff}^{\varphi}=\frac{1}{\Lambda^{2}}\varphi\varphi\left(\frac{v}{\sqrt{2}}\alpha_{e\phi}\bar{e}_{L}e_{R}+\frac{v}{\sqrt{2}}\alpha_{u\phi}\bar{u}_{L}u_{R}+\frac{v}{\sqrt{2}}\alpha_{d\phi}\bar{d}_{L}d_{R}\right).\label{eq:scalar:ssff}\end{equation}
In this work we consider two possible scenarios of the $\varphi\varphi f\bar{f}$
operators:
\begin{description}
\item[(A)] Universal $\alpha_{f\phi}$, i.e. $\alpha_{e\phi}=\alpha_{u\phi}=\alpha_{d\phi}.$
As to be shown later, such a case is ruled out by the current DM direct
search 
experiments. 
\item[(B)] The coefficients are proportional to the fermion mass, i.e. $\alpha_{f\phi}v=m_{f}\alpha_{f\phi}^{\prime}$,
and $\alpha_{f\phi}^{\prime}$ is universal for leptons and quarks,
i.e. $\alpha_{e\phi}^{\prime}=\alpha_{u\phi}^{\prime}=\alpha_{d\phi}^{\prime}$.
For simplicity, we will use $\alpha_{f\phi}$ instead of $\alpha_{f\phi}^{\prime}$
hereafter. 
\end{description}
Finally, the operator $\oo_{\phi1}$ induces the vertex $\varphi\varphi h$
\begin{equation}
\mathcal{L}_{\varphi\varphi h}^{(6)}=\frac{\alpha_{\phi1}}{\Lambda^{2}}\frac{v^{3}}{2}\varphi\varphi h,\label{eq:scalar-ssh}\end{equation}
which contributes to the processes of DM annihilation into the SM
Higgs bosons, vector bosons and fermions through the Higgs-mediated
$s$-channel processes. In summary, there are four operators contributing
to the scalar DM $\varphi$ annihilation: $\oo_{\phi1}$, $\oo_{\phi2}$,
$\oo_{\phi3}$ and $\oo_{f\phi}$.

\begin{figure}
\includegraphics[clip,scale=0.6]{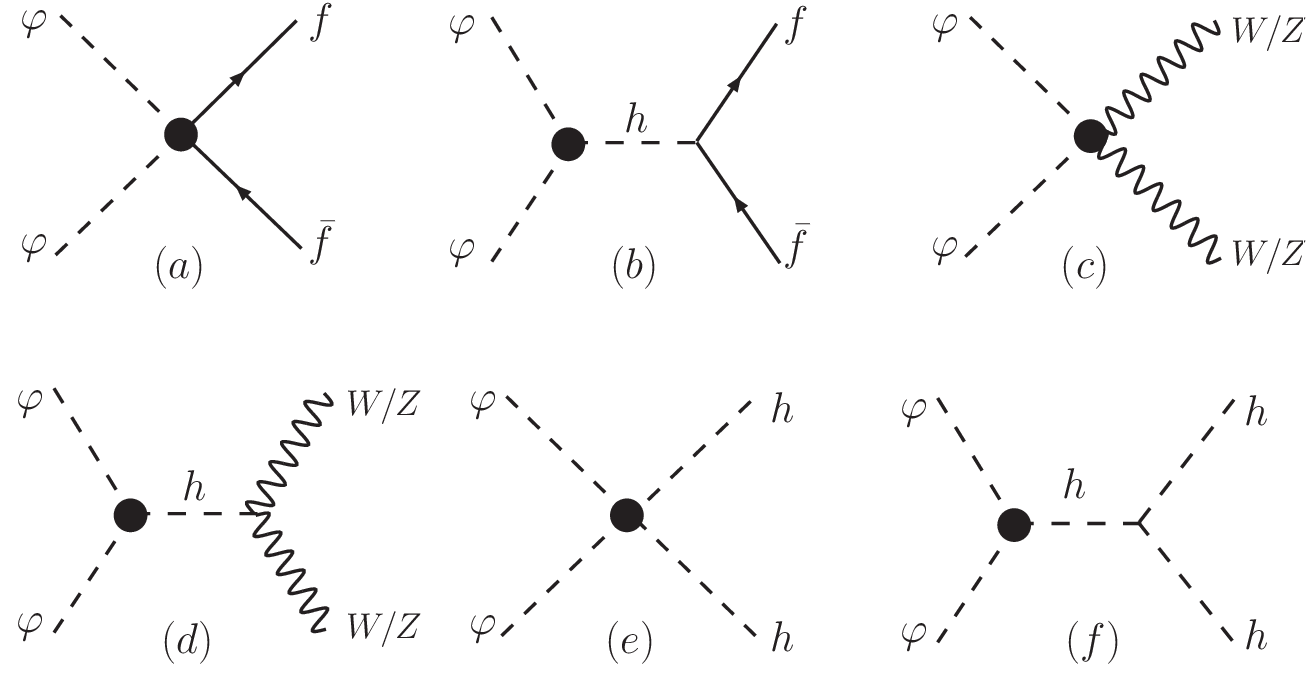}

\caption{Feynman diagrams for $\varphi\varphi$ annihilation. Blobs denote
the effective vertices induced by the dim-6 operators. \label{fig:feyn-ss}}

\end{figure}

\begin{table}
\caption{Sensitivities of various experiments to the dark operators where the
round brackets denote the Feynman diagrams in Fig.\ \ref{fig:feyn-ss}
while the square brackets denotes the dark operators given in Eqs.\ (\ref{eq:scalar:scalaronly}-\ref{eq:scalar-ssh}).\ \label{tab:sensitivities-scalar}}

\begin{tabular}{clllllllllll}
\hline 
 & $\Omega_{CDM}h^{2}$ & $\qquad$ & $\chi$-nucleon & $\qquad$ & cosmic $\gamma$-ray & $\qquad$ & $\gamma+\met$ & $\qquad$ & $j+\met$ & $\qquad$ & VBF\tabularnewline
\hline 
 & (a) {[}$\alpha_{f\phi}${]} &  & (a) {[}$\alpha_{f\phi}${]} &  & (a) {[}$\alpha_{f\phi}${]} &  & (a) {[}$\alpha_{f\phi}${]} &  & (a) {[}$\alpha_{f\phi}${]} &  & (c) {[}$\alpha_{\phi3}${]}\tabularnewline
 & (c) {[}$\alpha_{\phi3}${]} &  & (b) {[}$\alpha_{\phi1}${]} &  & (c) {[}$\alpha_{\phi3}${]} &  &  &  &  &  & (d) {[}$\alpha_{\phi1}${]}\tabularnewline
 & (e) {[}$\alpha_{\phi1,\phi2,\phi3}${]} &  &  &  & (e) {[}$\alpha_{\phi1,\phi2,\phi3}${]} &  &  &  &  &  & \tabularnewline
\hline
\end{tabular}
\end{table}

When the scalar $\varphi$ is the DM, we must consider a large number
of annihilation processes, e.g. $\varphi\varphi\to f\bar{f}/WW/ZZ/hh$
and $\varphi\varphi\to h\to f\bar{f}/WW/ZZ/hh$, see Fig.\ \ref{fig:feyn-ss}.
Sensitivities of these diagrams on various experiments are summarized
in Table\ \ref{tab:sensitivities-scalar}. The contributions from
the diagrams (b,d,f) are large only around the Higgs resonance region,
i.e. $m_{\varphi}\sim m_{h}/2$, but for the annihilation of heavy
$\varphi$ pair their contributions are highly suppressed. As we are
interested in the region of $m_{\varphi}>100\,{\rm GeV}$, only the
diagrams (a,c,e) need to be considered. For the elastic scattering
from the nucleus, only diagrams (a) and (b) contribute, whereas for
the cosmic gamma-ray detection, only diagrams (a,c,e) contribute.
Moreover, diagrams (a,b) can be probed at the LHC using the signature
of mono-photon plus missing energy while diagrams (c,d) can be probed
in the vector-boson-fusion process $q\bar{q}\to VVq\bar{q}\to\varphi\varphi q\bar{q}$.
Diagrams (e,f) cannot be probed at the LHC.

\subsection{Relic abundance}

\begin{figure}
\includegraphics[clip,scale=0.6]{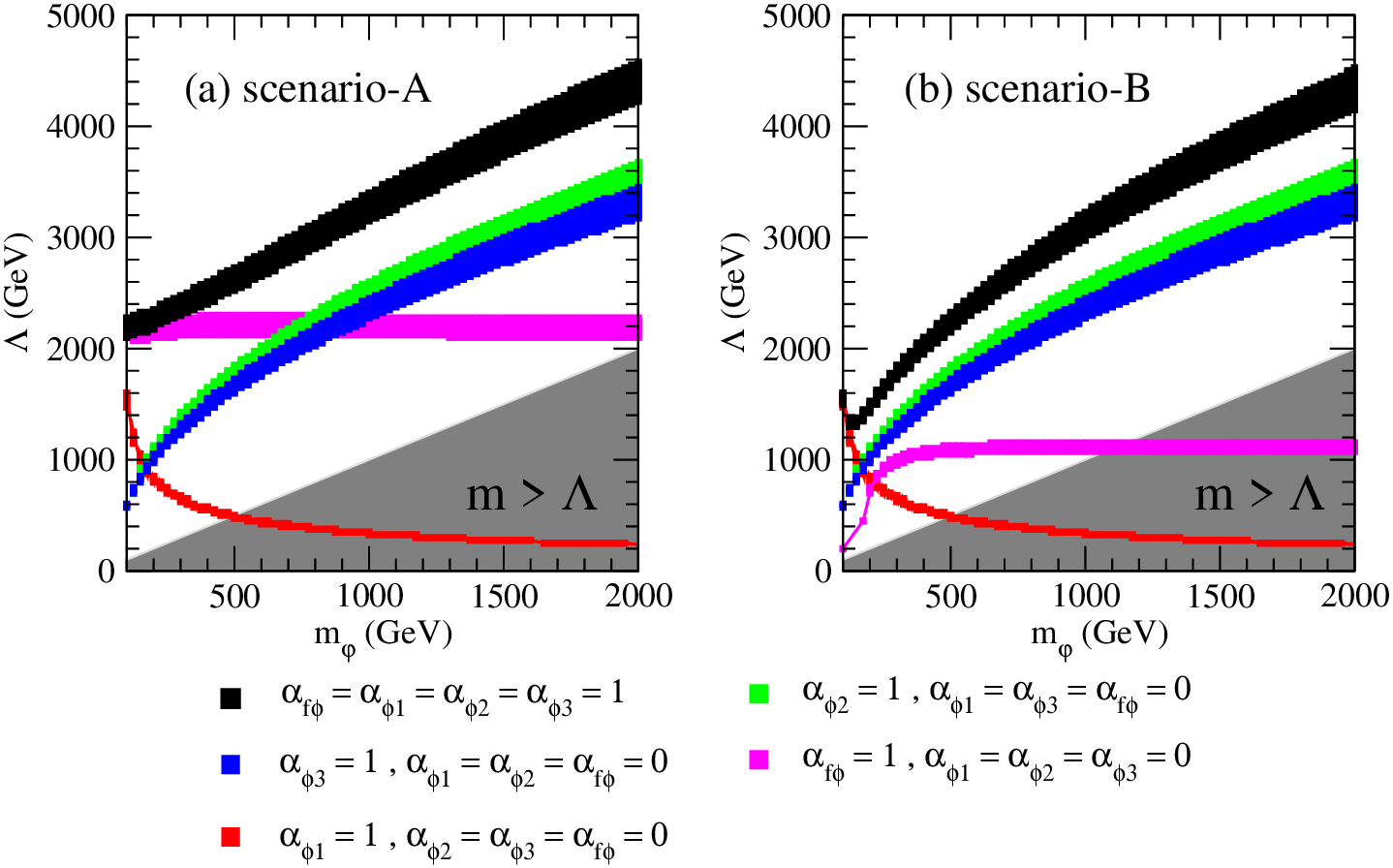}

\caption{(a) Allowed parameter set $\left(m_{\varphi},\Lambda\right)$ for
the scalar DM $\varphi$: (a) scenario-A (family universal coupling);
(b) scenario-B (family dependent coupling). The gray shaded region
is excluded as $m_{\varphi}>\Lambda$. The upper (lower) boundary
of each band corresponds to the upper (lower) limit of $\Omega_{DM}h^{2}$
given in Eq.\ (\ref{eq:omega_WMAP}). \label{fig:scalar}}

\end{figure}

The scalar DM can annihilate into light fermion pairs, or $WW$ and
$ZZ$ pairs if the vector boson channels open. The calculation of
the relic abundance is similar to the one of fermion DM $\chi$. We
present the annihilation cross sections and the leading terms in the
non-relativistic expansion in Appendix\ \ref{sec:Dark-matter-annihilation}.
The allowed parameter space of ($m_{\varphi}$, $\Lambda$) is shown
in Fig.\ \ref{fig:scalar} for both scenario-A and scenario-B. 

If the DM annihilation is purely induced by the operator $\oo_{\phi1}$, the scale $\Lambda$ decreases rapidly with increasing $m_{\varphi}$
in order to respect the WMAP data; see the red band. It can be easily
understood from dimensional counting as follows. For a heavy $m_{\varphi}$
the effective vertices induced by $\oo_{\phi1}$ give rise to the
leading terms $a$ and $b$ as following; see Eqs.\ (\ref{eq:dump1}-\ref{eq:dump4}),
\begin{equation}
a,\, b\propto\frac{v^{4}}{m_{\varphi}^{2}\Lambda^{4}}.\end{equation}
Since $a$ and $b$ are fixed by the well-measured relic abundance,
\begin{equation}
\Lambda\propto m_{\varphi}^{-1/2},\end{equation}
which results in the inverse behavior of the red band. Moreover, $m_{\varphi}$
is about equal to the cutoff scale $\Lambda$ around $500\,{\rm GeV}$.
We thus conclude that if $\oo_{\phi1}$ is the only source for DM
annihilation, then $\varphi$ should be very light, say $m_{\varphi}<500\,{\rm GeV}$.
Otherwise, other dark operators have to be considered to explain the
relic abundance. 

The bound on the operator $\oo_{f\phi}$ is not sensitive to $m_{\varphi}$;
see the magenta bands. It can also be understood from dimensional
counting. For a heavy $\varphi$, the effective vertices induced by
$\oo_{f\phi}$ give rise to the following leading terms; see Eqs.\ (\ref{eq:ssff-dump1}-\ref{eq:ssff-dump4}):
\begin{equation}
a,\, b\propto\begin{cases}
{\displaystyle \frac{v^{2}}{\Lambda^{4}}}, & \qquad({\rm scenario-A}),\\
{\displaystyle \frac{m_{f}^{2}}{\Lambda^{4}}}, & \qquad({\rm scenario-B}),\end{cases}\end{equation}
therefore, the relic abundance only depends on the scale $\Lambda$
for both scenarios. We found that for all the mass region concerned,
$\Lambda\simeq2\,{\rm TeV}$ in the scenario-A, whereas $\Lambda\simeq1.1\,{\rm TeV}$
for scenario-B. Again in order to satisfy the condition 
$m_{\varphi}<\Lambda$,
$\varphi$ cannot be very heavy. For example, $m_{\varphi}$ should
be less than $1\,{\rm TeV}$ for scenario-B. 

Let's consider now both $\oo_{\phi2}$ and $\oo_{\phi3}$ simultaneously,
 an interference between them occurs. From the annihilation cross
sections given in Eqs.\ (\ref{eq:xsec-ssvv})\ and\ (\ref{eq:xsec-sshh}),
we derive the coefficients in the thermal average taking $m\gg v$ and find\begin{eqnarray}
a & = & \frac{m_{\varphi}^{2}}{64\pi\Lambda^{4}}\left(16\alpha_{\phi2}^{2}+8\alpha_{\phi2}\alpha_{\phi3}+13\alpha_{\phi3}^{2}\right),\\
b & = & \frac{m_{\varphi}^{2}}{256\pi\Lambda^{4}}\left(16\alpha_{\phi2}^{2}+8\alpha_{\phi2}\alpha_{\phi3}+7\alpha_{\phi3}^{2}\right).\end{eqnarray}
The allowed parameter set $(m_{\varphi},\,\Lambda)$ for each operator
is shown in Fig.\ \ref{fig:scalar}; see the green band ($\alpha_{\phi2}$)
and the blue band ($\alpha_{\phi3}$). Similar to the case of fermionic
DM, we find that $x_{F}$ barely varies in the entire allowed parameter
space, $x_{F}\simeq25$, but the freeze-out temperature $T_{F}$ varies
from $20-80\,{\rm GeV}$ for $m_{\varphi}\sim500-2000\,{\rm GeV}$.
Choosing $x_{F}=25$, we obtain the following relation between $m_{\varphi}$
and $\Lambda$,\begin{equation}
\frac{6.98\times10^{-2}}{\left(\Omega_{CDM}h^{2}\right)\left(0.273\alpha_{\phi2}^{2}+0.136\alpha_{\phi2}\alpha_{\phi3}+0.213\alpha_{\phi3}^{2}\right)}=\left(\frac{m_{\varphi}}{100\,{\rm GeV}}\right)^{2}\left(\frac{{\rm TeV}}{\Lambda}\right)^{4}.\label{eq:scalar-relation}\end{equation}

Finally, we consider the case of all the operators contributing to
the DM annihilation; see the black band. It is clear that for a heavy
$\varphi$ the DM annihilation is dominated by $\oo_{\phi2}$ and
$\oo_{\phi3}$ given by Eq.\ (\ref{eq:scalar-relation}). Note that
the sign of the coefficients $\alpha_{\phi2}$ and $\alpha_{\phi3}$
is very important. For simplicity, we only consider the case of constructive
interference, i.e., both $\alpha_{\phi2}$ and $\alpha_{\phi3}$ being
positive, in this study. But one can easily get the result of destructive
interference by lowering the black band by $\sim25\%$. For a light
$\varphi$, however,  the DM annihilation is dominated by
$\oo_{f\phi}$ in scenario-A while by $\oo_{\phi1}$ in scenario-B.

\subsection{Direct Search}

When the DM is a scalar $\varphi$, it can be detected in the spin-independent
experiments via the scalar interaction with nucleus 
\begin{equation}
\mathcal{L}_{\varphi\varphi qq}=\frac{1}{\Lambda^{2}}\frac{\alpha_{f\phi}v}{\sqrt{2}}\varphi\varphi\bar{q}q+\frac{1}{\Lambda^{2}}\frac{\alpha_{\phi1}v^{2}}{2}\frac{m_{q}}{m_{h}^{2}}\varphi\varphi\bar{q}q,\label{eq:SDM-nucleon-interaction}
\end{equation}
which leads to a total spin-independent $\varphi$-nucleon cross section
as follows:
\begin{equation}
\sigma_{\varphi p}^{SI}=\frac{m_{p}^{2}}{4\pi\left(m_{\varphi}+m_{p}\right)^{2}}\left[f_{\varphi p}^{(p)}\right]^{2},\qquad\sigma_{\varphi p}^{SI}=\frac{m_{n}^{2}}{4\pi\left(m_{\varphi}+m_{n}\right)^{2}}\left[f_{\varphi n}^{(n)}\right]^{2},
\end{equation}
where the effective $\varphi$-nucleon couplings, $f_{\varphi P}^{(p)}$
and $f_{\varphi n}^{(n)}$, are given in Appendix\ \ref{sub:Scalar-DM-nucleon-interaction}.
In practice, we found $\sigma_{\varphi p}^{SI}\simeq\sigma_{\varphi n}^{SI}$. 

\begin{figure}
\includegraphics[clip,scale=0.4]{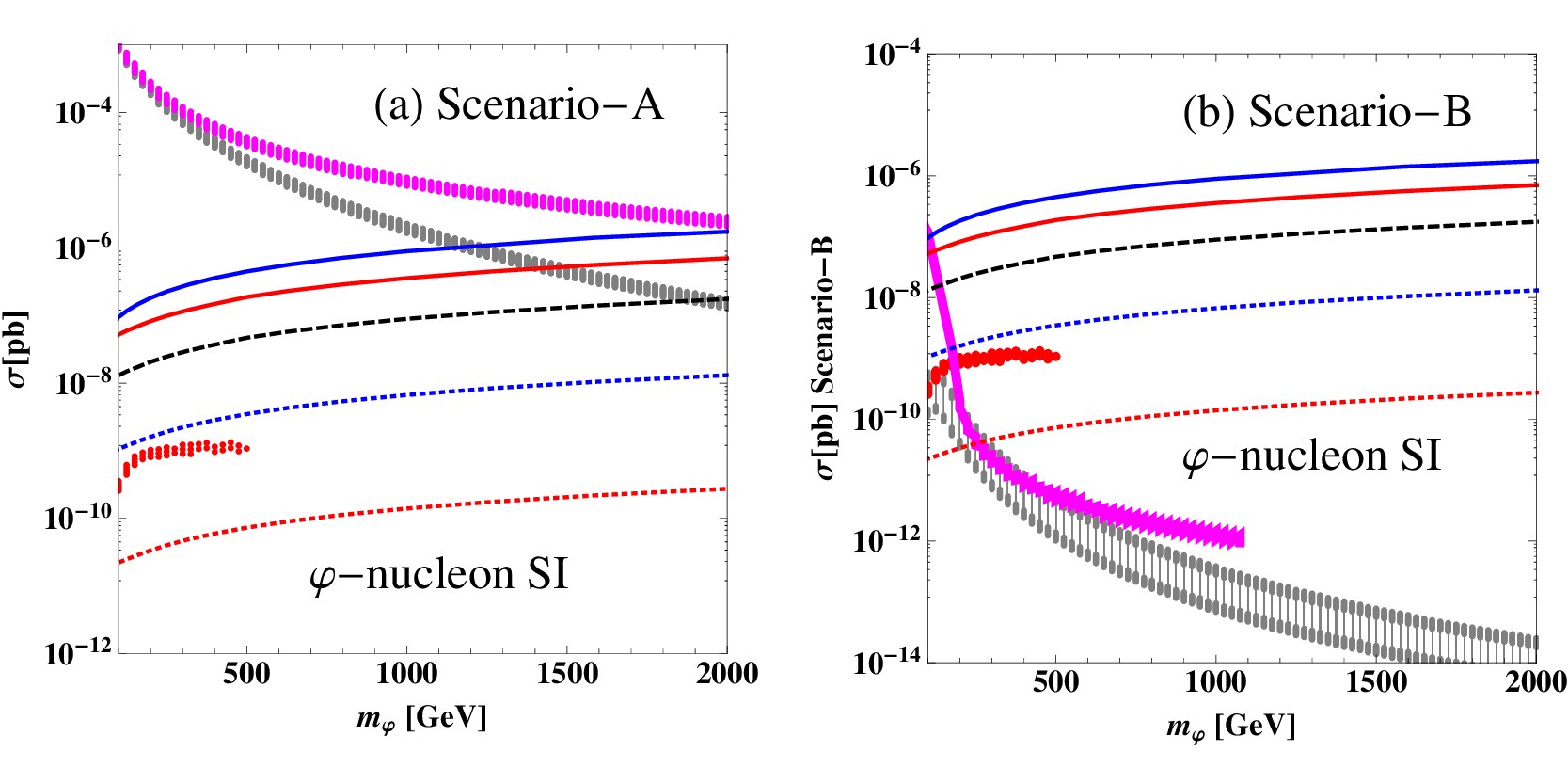}

\caption{SI $\varphi$-nucleon scattering cross sections as a function of $m_{\varphi}$.
The red band represents the cross section for $\alpha_{\phi1}=1$,
the magenta band denotes the cross section for $\alpha_{f\phi}=1$,
and the gray band denotes the cross section for $\alpha_{f\phi}=\alpha_{\phi1}=\alpha_{\phi2}=\alpha_{\phi3}=1$.
See Fig.\ \ref{fig:fermion-SI-SD} for description of other curves.
\label{fig:scalar-SI}}

\end{figure}

The SI elastic $\varphi$-nucleon scattering cross sections are plotted
in Fig.\ \ref{fig:scalar-SI}
along with the current limits placed by the CDMS II 2009 (red-solid)
and XENON10 (blue-solid) collaborations, the projected CDMS II sensitivity
(black-dashed), and the projected future sensitivity of SuperCDMS,
Stage-A (blue-dotted) and Stage-C (black-dotted). If the scalar DM
annihilation occurs purely through the $\oo_{\phi1}$ operator (see
the red band), then the current direct detection from CDMS and XENON
has no constraint on the DM as the elastic scattering cross section
is far below the current bounds, almost by two order of magnitude.
However, the operator can be probed completely in the projected future
SuperCDMS at Stage-C. One should keep in mind that the mass of $\varphi$
should be less than $500\,{\rm GeV}$, otherwise the mass of $\varphi$
is larger than the NP scale $\Lambda$. However, the positive CDMS
II would rule out $\mathcal{O}_{\phi1}$ when the DM signal is observed
in the near future.

The magenta band shows the cross section for $\alpha_{f\phi}=1$. Consider
scenario-A first. The cross section is so large that the entire
parameter space  $100\,{\rm GeV}\leq m_{\varphi}\leq2\,{\rm TeV}$ is excluded
by the CDMS and XENON data. If all the operators contribute to the
DM annihilation, the relic abundance allowed scale $\Lambda$ becomes
large so that the SI scattering cross section is suppressed, see the
gray band. The CDMS and XENON measurements excluded the region of
$m_{\varphi}\leq1450\,{\rm GeV}$, and the projected CDMS II sensitivity
covers $m_{\varphi}$ up to $2000\,{\rm GeV}$. On the contrary, in
scenario-B, the SI scattering cross section is well below the current
CDMS and XENON data. Even the long-term SuperCDMS at Stage-C can only
probe $m_{\varphi}$ up to $300\,{\rm GeV}$. Similar to the case
of fermion DM $\chi$, the broad gray band is due to the different sign assignments
of $\alpha_{f\phi}$ and $\alpha_{\phi1}$. Hence, we find following
interesting points: 
\begin{enumerate}
\item When the CDMS II observe dark matter signal in the near future, the
dark matter in the scenario-A (family universal coupling) lie between
the mass window (1450~GeV - 2000~GeV) when $\alpha_{f\phi}=\alpha_{\phi1}=\alpha_{\phi2}=\alpha_{\phi3}=1$.
The dark matter in the scenario-B (family dependent coupling), however,
is very light, say $100\,{\rm GeV}<m_{\varphi}<150\,{\rm GeV}$ with
$\Lambda\sim{\rm TeV}$. 
\item When no dark matter signal is observed within the projected CDMS II
sensitivity, the scenario-A is almost excluded for $m_{\varphi}<2\,{\rm TeV}$.
However, most of the parameter space of scenario-B is still allowed
and is hard to probe even with the long-term SuperCDMS at Stage-C.
\end{enumerate}

\subsection{Indirect search}

\begin{figure}
\includegraphics[clip,scale=0.35]{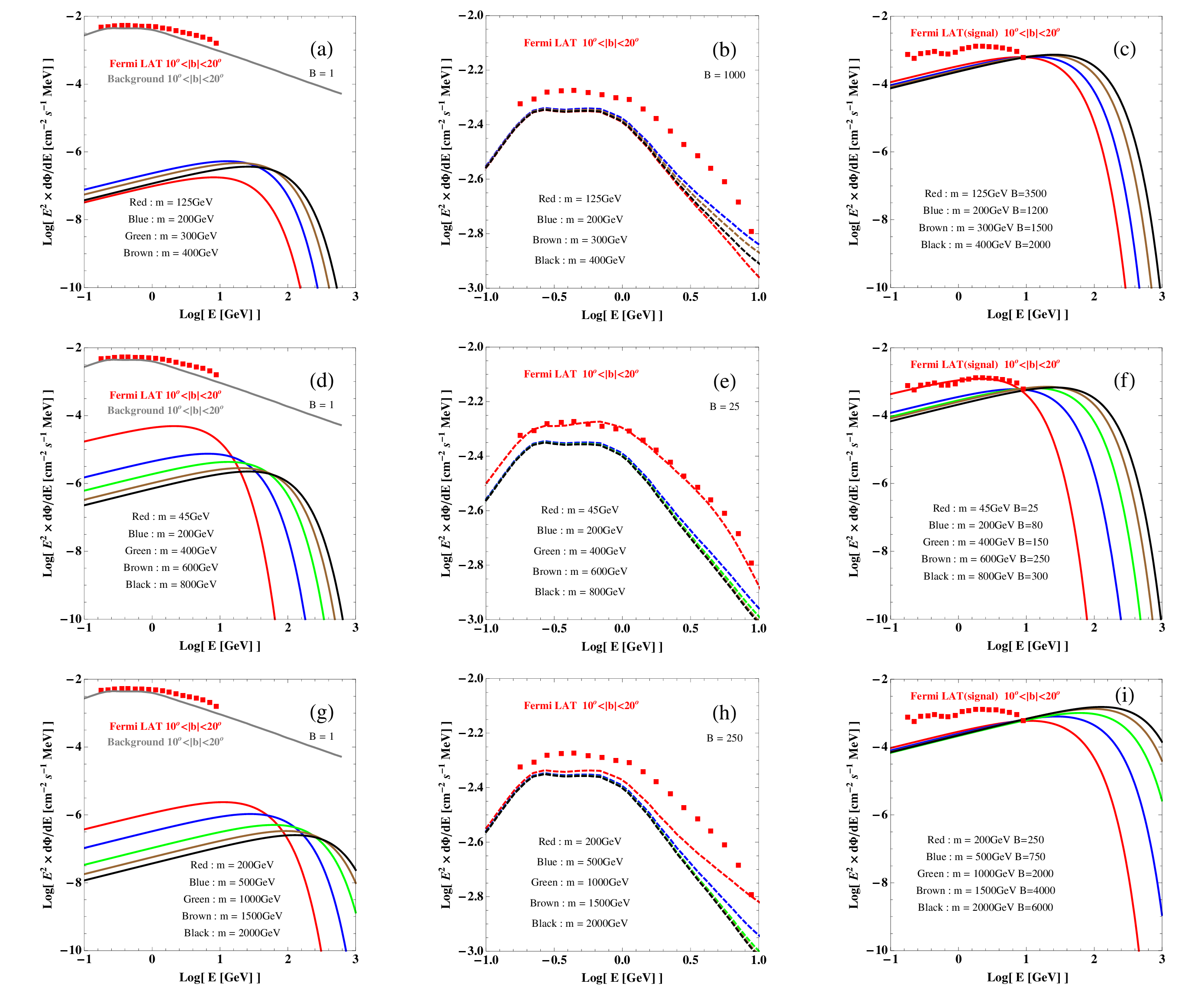}

\caption{Predicted gamma-ray spectra for the scalar DM $\varphi$ with the
NFW density profile: (a,~b,~c) $\alpha_{\phi1}=1$, (d,~e,~f)
$\alpha_{f\varphi}=1$, and (g,~h,~i) $\alpha_{\phi2}=$1 or $\alpha_{\phi3}=1$
or $\alpha_{\phi1}=\alpha_{\phi2}=\alpha_{\phi3}=\alpha_{f\phi}=1$.
The left panel shows the predicted gamma-ray spectra for the scalar
DM $\varphi$.
The middle panel shows the the summed gamma-ray spectra of signal plus background with the Fermi LAT observation. Note that a common boost factor is applied to all the DM signals.
The right panel shows the DM signal with the difference of Fermi LAT observation and background. Different boost factors are adopted for different DM masses 
so that the DM signal will not exceed the data.
\label{fig:dflux-scalar}}

\end{figure}

We now consider the indirect search of $\varphi$ via cosmic gamma-rays.
As most of the parameter space of scenario-A is excluded, we focus
on scenario-B hereafter. Similar to the fermionic dark matter in previous section, we plot the differential distribution of
the gamma-ray flux for various scalar DM masses and boost factors in Fig.\ \ref{fig:dflux-scalar}.
Figs.\ (a,~b,~c) shows the distributions of $\alpha_{\phi1}=1$;
Figs.\ (d,~e,~f) display the distributions of $\alpha_{f\phi}=1$;
Figs.\ (g,~h,~i) denote the distributions of $\alpha_{\phi2}=1$,
or $\alpha_{\phi3}=1$, or $\alpha_{\phi1}=\alpha_{\phi2}=\alpha_{\phi3}=\alpha_{f\phi}=1$
as their distributions are almost the same. 
It is clear that the allowed boost factors for 
the operator $\alpha_{f\phi}$ is much smaller than  that for $\alpha_{\phi1}$ and $\alpha_{\phi2}$ (or $\alpha_{\phi3}=1$, or $\alpha_{\phi1}=\alpha_{\phi2}=\alpha_{\phi3}=\alpha_{f\phi}=1$). 

\begin{figure}
\includegraphics[clip,scale=0.7]{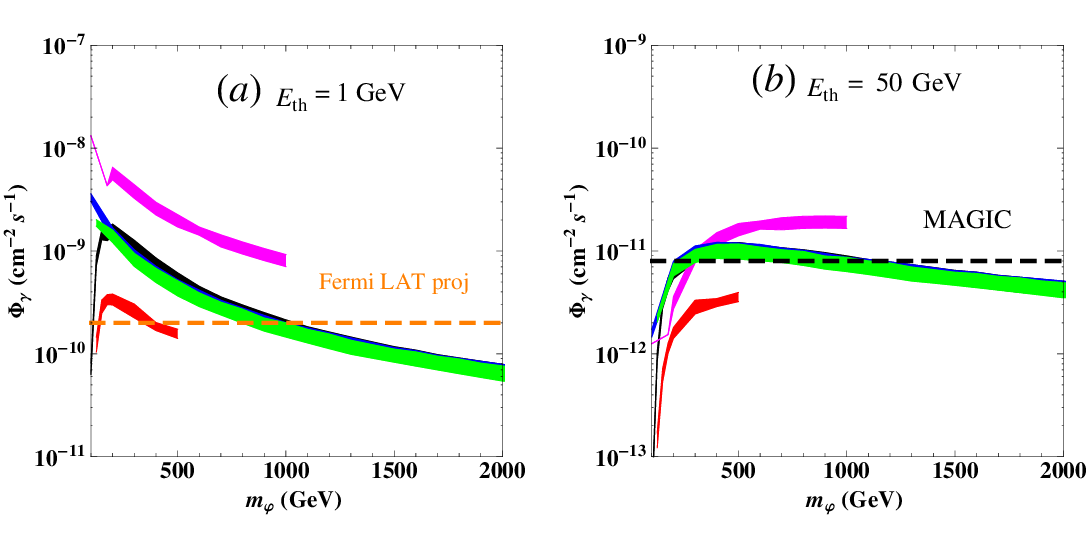}

\caption{Integrated photon flux as a function of $m_{\varphi}$ for energy
threshold of $1\,{\rm GeV}$ (a) and $50\,{\rm GeV}$ (b) (notation
of the color bands is the same as Fig.\ \ref{fig:scalar}). The plot
assume $\bar{J}\left(\Psi,\Delta\Omega\right)\Delta\Omega=1$; all
fluxes scale linearly with this parameter. \label{fig:Integrated-flux-scalar}}

\end{figure}

The integrated cosmic photon-flux as a function of $m_{D}$ is plotted
in Fig.\ \ref{fig:Integrated-flux-scalar}. We choose two energy
thresholds $E_{th}=1\,{\rm GeV}$ and $E_{th}=50\,{\rm GeV}$ to mimic
the space-based and ground-based telescopes. For simplicity we assume
$\bar{J}\left(\Delta\Omega\right)\Delta\Omega=1$ and all the fluxes
scale linearly to account for other DM density profiles. Fermi LAT
and MAGIC have great potential to observe anomalous gamma-rays. With
an enhancement factor, say $\sim10$, from either the DM density profile
or the boost factor, almost all the allowed mass region ($100\,{\rm GeV}-2000\,{\rm GeV}$)
can be explored by Fermi LAT and MAGIC. 

\subsection{Collider search }

\begin{figure}
\includegraphics[clip,scale=0.6]{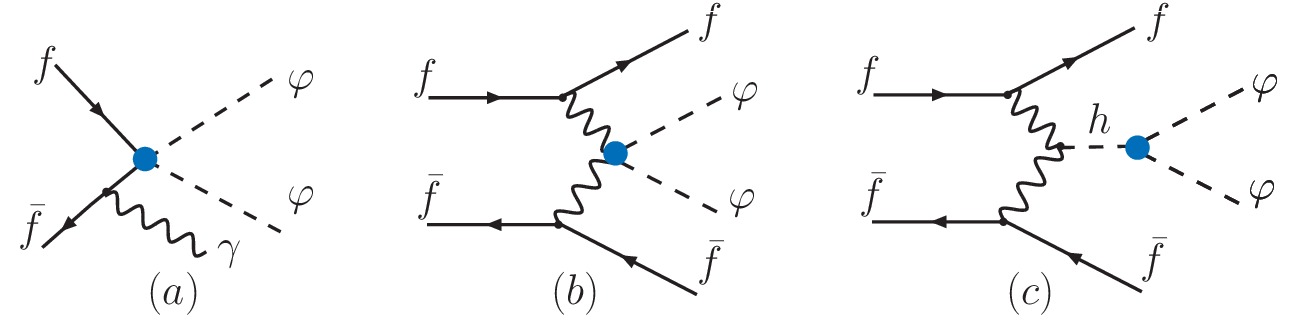}

\caption{Schematic diagrams of the searching channels of DM $\chi$ at the
colliders.\label{fig:scalar-lhc-feyn}}

\end{figure}

Now consider the direct search of $\varphi$ at the LHC. The scalar
DM $\varphi$ can be detected in the following processes\begin{eqnarray}
q\bar{q} & \to & \gamma\varphi\varphi,\label{eq:lhc-scalar-1}\\
q\bar{q} & \to & VVq^{\prime}\bar{q}^{\prime}\to q^{\prime}\bar{q}^{\prime}\varphi\varphi,\label{eq:lhc-scalar2}\end{eqnarray}
where the former is from the four-particle interaction; see Fig.\ \ref{fig:scalar-lhc-feyn}(a),
while the latter is from the so-called vector boson fusion (VBF) process;
see Fig.\ \ref{fig:scalar-lhc-feyn}(b)\ and\ (c). 

The collider signature of the process in Eq.\ (\ref{eq:lhc-scalar-1})
consists of a hard photon plus large $\met$ due to the missing DM
particles. Furthermore, $\varphi$ can also be probed with a collider
signature of a mono-jet plus large $\met$. The corresponding processes
are as follows:\[
q\bar{q}\to g\varphi\varphi,\qquad qg\to q\varphi\varphi.\]
The signal of the VBF process is characterized by two quark jets,
which typically stay in the forward and backward regions of the detector
and are widely separated in pseudo-rapidity, and also by a large missing
transverse momentum ($\met$), due to the two missing DM particles.

\begin{figure}
\includegraphics[scale=0.5]{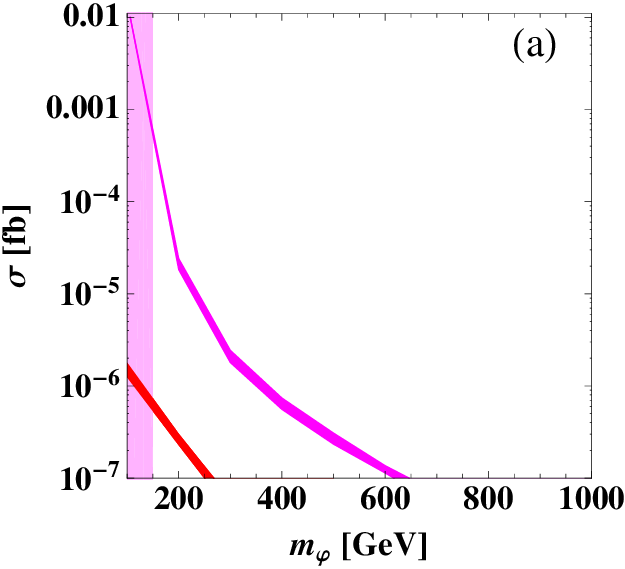}\includegraphics[scale=0.5]{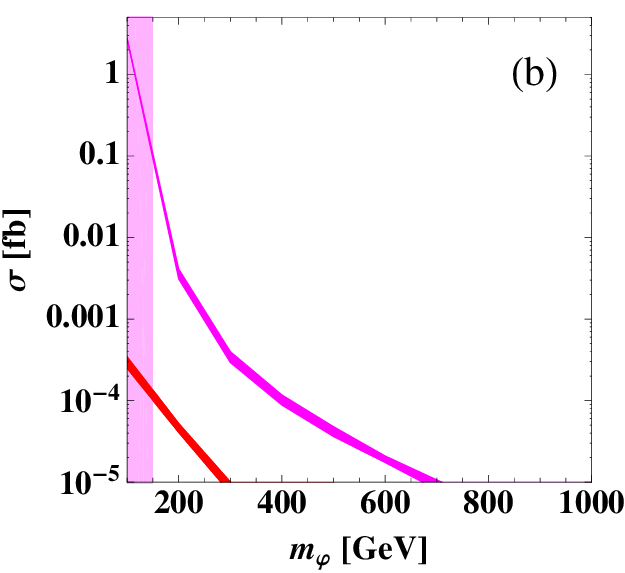}\includegraphics[scale=0.5]{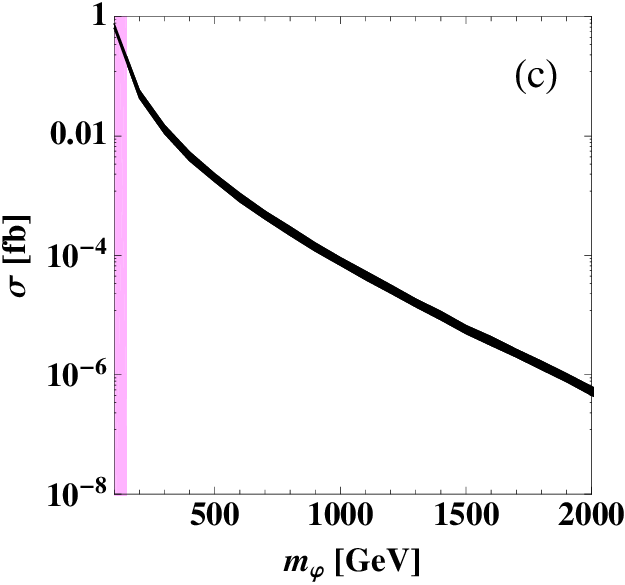}

\caption{Production cross sections (in the unit of fb) of the $\varphi$ pair
together with a SM particle at the LHC: (a) mono-photon plus $\met$,
(b) mono-jet plus $\met$, and (c) VBF. Here, a kinematics cut $p_{T}^{\gamma}>5\,{\rm GeV}$
(or $p_{T}^{j}>5\,{\rm GeV}$) is imposed in order to avoid the collinear
singularities arising from the light quark propagators. In (a) and
(b), the lower band (red) denotes the cross section for $\alpha_{\phi1}=1$
while the upper bands (magenta) denotes the cross section for $\alpha_{f\phi}=1$.
The shaded regions denote the preferred DM mass region by the CDMS
II positive signal. \label{fig:lhc-xsec-scalar}}

\end{figure}

In Fig.\ \ref{fig:lhc-xsec-scalar} we plot the cross sections of
the above processes as a function of $m_{\varphi}$ with the corresponding
$\Lambda$ consistent with the DM relic abundance. As only $\oo_{\phi1}$
and $\oo_{f\phi}$ operators can contribute to the signatures of $\gamma+\met$
and $j+\met$, we plot their effects separately in Fig.\ \ref{fig:lhc-xsec-scalar}\ (a)\ and\ (b)
where the red curve denotes $\oo_{\phi1}$ while the magenta curve
denotes $\oo_{f\phi}$. Although both $\oo_{\phi3}$ and $\oo_{f\phi}$
are involved in the VBF process, the latter contributes much
less than the former, especially for a heavy $\varphi$. Therefore,
we only present the cross section of $\oo_{\phi3}$ in Fig.\ \ref{fig:lhc-xsec-scalar}(c).
We note that the cross sections are generally small for all the processes.
Note also that various kinematics cuts are needed in order to suppress
the SM background. 
We hence conclude that it is very challenging 
to directly detect the DM $\varphi$ signal through those processes
at the LHC. 

\section{Vector dark matter $\mathcal{Z}$\ \label{sec:Vector}}

The DM candidate can also be a vector boson. 
In this work we consider
a simple extension of the SM electroweak gauge group, i.e.
adding an Abelian gauge group $U(1)_{X}$ to the SM gauge group, extending
it to $SU(2)_{L}\times U(1)_{Y}\times U(1)_{X}$. Note that the $SU(2)_{L}\times U(1)_{Y}\times U(1)_{X}$
is only an effective theory below the scale $\Lambda$ where other
fields which 
interact with both the SM and the $U(1)_{X}$ in the underlying
theory decouple. The Abelian vector field for the $U(1)_{X}$ is denoted
as $\mathcal Z$ with a field strength $C_{\mu\nu}\equiv\partial_{\mu}\mathcal{Z}_{\nu}-\partial_{\nu}\mathcal{Z}_{\mu}$.
We further assume that the quarks, leptons and the Higgs field of
the SM do not carry $U(1)_{X}$ quantum numbers, and the field $\mathcal{Z}$
does not carry quantum numbers of the SM gauge group so that it can
be the DM candidate. 

\begin{figure}
\includegraphics[clip,scale=0.7]{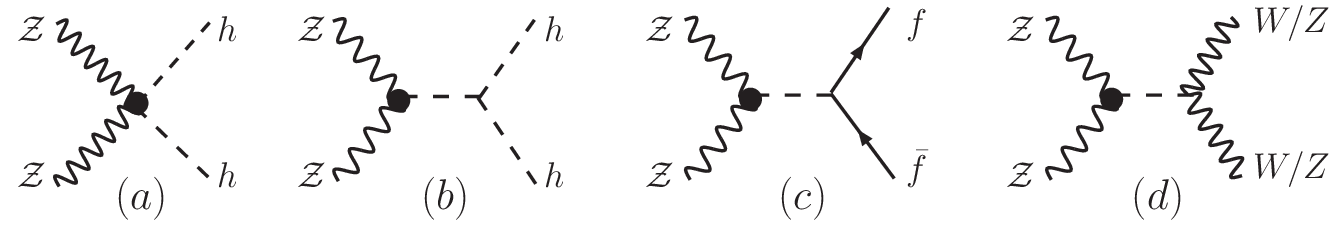}

\caption{Feynman diagrams for $\mathcal{Z}\mathcal{Z}$ annihilation. Blobs
denote the effective vertices induced by the dim-6 operators. \label{fig:feyn-vv}}

\end{figure}

We can write down the following dim-6 effective operator\[
\oo_{\phi\mathcal{Z}}=\frac{1}{2}\left(\phi^{+}\phi\right)C_{\mu\nu}C^{\mu\nu},\]
which induces the $\mathcal{ZZ}\to hh$ annihilation via the following
vertices \[
\mathcal{L}_{\mathcal{ZZ}hh}^{(6)}=\frac{\alpha_{\phi\mathcal{Z}}}{4\Lambda^{2}}hhC_{\mu\nu}C^{\mu\nu},\qquad\mathcal{L}_{\mathcal{ZZ}h}^{(6)}=\frac{\alpha_{\phi\mathcal{Z}}\, v}{2\Lambda^{2}}hC_{\mu\nu}C^{\mu\nu}.\]
In the dimension-six operator, the vector DM $\mathcal{Z}$ can either
directly annihilate into the Higgs boson pair or annihilate via the
Higgs-mediated $s$-channel process into the fermion, Higgs boson
and vector boson pair; see Fig.\ \ref{fig:feyn-vv}.  It is much
simpler as only one operator is present now. Needless to say, for
a heavy $\mathcal{Z}$, the diagram Fig.\ \ref{fig:feyn-vv}(c) is negligible.

\subsection{Relic abundance}

\begin{figure}
\includegraphics[clip,scale=0.5]{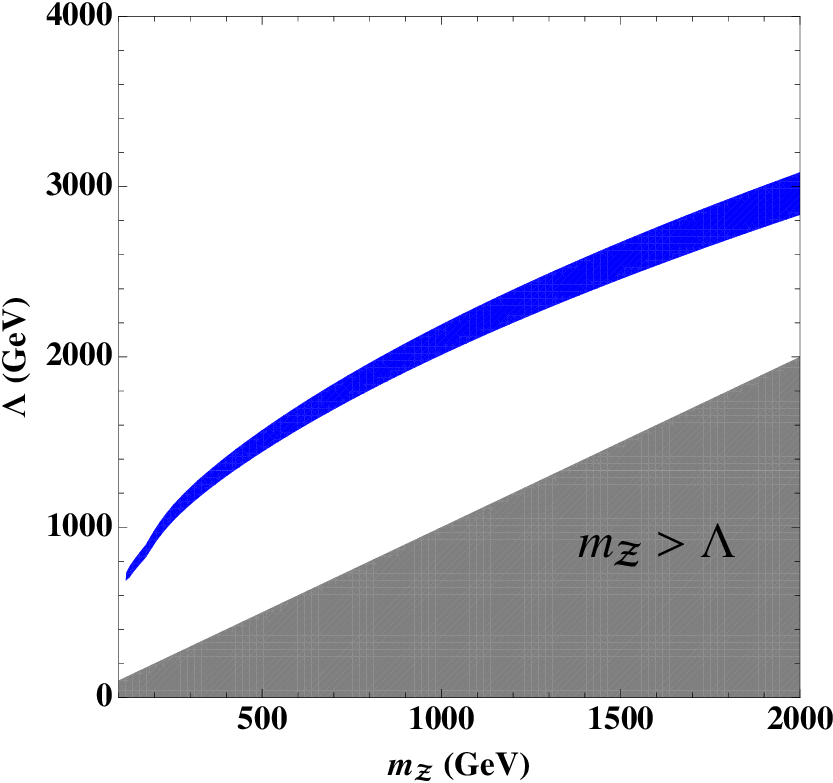}

\caption{Allowed parameter set $\left(m_{\mathcal{Z}},\Lambda\right)$ for
a vector
DM $\mathcal{Z}$. The upper (lower) boundary of each band
corresponds to the upper (lower) limit of $\Omega_{DM}h^{2}$ given
in Eq.\ (\ref{eq:omega_WMAP}). \label{fig:vector}}

\end{figure}

Since the cross section of $\mathcal{ZZ}\to f\bar{f}$ is proportional
to the fermion mass square, only the top quark can contribute significantly
to the DM annihilation. When the DM $\mathcal{Z}$ is light, say $80\,{\rm GeV}<m<150\,{\rm GeV}$,
it predominantly annihilates into the vector bosons. In the very heavy
limit, i.e. $m\gg m_{t}$, the DM $\mathcal{Z}$ will annihilate into
$WW$, $ZZ$ and Higgs boson pairs, giving rise to the following simple
form of $a$ and $b$,
\begin{equation}
a=\frac{39\alpha_{\phi\mathcal{Z}}^{2}m^{2}}{256\pi\Lambda^{4}}\,,\qquad b=\frac{169\alpha_{\phi\mathcal{Z}}^{2}m^{2}}{2048\pi\Lambda^{4}}\,.\end{equation}
We calculate the relic abundance and
plot the allowed parameter space of $\left(m_{\mathcal{Z}},\Lambda\right)$
in Fig.\ \ref{fig:vector}; for instance, $\Lambda\simeq700\,{\rm GeV}$ for $m_{\mathcal{Z}}\simeq100\,{\rm GeV}$
while $\Lambda\simeq3000\,{\rm GeV}$ for $m_{\mathcal{Z}}\simeq2000\,{\rm GeV}$.
Again, since $x_{F}$ barely changes in the whole region, we choose
$x_{F}=25$ and obtain an interesting relation between $m$ and $\Lambda$
as follows:
\begin{equation}
\frac{0.43}{\alpha_{\phi\mathcal{Z}}^{2}\,\left(\Omega_{CDM}h^{2}\right)}=\left(\frac{m}{100\,{\rm GeV}}\right)^{2}\left(\frac{{\rm TeV}}{\Lambda}\right)^{4},\label{eq:vector-relation}
\end{equation}
which agrees with the exact scanning results 
shown in Fig.\ \ref{fig:vector}.
But such a simple relation is no longer valid when the
SM Higgs boson is very heavy because of the large enhancement at the
threshold $m_{\mathcal{Z}}\approx m_{h}/2$. 

\subsection{Direct Detection}

Since the vector DM $\mathcal{Z}$ only interacts with the SM Higgs
boson in our model, it can be detected in spin-independent experiments.
The relevant Lagrangian is given by
\begin{equation}
\mathcal{L}_{\mathcal{ZZ}qq}=\frac{\alpha_{\phi\mathcal{Z}}}{2\Lambda^{2}}\frac{m_{q}}{m_{h}^{2}}C_{\mu\nu}C^{\mu\nu}\bar{q}q,
\end{equation}
which gives rise to the effective $\mathcal{Z}$-nucleon cross sections
$\sigma_{\mathcal{Z}p}^{SI}$ and $\sigma_{\mathcal{Z}n}^{SI}$ as
follows:
\begin{equation}
\sigma_{\mathcal{Z}p}^{SI}=\sigma_{\mathcal{Z}n}^{SI}\approx\left(2.88\times10^{-10}\,{\rm pb}\right)\alpha_{\phi\mathcal{Z}}^{2}\left(\frac{m_{\mathcal{Z}}}{100\,{\rm GeV}}\right)^{2}\left(\frac{{\rm TeV}}{\Lambda}\right)^{4}\left(\frac{100\,{\rm GeV}}{m_{h}}\right)^{4}.\label{eq:vector-nucleon-SI}
\end{equation}

\begin{figure}
\includegraphics[scale=0.5]{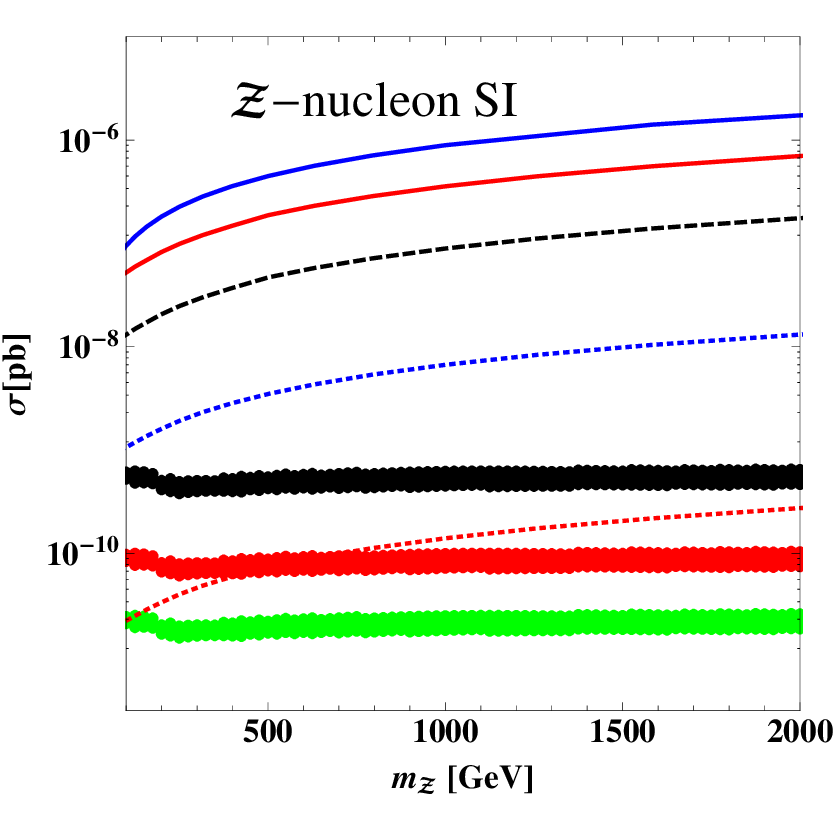}

\caption{Vector WIMP-nucleon cross sections for spin-independent dark matter
search. The shaded region is our parameter space which respects
the observed WMAP data for $\alpha_{\phi\mathcal{Z}}=1$: the black
band for $m_{h}=120\,{\rm GeV}$, the red band for $m_{h}=300\,{\rm GeV}$,
and the green band for $m_{h}=600\,{\rm GeV}$. See Fig.\ \ref{fig:fermion-SI-SD}
for descriptions of other curves. \label{fig:vector-SI}}

\end{figure}

The SI elastic $\mathcal{Z}$-nucleon scattering cross sections are
plotted in Fig.\ \ref{fig:vector-SI} for three choices of the SM
Higgs boson mass, along with the projected future sensitivity of SuperCDMS
Stage-A (blue dashed line) and Stage-C (red dashed curve). The cross
sections are well below all the current bounds from CDMS collaboration,
but can be probed at the future SuperCDMS experiment at Stage-C. The
almost flat behavior of the black band can be easily understood from
Eqs.\ (\ref{eq:vector-relation})\ and\ (\ref{eq:vector-nucleon-SI})
as 
\begin{equation}
\sigma_{\mathcal{Z}p}^{SI}\propto\alpha_{\phi\mathcal{Z}}^{2}\left(\frac{m_{\mathcal{Z}}}{100\,{\rm GeV}}\right)^{2}\left(\frac{{\rm TeV}}{\Lambda}\right)^{4}\propto\frac{1}{\Omega_{CDM}h^{2}}.
\end{equation}
Since $\sigma_{\mathcal{Z}p}^{SI}$ does not depend upon $m_{\mathcal{Z}}$,
one can not determine $m_{\mathcal{Z}}$ from the SuperCDMS experiment
even if such an excess is indeed observed. But it might be possible
to measure $m_{\mathcal{Z}}$ via indirect cosmic gamma-ray measurements
as discussed below. Needless to say, the positive signal of CDMS II
would exclude the possibility of vector dark matter.

\subsection{Indirect search}

\begin{figure}
\includegraphics[scale=0.4]{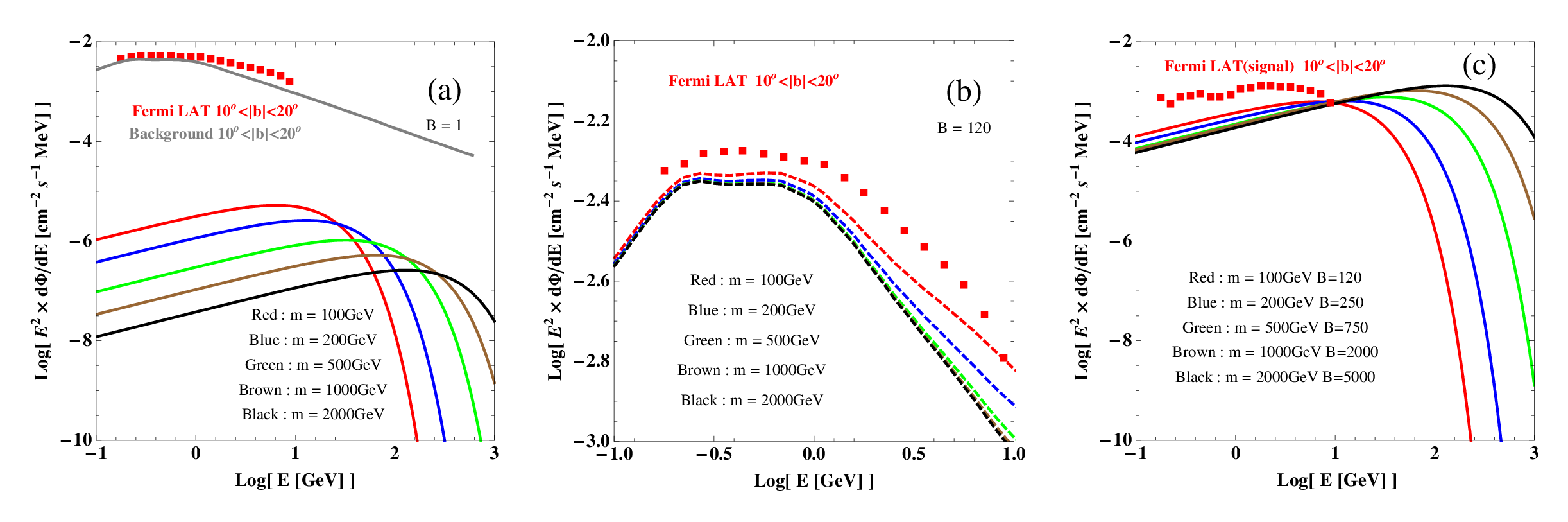}
\caption{(a) Predicted gamma-ray spectra for the DM $\mathcal{Z}$  with the
NFW density profile, where the Fermi LAT observation (background)
is presented by the red-box (grey-solid) line; (b) Comparison between the DM signal plus background and the Fermi LAT observation. Note that a common boost factor of 120 is applied to all the DM signals; (c) The comparison between the DM signal with the difference of Fermi LAT observation and background. Different boost factors are adopted for different DM masses 
so that the DM signal will not exceed the data.
\label{fig:dfluxx-vector}}
\end{figure}

The differential cosmic gamma-ray distribution is given by\begin{eqnarray}
\frac{d\Phi}{dE_{\gamma}} & \approx & \left(0.76\times10^{-12}{\rm s}^{-11}{\rm cm}^{-2}{\rm GeV}^{-1}\right)\bar{J}\left(\Psi,\Delta\Omega\right)\Delta\Omega\nonumber \\
 & \times & \alpha_{\phi\mathcal{Z}}^{2}\,\left(\frac{100\,{\rm GeV}}{m_{\chi}}\right)\left(\frac{{\rm TeV}}{\Lambda}\right)^{4}x^{-1.5}e^{-7.76x}.\label{eq:vector-diff-flux}\end{eqnarray}
We plot the differential distributions as a function of $E_{\gamma}$
in Fig.\ \ref{fig:dfluxx-vector}(a) for various $m_{\mathcal{Z}}$
with the NFW density profile. For $m_{\mathcal{Z}}\sim100-2000\,{\rm GeV}$
the distributions are well below the current Fermi LAT data. 
The discovery of the dark matter signal could be possible in the high energy region after including an allowed boost factor;
see Figs.\ \ref{fig:dfluxx-vector}(b) and (c).

\begin{figure}
\includegraphics[scale=0.8]{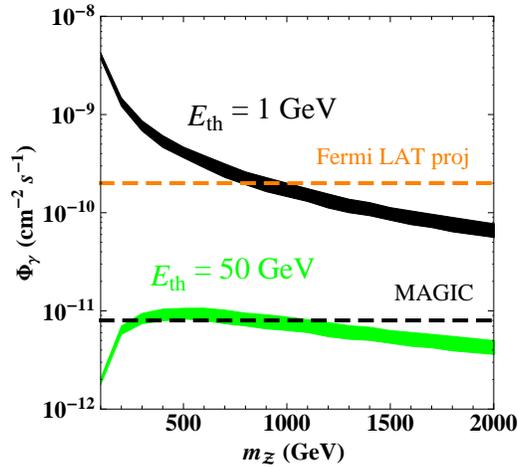}

\caption{Integrated photon flux as a function of $m_{\mathcal{Z}}$ for energy
threshold of $1\,{\rm GeV}$ (a) and $50\,{\rm GeV}$ (b) (notation
of the color bands is the same as Fig.\ \ref{fig:vector}). The plot
assumes $\bar{J}\left(\Psi,\Delta\Omega\right)\Delta\Omega=1$; all
fluxes scale linearly with this parameter. \label{fig:Integrated-flux-vector}}

\end{figure}

We plot the integrated cosmic photon-flux as a function of $m_{\mathcal{Z}}$
in Fig.\ \ref{fig:Integrated-flux-vector}. Similar to the study
of the fermion and scalar DM, we choose two energy thresholds $E_{th}=1\,{\rm GeV}$
and $E_{th}=50\,{\rm GeV}$ to mimic the space-based and ground-based
telescopes. Assuming $\bar{J}\left(\Psi,\Delta\Omega\right)\Delta\Omega=1$
and without the boost factor effect, Fermi LAT can probe $m_{\mathcal{Z}}\sim100\,{\rm GeV}-1000\,{\rm GeV}$
while MAGIC can probe $m_{\mathcal{Z}}\sim250\,{\rm GeV}-1000\,{\rm GeV}$. 

\subsection{Collider search}

\begin{figure}[b]
\includegraphics[scale=0.7]{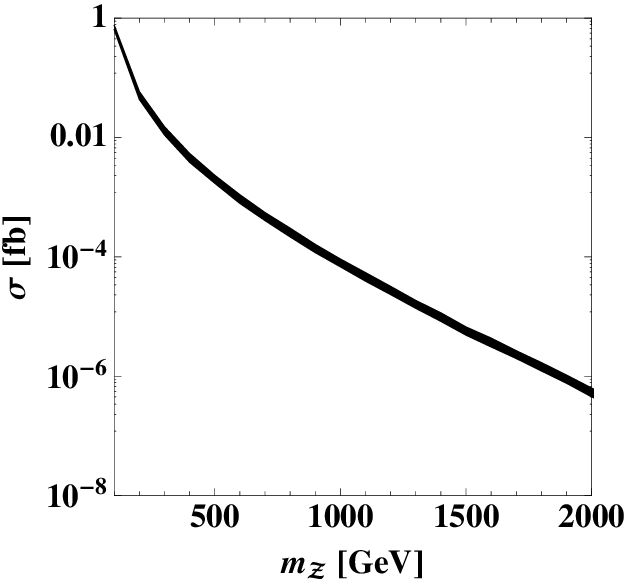}

\caption{Cross section of the vector DM $\mathcal{Z}$ pair in the VBF process.\label{fig:lhc-xsec-vector}}

\end{figure}

The detection of the vector $\mathcal{Z}$ can be probed in the VBF
process at the LHC as $\mathcal{Z}$ only interacts with the SM Higgs
boson. The production process is given by \[
q\bar{q}\to q\bar{q}VV\to q\bar{q}\mathcal{Z}\mathcal{Z}.\]
The cross section of the production is plotted in Fig.\ \ref{fig:lhc-xsec-vector},
which is too small to be detected at the LHC.

\section{Conclusion\ \label{sec:Conclusion}}

Despite its great success, the Standard Model of particle physics
needs to be extended to explain the dark matter observation. So far
many new physics models have been introduced to address DM physics,
but our knowledge of the DM candidate is still subject to cosmological
observations. For example, the properties of the DM candidate, e.g.,
its mass and spin, remain mysterious.  
In this
work we present a model-independent study of DM physics using effective
field theory. We add to the SM a new DM field $D$ whose stability
is guaranteed by a discrete $Z_{2}$ symmetry. The interactions between
the dark matter and the SM fields are assumed to be induced by other
heavy particles which decouple at the scale $\Lambda$. 
After writing down the subset of dim-6 operators (named dark operators) we are interested in, 
we examine the constraints on those dark operators from the relic abundance measurements
and direct/indirect searches. The discovery potential for those dark
operators at the LHC is also studied. 

The dark matter candidate can be a scalar $\varphi$, a fermion $\chi$
or a vector $\mathcal{Z}$. We found that the fermion DM annihilation
can be described by three independent operators ($\oo_{f\chi}^{L},\,\oo_{f\chi}^{R},\,\oo_{\phi\chi}$),
the scalar DM annihilation by four independent operators ($\oo_{\phi1},\,\oo_{\phi2},\,\oo_{\phi3},\,\oo_{f\phi}$)
and the vector DM annihilation by one operator $\oo_{\phi\mathcal{Z}}$.
Owing to the different setups of various experiments, we can probe
those dark operators separately. 

\emph{Fermion DM $\chi$}: A positive signal at the CDMS II 
in the near future will indcate that
the dark matter lie between 
 $[1000,\,2000]\,{\rm TeV}$
for the operators $\alpha_{f\chi}^{L/R}$ while $[900,\,1700]\,{\rm GeV}$
for the operator $\alpha_{\phi\chi}$.
Each individual fermion DM
operator could satisfy the cosmic gamma-ray spectrum observed by the Fermi
LAT since the contribution from the dark matter is far below the data. The allowed boost factors can be derived Galactic background agrees with the Fermi LAT measurement. For $m_{\chi}= 30\,{\rm GeV} \sim 1000\,{\rm GeV}$ the allowed boost factor  $B= 20 \sim 600$.
It is possible to detect the heavy dark matter signal from cosmic gamma-ray in the higher energy region due to the fact that the dark matter annihilation (including a boost factor) produces much more gamma-ray than the background, which is also true in the cases of scalar dark matter and vector dark matter discussed below.
Furthermore, the fermion DM
operator might be probed at the LHC via 
the mono-photon plus missing energy and mono-jet plus missing energy signatures.

\emph{Scalar DM $\varphi$: }The family universal operators are almost
ruled out by the CDMS II exclusion limit, while the family dependent operators
is still allowed with a light scalar dark matter. 
The allowed boost factor is about $25 \sim 300$ for 
the scalar dark matter $m_{\varphi} = 45\,{\rm GeV} \sim 800$ GeV), if only 
$\alpha_{f\varphi}$ is non-zero. For non-vanishing $\alpha_{\phi1}$ and non-vanishing $\alpha_{\phi2}$ (or $\alpha_{\phi3}=1$, or $\alpha_{\phi1}=\alpha_{\phi2}=\alpha_{\phi3}=\alpha_{f\phi}=1$), the constraints are weaker, i.e. the allowed boost factor can be much larger. It is very challenging to search for such scalar dark matter at the LHC.

\emph{Vector DM $\mathcal{Z}$ }: 
There is no constraint from the CDMS II on  vector DM operator as the scattering is
mediated by the SM Higgs boson which is very small. 
The allowed boost factor is $120 \sim 2000$ for 
the dark matter $m_{\varphi} = 100\,{\rm GeV} \sim 1000$ GeV. 
The cross section of vector dark matter signal is too small to be detected at the LHC.

\begin{sidewaystable}
\caption{Constraints and potential reaches of the dark operators of fermion
and vector dark matter in various experiments. Shorthand notations:
CII= CDMS II ; CIIp = projected CDMS II; SC(A/C)=SuperCDMS(Stage-A/C);
$B$: boost factor need to fit Fermi LAT; $(\surd\surd)$: full sensitivity;
$(\surd)$: good sensitivity; $(-)$: not applicable or too small;
$(\times)$: no constraint; $(\times\times)$: ruled out. DM mass
and scale $\Lambda$ are in the unit of GeV.\label{tab:summary1}}

\begin{tabular}{c|c|c|cccc|c|c|c|c}
\hline 
 &  & $\Omega_{CDM}h^{2}$ & \multicolumn{4}{c|}{direct search} & \multicolumn{1}{c|}{Fermi} & \multicolumn{3}{c}{LHC (fb)}\tabularnewline
\hline 
 &  & constraint & CII & CIIp & SC(A) & SC(C) & $\gamma$-ray & $\gamma+\met$ & $j+\met$ & VBF\tabularnewline
\hline
\hline 
 & $\alpha_{f\chi}^{L}$ & $m_{\chi}\sim\left[100,\,2000\right]$ & $m_{\chi}>1000$ & $m_\chi<1700$ & $\surd $ & $\surd$ & $\surd\surd$: $m_{\chi}\sim30$, $B\sim20$ & $\sigma\sim[20,\,1]$  &  & $-$\tabularnewline
 &  & $\Lambda\sim[1000,\,4000]$ &  &  &  &  & $\surd$ : $m_{\chi}>100$ & for $m_{\chi}\sim[100,\,400]$ &  & \tabularnewline
 & $\alpha_{f\chi}^{R}$ & $m_{\chi}\sim\left[100,\,2000\right]$ & $m_{\chi}>1100$ & $m_\chi<1800 $ & $\surd $ & $\surd$ & $\surd\surd$: $m_{\chi}\sim30$, $B\sim20$ & $\sigma\sim[20,\,1]$ for  &  & $-$\tabularnewline
$\chi$ &  & $\Lambda\sim[1000,\,4000]$ &  &  &  &  & $\surd$ : $m_{\chi}>100$ & $m_{\chi}\sim[100,400]$ &  & \tabularnewline
 & $\alpha_{\phi\chi}$ & $m_{\chi}\sim\left[100,\,2000\right]$ & $m_{\chi}>850$ & $m_{\chi}<1400$ & $\surd$ & $\surd$ & $\surd\surd$: $m_{\chi}\sim30$, $B\sim20$ & $-$ & $-$ & $-$\tabularnewline
 &  & $\Lambda\sim[400,\,2000]$ &  &  &  &  & $\surd$ : $m_{\chi}>100$ &  &  & \tabularnewline
\cline{2-11} 
 & all & $m_{\chi}\sim\left[100,\,2000\right]$ & $m_{\chi}>250$ & $m_{\chi}<400$ & $m_\chi<1000$ & $\surd$ & $\surd\surd$: $m_{\chi}\sim30$, $B\sim20$ & $\sigma\sim[20,\,0.001]$ for  &  & $-$\tabularnewline
 &  & $\Lambda\sim[1000,\,5500]$ &  &  &  &  & $\surd$ : $m_{\chi}>100$ & $m_{\chi}\sim[100,\,20000]$ &  & \tabularnewline
\hline
\hline 
$\mathcal{Z}$ & $\alpha_{\phi\mathcal{Z}}$ & $m_{\chi}\sim\left[100,\,2000\right]$ & $\times$ & $\times$ & $\times$ & $\surd$ & $\surd$ & $-$ & $-$ & $-$\tabularnewline
 &  & $\Lambda\sim[600,\,3000]$ &  &  &  & $m_{h}\sim120$ & large $B$ &  &  & \tabularnewline
\hline
\end{tabular}
\end{sidewaystable}

\begin{sidewaystable}
\caption{Constraints and potential reaches of the dark operators of scalar
dark matter in various experiments. Shorthand notations: CII= CDMS
II ; CIIp = projected CDMS II; SC(A/C)=SuperCDMS(Stage-A/C); $B$:
boost factor need to fit Fermi LAT; $(\surd\surd)$: full sensitivity;
$(\surd)$: good sensitivity; $(-)$: not applicable or too small;
$(\times)$: no constraint; $(\times\times)$: ruled out. DM mass
and scale $\Lambda$ are in the unit of GeV.\label{tab:summary2}}

\begin{tabular}{c|c|c|cccc|c|c|c|c}
\hline 
 &  & $\Omega_{CDM}h^{2}$ & \multicolumn{4}{c|}{direct search} & \multicolumn{1}{c|}{Fermi} & \multicolumn{3}{c}{LHC (fb)}\tabularnewline
\hline 
 &  & constraint & CII & CIIp & SC(A) & SC(C) & $\gamma$-ray & $\gamma+\met$ & $j+\met$ & VBF\tabularnewline
\hline
\hline 
 & $\alpha_{f\phi}$ & $m_{\chi}\sim\left[100,\,2000\right]$ & $\times\times$ & $\times\times$ & $\times\times$ & $\times\times$ & $\times\times$ & $\times\times$ & $\times\times$ & $\times\times$\tabularnewline
 &  & $\Lambda\sim2000$ &  &  &  &  &  &  &  & \tabularnewline
 & $\alpha_{\phi1}$ & $m_{\chi}\sim\left[100,\,500\right]$ & $\times\times$ & $\times\times$ & $\times\times$ & $\times\times$ & $\times\times$ & $\times\times$ & $\times\times$ & $\times\times$\tabularnewline
 &  & $\Lambda\sim[1500,\,500]$ &  &  &  &  &  &  &  & \tabularnewline
$\varphi$ & $\alpha_{\phi2}$ & $m_{\chi}\sim\left[100,\,2000\right]$ & $\times\times$ & $\times\times$ & $\times\times$ & $\times\times$ & $\times\times$ & $\times\times$ & $\times\times$ & $\times\times$\tabularnewline
{[}scenario A{]} &  & $\Lambda\sim[600,\,3500]$ &  &  &  &  &  &  &  & \tabularnewline
 & $\alpha_{\phi3}$ & $m_{\chi}\sim\left[100,\,2000\right]$ & $\times\times$ & $\times\times$ & $\times\times$ & $\times\times$ & $\times\times$ & $\times\times$ & $\times\times$ & $\times\times$\tabularnewline
 &  & $\Lambda\sim[600,\,3400]$ &  &  &  &  &  &  &  & \tabularnewline
\cline{2-11} 
 & all & $m_{\chi}\sim\left[100,\,2000\right]$ & $m_{\varphi}>1500$ & $m_{\varphi}<2000$ & $\surd\surd$ & $\surd\surd$ & $\surd$ &  &  & \tabularnewline
 &  & $\Lambda\sim[2000,\,4500]$ &  &  &  &  & large $B$ &  &  & \tabularnewline
\hline
\hline 
 & $\alpha_{f\phi}$ & $m_{\chi}\sim\left[100,\,2000\right]$ & $m_{\varphi}>100$ & $m_{\varphi}<150$ & $m_{\varphi}\to200$ & $m_{\varphi}\to300$ & $\surd$ & $-$ & $-$ & $-$\tabularnewline
 &  & $\Lambda\sim[400,\,1000]$ &  &  &  &  & large $B\sim100$ &  &  & \tabularnewline
 & $\alpha_{\phi1}$ & $m_{\chi}\sim\left[100,\,500\right]$ & $-$ & $-$ & $-$ & $\surd$ & $\surd$ & $-$ & $-$ & $-$\tabularnewline
 &  & $\Lambda\sim[1400,\,500]$ &  &  &  &  & huge $B>1000$ &  &  & \tabularnewline
$\varphi$ & $\alpha_{\phi2}$ & $m_{\chi}\sim\left[100,\,2000\right]$ & $-$ & $-$ & $-$ & $-$ & $\surd$ & $-$ & $-$ & $-$\tabularnewline
{[}scenario B{]} &  & $\Lambda\sim[600,\,3600]$ &  &  &  &  & huge $B>1000$ &  &  & \tabularnewline
 & $\alpha_{\phi3}$ & $m_{\chi}\sim\left[100,\,2000\right]$ & $-$ & $-$ & $-$ & $-$ & $\surd$ & $-$ & $-$ & $-$\tabularnewline
 &  & $\Lambda\sim[400,\,3000]$ &  &  &  &  & huge $B>1000$ &  &  & \tabularnewline
\cline{2-11} 
 & all & $m_{\chi}\sim\left[100,\,2000\right]$ & $m_{\varphi}>100$ & $m_{\varphi}<150$ & $m_{\varphi}\to200$ & $m_{\varphi}\to300$ & $\surd$ & $-$ & $-$ & $-$\tabularnewline
 &  & $\Lambda\sim[1400,\,4500]$ &  &  &  &  & large $B>100$ &  &  & \tabularnewline
\hline
\end{tabular}
\end{sidewaystable}

\begin{acknowledgments}
We thank J. L. Rosner, Tim Tait, Jose Wudka and C.-P. Yuan for useful
discussions. Q.-H.~C. is supported by the Argonne National Laboratory
and University of Chicago Joint Theory Institute (JTI) Grant 03921-07-137,
and by the U.S.~Department of Energy under Grants No.~DE-AC02-06CH11357
and No.~DE-FG02-90ER40560. C.~R.~C. is supported by the World Premier
International Research Center Initiative (WPI initiative) by MEXT,
Japan. C.~S.~Li is supported by the National Natural Science Foundation
of China, under Grants No.10721063, No.10975004 and No.10635030. H.~Z.
is supported in part by the National Natural Science Foundation of
China under Grants 10975004 and the China Scholarship Council (CSC)
File No. 2009601282, and U.~S.~Department of Energy under Grants
No.~DE-AC02-06CH11357. C.~R.~C. thanks Institute
of Physics, Academia Sinica in Taiwan for its hospitality during the final stages
of this work. 
\newpage{}
\end{acknowledgments}
\appendix

\section{Relic abundance\label{sec:Relic-abundance}}

In the computation of the 
DM ($\chi)$  relic density, one assumes that 
$\chi$
 was in thermal equilibrium with the SM particles in the early universe
and decouple when it was non-relativistic. Once 
$\Gamma=n_{\chi}\left\langle \sigma^{ann}v\right\rangle <H$
($H=\left(8\pi\rho/3M_{Pl}\right)^{1/2}$ is the Hubble expansion rate), 
$\chi$ stopped annihilating and fell out of equilibrium, and its
density remains intact till now. The number density 
$n_{\chi}$ 
is governed by the Boltzmann
equation and the law of entropy
conservation:
\begin{eqnarray}
\frac{d}{dt}n_{\chi} & = & -3Hn_{\chi}-\left\langle \sigma v\right\rangle _{SA}\left(n_{\chi}^{2}-n_{\chi,eq}^{2}\right)-\left\langle \sigma v\right\rangle _{CA}\left(n_{\chi}n_{\phi}-n_{\chi,eq}n_{\phi,eq}\right)+C_{\Gamma},\label{eq:density_evolution}\\
\frac{ds}{dt} & = & -3Hs\label{eq:entropy_conservation}
\end{eqnarray}
where $\left\langle \sigma v\right\rangle $ is the thermally averaged
annihilation cross section times the relative velocity, the subscripts
SA and CA on $\left\langle \sigma v\right\rangle $ denote self-annihilation
and co-annihilation (annihilation with another species $\phi$) respectively,
$n_{\chi,eq}$ 
denotes the equilibrium number density, and $C_{\Gamma}$
is the contribution due to the decay of heavier particles into 
$\chi$.
Finally, $s$ is the entropy density. For a massive cold dark matter candidate, its equilibrium
density is given by the non-relativistic limit:
\begin{equation}
n_{\chi,eq}=g_{\chi}\left(\frac{mT}{2\pi}\right)^{3/2}e^{-m/T},\label{eq:density_equilibrium}\end{equation}
where $m$ is the mass of the particle species in question. The physics
of Eq.~(\ref{eq:density_evolution}) is as following: At early times,
when the temperature was higher than the mass of the particle, the
number density was $n_{eq}\propto T^{3}$, $\chi$ annihilated with
its own or other particle species $\phi$ into lighter states and
vice verse. As the temperature decreased below the mass, $n_{\chi}$
dropped exponentially as indicated in Eq.~(\ref{eq:density_equilibrium})
and the annihilation rate $\Gamma=n\left\langle \sigma v\right\rangle $
dropped below $H$. The $\chi$ can no longer annihilate
and its density per co-moving volume remains fixed. The temperature
at which the particle decouples from the thermal bath is denoted $T_{F}$
(freeze-out temperature) and roughly corresponds to the time when
$\Gamma$ is of the same order as $H$. 

It is usually useful to scale out the effects of the expansion of
the universe by considering the evolution of the number of particles
in a co-moving volume. This is done by using the entropy density,
$s$, as a fiducial quantity, and by defining as the dependent variable,
$Y=n/s$ with $Y_{eq}=n_{\chi,eq}/s$. In this case, Eq.~(\ref{eq:density_evolution})
can be rewritten as\begin{equation}
\frac{dY}{dx}=\frac{1}{3H}\frac{ds}{dx}\left\langle \sigma v\right\rangle _{SA}\left(Y_{\chi}^{2}-Y_{\chi,eq}^{2}\right)+\frac{1}{3H}\frac{ds}{dx}\left\langle \sigma v\right\rangle _{CA}\left(Y_{\chi}Y_{\phi}-Y_{\chi,eq}Y_{\phi,eq}\right),\end{equation}
where $x=m/T$. In radiation domination era, the entropy, as function
of the temperature, is given by \[
s=\frac{2\pi^{2}}{45}g_{*s}\left(x\right)\left(\frac{m}{x}\right)^{3}\equiv k_{1}x^{-3},\]
which is deduced from the fact that $s=\left(\rho+P\right)/T$ and
$g_{*s}$ is the effective degrees of freedom for the entropy density.
Therefore, one finds\begin{equation}
\frac{ds}{dx}=-\frac{3s}{x}.\end{equation}
Hence, the Boltzmann equation for the DM number density is obtained
\begin{equation}
\frac{dY}{dx}=-\frac{s}{Hx}\left\{ \left\langle \sigma v\right\rangle _{SA}\left(Y_{\chi}^{2}-Y_{\chi,eq}^{2}\right)\right\} .\label{eq:boltzmann-density}\end{equation}

As is well known, $\left\langle \sigma v\right\rangle $ is well approximated
by a non-relativistic expansion (obtained by replacing the square
of the energy in the center of mass frame by $s=4m^{2}+m^{2}v^{2}$):\begin{equation}
\left\langle \sigma v\right\rangle =a+b\left\langle v^{2}\right\rangle +\mathcal{O}\left(\left\langle v^{4}\right\rangle \right)\approx a+6\frac{b}{x}.\label{eq:ab}\end{equation}
The freeze-out temperature is defined by solving the following equation\begin{equation}
x_{F}=\ln\left(c\left(c+2\right)\sqrt{\frac{45}{8}}\frac{g}{2\pi^{3}}\frac{mM_{Pl}\left(a+6b/x_{F}\right)}{g_{*}^{1/2}x_{F}^{1/2}}\right),\label{eq:xf}\end{equation}
where $c$ is a constant of order one determined by matching the later-time
and early-time solutions. The result does not depend dramatically
on the precise value of $c$ which we will choose the usual value
$c=1/2$.

\section{Dark matter annihilation\label{sec:Dark-matter-annihilation}}

\subsection{Fermionic dark matter annihilation}

The relevant cross sections for pair of $\chi$ to annihilate have
final states into fermions, into Higgs bosons, or into pair of Higgs
boson and $Z$-boson. The Feynman diagrams of the annihilations are
shown in Fig.\ \ref{fig:feyn-ff}. Summing/averaging over final/initial
spins and integrating over the phase space of the final state particles,
we obtain the cross sections of $\chi\bar{\chi}$ annihilation shown
below.

\subsubsection{$\chi\bar{\chi}\to f\bar{f}$}

The annihilation cross section into a fermion pair is given by \begin{eqnarray}
 &  & \sigma\left(\chi\bar{\chi}\to f\bar{f}\right)\nonumber \\
 & = & \frac{N_{C}\,\beta_{f}}{12\pi s\beta\Lambda^{4}}\Biggl\{\left(g_{L}^{2}+g_{R}^{2}\right)\left[s^{2}-s\left(m^{2}+m_{f}^{2}\right)+4m^{2}m_{f}^{2}\right]+6g_{L}g_{R}\left(s-2m^{2}\right)m_{f}^{2}\Biggr\},\end{eqnarray}
where $\beta\equiv\sqrt{1-4m^{2}/s}$ , $\beta_{f}\equiv\sqrt{1-4m_{f}^{2}/s}$
and the factor $N_{C}$ sums over the different color combinations
allowed in the final state, $N_{C}=3$ for quarks and $N_{C}=1$ for
leptons. The couplings $g_{L}$ and $g_{R}$ are given in Eq.\ \ref{eq:coeff:ffff}.
Expanding $\sigma v_{rel}$ in powers of the relative speed between
the $\chi$ fermions, $v_{rel}$, gives\begin{eqnarray}
a & = & N_{C}\frac{m^{2}}{2\pi\Lambda^{4}}\sqrt{1-r_{f}^{2}}\left(g_{L}^{2}+g_{R}^{2}+r_{f}^{2}\, g_{L}g_{R}\right),\\
b & = & N_{C}\frac{m^{2}}{48\pi\Lambda^{4}\sqrt{1-r_{f}^{2}}}\Biggl\{\left(g_{L}^{2}+g_{R}^{2}\right)\left(2r_{f}^{4}-10r_{f}^{2}+11\right)+3g_{L}g_{R}\, r_{f}^{2}\left(3-2r_{f}^{2}\right)\Biggr\},\end{eqnarray}
where $r_{f}\equiv m_{f}/m$.

\subsubsection{$\chi\bar{\chi}\to Z\to f\bar{f}$}

The annihilation cross section into a fermion pair via the $Z$-mediated
$s$-channel process is given by \begin{equation}
\sigma\left(\chi\bar{\chi}\to Z\to f\bar{f}\right)=N_{C}\frac{\alpha_{\phi\chi}^{2}}{48\pi\beta\Lambda^{4}}\left(g_{Z\, L}^{2}+g_{Z\, R}^{2}\right)\frac{v^{2}\left(s-m^{2}\right)m_{Z}^{2}}{\left(s-m_{Z}^{2}\right)^{2}+m_{Z}^{2}\Gamma_{Z}^{2}},\end{equation}
where $g_{Z\, L/R}$ denotes the SM couplings of the $Z$-boson to
fermions and $m_{Z}\left(\Gamma_{Z}\right)$ denotes the mass (width)
of the $Z$-boson. We have ignored the masses of the SM fermions in
the annihilation cross section of the $\chi\bar{\chi}\to Z\to f\bar{f}$
process because it only contributes around the $Z$-pole region, i.e.
$m_{DM}\approx\frac{1}{2}m_{Z}$, where the masses of the SM lepton
and light quarks are negligible. The corresponding leading terms in
the non-relativistic expansion are \begin{eqnarray}
a & = & N_{C}\frac{\alpha_{\phi\chi}^{2}\, m^{2}}{8\pi\Lambda^{4}}\left(g_{Z\, L}^{2}+g_{Z\, R}^{2}\right)\frac{m_{Z}^{2}v^{2}}{\left(4m^{2}-m_{Z}^{2}\right)^{2}+m_{Z}^{2}\Gamma_{Z}^{2}},\\
b & = & -N_{C}\frac{\alpha_{\phi\chi}^{2}\, m^{2}}{192\pi\Lambda^{4}}\left(g_{Z\, L}^{2}+g_{Z\, R}^{2}\right)\frac{16m^{4}+40m_{Z}^{2}m^{2}-11m_{Z}^{4}}{\left(4m^{2}-m_{Z}^{2}\right)^{2}+m_{Z}^{2}\Gamma_{Z}^{2}}.\end{eqnarray}
As to be shown later, this annihilation channel is dominant around
the $Z$-pole region, i.e. $m\sim\frac{1}{2}m_{Z}$, but its contribution
decreases so rapidly, once $m_{\chi}$ goes beyond the $Z$-pole,
that it is overwhelmed by other annihilation channels.

\subsubsection{$\chi\bar{\chi}\to hZ$}

The annihilation cross section into a pair of $Z$-boson and Higgs
scalar shown in Fig.\ \ref{fig:feyn-ff}(c) is given by\begin{eqnarray}
 &  & \sigma\left(\chi\bar{\chi}\to hZ\right)\nonumber \\
 & = & \frac{\alpha_{\phi\chi}^{2}}{96\pi s^{3}\beta\Lambda^{4}}\sqrt{\lambda\left(s,m_{Z}^{2},m_{h}^{2}\right)}\Biggl\{\left(s+2m^{2}\right)\lambda\left(s,m_{Z}^{2},m_{h}^{2}\right)+12s\, m_{Z}^{2}\left(3m^{2}-s\right)\Biggr\},\end{eqnarray}
where the function $\lambda$ being defined as\[
\lambda\left(x,y,z\right)=\left(x-y-z\right)^{2}-4yz.\]
The non-relativistic expansion gives rise to the following leading
terms\begin{eqnarray}
a & = & \frac{\alpha_{\phi\chi}^{2}\, m^{2}}{512\pi\Lambda^{4}}\lambda^{1/2}\left(4,r_{h}^{2},r_{Z}^{2}\right)\left[\lambda\left(4,r_{h}^{2},r_{Z}^{2}\right)-8r_{Z}^{2}\right]\\
b & = & \frac{\alpha_{\phi\chi}^{2}\, m^{2}}{12288\pi\Lambda^{4}}\lambda^{-1/2}\left(4,r_{h}^{2},r_{Z}^{2}\right)\Biggl[1792-128\left(5r_{h}^{2}+26r_{Z}^{2}\right)\nonumber \\
 &  & \qquad-64\left(3r_{h}^{4}-16r_{h}^{2}r_{Z}^{2}-16r_{Z}^{4}\right)+8\left(13r_{h}^{2}-2r_{Z}^{2}\right)\left(r_{h}^{2}-r_{Z}^{2}\right)-11\left(r_{h}^{2}-r_{Z}^{2}\right)^{4}\Biggr].\end{eqnarray}
where $r_{h(Z)}\equiv m_{h(Z)}/m$.

\subsubsection{$\chi\bar{\chi}\to Z\to hZ$}

The annihilation into a pair of $Z$-boson and Higgs scalar can also
be induced by the $Z$-mediated $s$-channel process, see Fig.\ \ref{fig:feyn-ff}(d).
This annihilation channel opens only if $2m>m_{Z}+m_{h}$, which implies
$s\approx4m^{2}>4m_{Z}^{2}$, thus its contribution is highly suppressed
so that we can ignore it in this study. 

\subsubsection{$\chi\bar{\chi}\to Z\to W^+W^-$}
For this channel, 
\begin{eqnarray}
a&=&\frac{\alpha_{\phi\chi}^2m^2}{16\pi\Lambda^4}\left(1-r_W^2\right)^{1/2}\left(1+4r_W^2-\frac{17}{4}r_W^4-\frac{3}{4}r_W^6\right)\left(1-\frac{r_Z^2}{4}\right)^{-2},\\
b&=&\frac{7\alpha_{\phi\chi}^2m^2}{384\pi\Lambda^4}\left(1-r_W^2\right)^{1/2}\Biggl[1+r_W^2+\frac{145}{28}r_W^4+\frac{3}{2}r_W^6\nonumber\\
&&-\frac{r_Z^2}{28}\left(19+55r_W^2-\frac{59}{4}r_W^4+6r_W^6\right)\Biggr]\left(1-\frac{r_Z^2}{4}\right)^{-3},
\end{eqnarray}
where $r_{W(Z)}=m_{W(Z)}/m$.

\subsection{Scalar dark matter annihilation}

When the scalar $\varphi$ is the DM, we must consider a large number
of annihilation processes, e.g. $\varphi\varphi\to f\bar{f}/WW/ZZ/hh$
and $\varphi\varphi\to h\to f\bar{f}/WW/ZZ/hh$, see Fig.\ \ref{fig:feyn-ss}.
The latter, Higgs-mediated $s$-channel processes, contributes significantly
around the Higgs resonance region only.

\subsubsection{$\varphi\varphi\to f\bar{f}$}

In the scenario-A, the annihilation cross sections of $\varphi\varphi\to f\bar{f}$
read as \begin{equation}
\sigma\left(\varphi\varphi\to f\bar{f}\right)=N_{C}\frac{\alpha_{f\phi}^{2}v^{2}}{16\pi\beta\Lambda^{4}}\beta_{f}^{3/2}+N_{C}\frac{\alpha_{\phi1}^{2}v^{2}}{32\pi\beta\Lambda^{4}}\beta_{f}^{3/2}\frac{v^{2}\, m_{f}^{2}}{\left(s-m_{h}^{2}\right)^{2}+m_{h}^{2}\Gamma_{h}^{2}},\label{eq:xsec-ssff}\end{equation}
where the first and second term corresponds to Fig.\ \ref{fig:feyn-ss}(a)
and (b), respectively. Expanding $\sigma v_{rel}$ in powers of the
relative speed $v_{rel}$ gives the following leading terms\begin{eqnarray}
a & = & N_{C}\frac{v^{2}}{16\pi\Lambda^{4}}\left(1-r_{f}^{2}\right)^{3/2}\left[2\alpha_{f\phi}^{2}+\alpha_{\phi1}^{2}\frac{m_{f}^{2}v^{2}}{\left(4m^{2}-m_{h}^{2}\right)^{2}+m_{h}^{2}\Gamma_{h}^{2}}\right],\label{eq:ssff-dump1}\\
b & = & N_{C}\frac{v^{2}}{128\pi\Lambda^{4}}\sqrt{1-r_{f}^{2}}\;\Biggl\{2\alpha_{f\phi}^{2}\left(2r_{f}^{2}+1\right)\nonumber \\
 &  & +\alpha_{\phi1}^{2}\, r_{f}^{2}\,\frac{v^{2}\left[-48m^{6}+8\left(12m_{f}^{2}+m_{h}^{2}\right)m^{4}+m_{h}^{2}\left(m_{h}^{2}-32m_{f}^{2}\right)m^{2}+2m_{f}^{2}m_{h}^{4}\right]}{\left(\left(4m^{2}-m_{h}^{2}\right)^{2}+m_{h}^{2}\Gamma_{h}^{2}\right)^{2}}\Biggr\}.\label{eq:ssff-dump2}\end{eqnarray}
Similarly, in the scenario-B, the leading terms are given by\begin{eqnarray}
a & = & N_{C}\frac{v^{2}}{16\pi\Lambda^{4}}\left(1-r_{f}^{2}\right)^{3/2}\left[2\alpha_{f\phi}^{2}\left(\frac{m_{f}^{2}}{v^{2}}\right)+\alpha_{\phi1}^{2}\frac{m_{f}^{2}v^{2}}{\left(4m^{2}-m_{h}^{2}\right)^{2}+m_{h}^{2}\Gamma_{h}^{2}}\right],\label{eq:ssff-dump3}\\
b & = & N_{C}\frac{v^{2}}{128\pi\Lambda^{4}}\sqrt{1-r_{f}^{2}}\;\Biggl\{2\alpha_{f\phi}^{2}\left(2r_{f}^{2}+1\right)\left(\frac{m_{f}^{2}}{v^{2}}\right)\nonumber \\
 &  & +\alpha_{\phi1}^{2}\, r_{f}^{2}\,\frac{v^{2}\left[-48m^{6}+8\left(12m_{f}^{2}+m_{h}^{2}\right)m^{4}+m_{h}^{2}\left(m_{h}^{2}-32m_{f}^{2}\right)m^{2}+2m_{f}^{2}m_{h}^{4}\right]}{\left(\left(4m^{2}-m_{h}^{2}\right)^{2}+m_{h}^{2}\Gamma_{h}^{2}\right)^{2}}\Biggr\}.\label{eq:ssff-dump4}\end{eqnarray}

\subsubsection{$\varphi\varphi\to WW/ZZ$}

The annihilation cross sections of $\varphi\varphi$ annihilation
into vector bosons are given by\begin{eqnarray}
\sigma\left(\varphi\varphi\to VV\right) & = & \delta_{V}\frac{\beta_{V}\left(s^{2}-4s\, m_{V}^{2}+12m_{V}^{4}\right)}{64\pi s\beta\Lambda^{4}}\left(\alpha_{\phi3}^{2}+\frac{\alpha_{\phi1}^{2}v^{4}}{\left(s-m_{h}^{2}\right)^{2}+m_{h}^{2}\Gamma_{h}^{2}}\right)\label{eq:xsec-ssvv}\end{eqnarray}
where the subscript $V$ denotes the type of vector boson and $\beta_{V}=\sqrt{1-4m_{V}^{2}/s}$.
$\delta_{V}$ counts for the identical particle: $\delta_{W}=1$ and
$\delta_{Z}=1/2$. The first and second term in Eq.\ \ref{eq:xsec-ssvv}
corresponds to Fig.\ \ref{fig:feyn-ss}(c)\ and\ (d), respectively.
The non-relativistic expansion gives rise to the following leading
terms\begin{eqnarray}
a_{V} & = & \frac{\delta_{V}}{32\pi m^{3}\Lambda^{4}}\sqrt{m^{2}-m_{V}^{2}}\nonumber \\
 & \times & \left\{ \alpha_{\phi3}^{2}\left(4m^{4}-4m_{V}^{2}m^{2}+3m_{V}^{4}\right)+\alpha_{\phi1}^{2}\, v^{4}\,\frac{4m^{4}-4m_{V}^{2}m^{2}+3m_{V}^{4}}{\left(m_{h}^{2}-4m^{2}\right)^{2}+m_{h}^{2}\Gamma_{h}^{2}}\right\} ,\label{eq:dump1}\\
b_{V} & = & \frac{\delta_{V}}{256\pi m^{3}\Lambda^{4}}\frac{1}{\sqrt{m^{2}-m_{V}^{2}}}\nonumber \\
 & \times & \Biggl\{3\alpha_{\phi3}^{2}\left(4m^{6}-4m_{V}^{2}m^{4}-m_{V}^{4}m^{2}+2m_{V}^{6}\right)\nonumber \\
 &  & +\frac{\alpha_{\phi1}^{2}\, v^{4}\left(4m^{2}-m_{h}^{2}\right)}{\left(\left(m_{h}^{2}-4m^{2}\right)^{2}+m_{h}^{2}\Gamma_{h}^{2}\right)^{2}}\biggl[16m^{8}+4\left(3m_{h}^{2}-20m_{V}^{2}\right)m^{6}-4\left(3m_{h}^{2}m_{V}^{2}-31m_{V}^{4}\right)m^{4}\nonumber \\
 &  & \qquad\qquad\qquad\qquad\qquad\qquad+3m_{V}^{4}\left(m_{h}^{2}+24m_{V}^{2}\right)m^{2}+6m_{h}^{2}m_{V}^{6}\biggr]\Biggr\}.\label{eq:dump2}\end{eqnarray}

\subsubsection{$\varphi\varphi\to hh$}

The cross section of $\varphi\varphi\to hh$ reads as\begin{equation}
\sigma\left(\varphi\varphi\to hh\right)=\frac{\beta_{h}\left(c_{1}v^{2}+c_{2}s\right)^{2}}{32\pi s\beta\Lambda^{4}}+\frac{9v^{4}\alpha_{\phi1}^{2}}{128\pi s\beta\Lambda^{4}}\frac{\beta_{h}m_{h}^{4}}{\left(s-m_{h}^{2}\right)^{2}+m_{h}^{2}\Gamma_{h}^{2}},\label{eq:xsec-sshh}\end{equation}
where\[
c_{1}=\frac{3}{4}\alpha_{\phi1},\qquad c_{2}=\alpha_{\phi2}+\frac{\alpha_{\phi3}}{4}.\]
The first and second term in Eq.\ \ref{eq:xsec-sshh} corresponds
to Fig.\ \ref{fig:feyn-ss}(e)\ and\ (f), respectively. The non-relativistic
expansion gives the following leading terms\begin{eqnarray}
a & = & \frac{\sqrt{m^{2}-m_{h}^{2}}}{128\pi m^{3}\Lambda^{4}}\left[2\left(c_{1}v^{2}+4c_{2}m^{2}\right)^{2}+\frac{9\,\alpha_{\phi1}^{2}\, m_{h}^{4}\, v^{4}}{\left(4m^{2}-m_{h}^{2}\right)^{2}+m_{h}^{2}\Gamma_{h}^{2}}\right],\label{eq:dump3}\\
b & = & \frac{\left(4c_{2}m^{2}+c_{1}v^{2}\right)\left[4c_{2}m^{2}\left(3m^{2}-2m_{h}^{2}\right)-c_{1}v^{2}\left(m^{2}-2m_{h}^{2}\right)\right]}{512\pi m^{3}\Lambda^{4}\sqrt{m^{2}-m_{h}^{2}}}\nonumber \\
 & + & \frac{9\alpha_{\phi1}^{2}m_{h}^{4}\left(-80m^{6}+120m_{h}^{2}m^{4}-33m_{h}^{4}m^{2}+2m_{h}^{6}\right)v^{4}}{2048\pi m^{3}\Lambda^{4}\sqrt{m^{2}-m_{h}^{2}}\left[\left(m_{h}^{2}-4m^{2}\right)^{2}+m_{h}^{2}\Gamma_{h}^{2}\right]^{2}}.\label{eq:dump4}\end{eqnarray}

\subsection{Vector dark matter annihilation}

The vector DM $\mathcal{Z}$ can either directly annihilate into the
Higgs boson pair or annihilate into the fermion, Higgs scalar and
vector boson pair through the Higgs-mediated $s$-channel process,
see Fig.\ \ref{fig:feyn-vv}

\subsubsection{$\mathcal{Z}\mathcal{Z}\to hh$}

The annihilation cross section of $\mathcal{ZZ}\to hh$ is given by\begin{equation}
\sigma\left(\mathcal{ZZ}\to hh\right)=\frac{\alpha_{\phi\mathcal{Z}}^{2}\beta_{h}\left(6m^{4}-4s\, m^{2}+s^{2}\right)}{256s\beta\Lambda^{4}},\label{eq:xsec-vvhh}\end{equation}
which gives rise to the leading term in the non-relativistic expansion
as follows:\begin{equation}
a=\frac{3\alpha_{\phi\mathcal{Z}}^{2}\, m^{2}}{256\pi\Lambda^{4}}\sqrt{1-r_{h}^{2}},\qquad b=\frac{\alpha_{\phi\mathcal{Z}}^{2}\left(13m^{2}-10m_{h}^{2}\right)}{2048\pi\Lambda^{4}\sqrt{1-r_{h}^{2}}}.\end{equation}

\subsubsection{$\mathcal{ZZ}\to h\to hh$}

The annihilation cross section of $\mathcal{ZZ}$ pair into the Higgs
scalar pair via the Higgs-mediated $s$-channel process is given by\begin{equation}
\sigma\left(\mathcal{ZZ}\to h\to hh\right)=\frac{9\alpha_{\phi\mathcal{Z}}^{2}\,\beta_{h}}{64\pi s\beta\Lambda^{4}}\frac{m_{h}^{4}\left(6m^{4}-4s\, m^{2}+s^{2}\right)}{\left(s-m_{h}^{2}\right)^{2}+m_{h}^{2}\Gamma_{h}^{2}},\label{eq:xsec-vvhh2}\end{equation}
which yields the following leading non-relativistic expansion terms\begin{eqnarray}
a & = & \frac{27\alpha_{\phi\mathcal{Z}}^{2}}{64\pi\Lambda^{4}}\frac{m^{2}\, m_{h}^{4}\,\sqrt{1-r_{h}^{2}}}{\left(m_{h}^{2}-4m^{2}\right)^{2}+m_{h}^{2}\Gamma_{h}^{2}},\\
b & = & \frac{9\alpha_{\phi\mathcal{Z}}^{2}}{512\pi\Lambda^{4}}\frac{\left(4m^{2}-m_{h}^{2}\right)\left(4m^{4}-5m_{h}^{2}m^{2}+10m_{h}^{4}\right)}{\sqrt{1-r_{h}^{2}}\left(\left(m_{h}^{2}-4m^{2}\right)^{2}+m_{h}^{2}\Gamma_{h}^{2}\right)^{2}}.\end{eqnarray}

\subsubsection{$\mathcal{ZZ}\to h\to f\bar{f}$}

The annihilation cross section of $\mathcal{ZZ}\to h\to f\bar{f}$
is given by\begin{equation}
\sigma\left(\mathcal{ZZ}\to h\to f\bar{f}\right)=N_{C}\frac{\alpha_{\phi\mathcal{Z}}^{2}\beta_{f}^{3/2}}{16\pi\beta\Lambda^{4}}\frac{m_{f}^{2}\left(6m^{4}-4s\, m^{2}+s^{2}\right)}{\left(s-m_{h}^{2}\right)^{2}+m_{h}^{2}\Gamma_{h}^{2}},\label{eq:xsec-vvff}\end{equation}
which yields\begin{eqnarray}
a & = & N_{C}\frac{3\alpha_{\phi\mathcal{Z}}^{2}\, m_{f}^{2}}{4\pi\Lambda^{4}}\frac{m^{4}\left(1-r_{f}^{2}\right)^{3/2}}{\left(m_{h}^{2}-4m^{2}\right)^{2}+m_{h}^{2}\Gamma_{h}^{2}},\\
b & = & N_{C}\frac{\alpha_{\phi\mathcal{Z}}^{2}\, m_{f}^{2}}{32\pi\Lambda^{4}}\sqrt{1-r_{f}^{2}}\frac{m^{2}\left(4m^{2}-m_{h}^{2}\right)\left(28m^{4}+\left(8m_{f}^{2}-19m_{h}^{2}\right)m^{2}+10m_{f}^{2}m_{h}^{2}\right)}{\left(\left(m_{h}^{2}-4m^{2}\right)^{2}+m_{h}^{2}\Gamma_{h}^{2}\right)^{2}}.\end{eqnarray}
We note that both $a$ and $b$ are proportional to $m_{f}$ so that
they are negligible in the limit of $m\gg m_{f}$.

\subsubsection{$\mathcal{ZZ}\to h\to WW/ZZ$}

The annihilation cross section of $\mathcal{ZZ}$ into vector boson
pair is given by\begin{equation}
\sigma\left(\mathcal{ZZ}\to h\to VV\right)=\delta_{V}\frac{\alpha_{\phi\mathcal{Z}}^{2}\beta_{V}}{32\pi s\beta\Lambda^{4}}\frac{\left(6m^{4}-4s\, m^{2}+s^{2}\right)\left(12m_{V}^{2}-4sm_{V}^{2}+s^{2}\right)}{\left(s-m_{h}^{2}\right)^{2}+m_{h}^{2}\Gamma_{h}^{2}},\label{eq:xsec-vvvv}\end{equation}
where $\delta_{W}=1$ and $\delta_{Z}=1/2$. The non-relativistic
expansion gives rise to the two leading terms as follows:\begin{eqnarray}
a_{V} & = & \delta_{V}\frac{3\alpha_{\phi\mathcal{Z}}^{2}m^{2}}{8\pi\Lambda^{4}}\sqrt{1-r_{V}^{2}}\frac{4m^{4}-4m_{V}^{2}m^{2}+3m_{V}^{4}}{\left(4m^{2}-m_{h}^{2}\right)^{2}+m_{h}^{2}\Gamma_{h}^{2}},\\
b_{V} & = & \delta_{V}\frac{\alpha_{\phi\mathcal{Z}}^{2}}{64\pi\Lambda^{4}\sqrt{1-r_{V}^{2}}}\frac{\left(4m^{2}-m_{h}^{2}\right)}{\left(\left(4m^{2}-m_{h}^{2}\right)^{2}+m_{h}^{2}\Gamma_{h}^{2}\right)^{2}}\Biggl\{208m^{8}-4\left(25m_{h}^{2}+68m_{V}^{2}\right)m^{6}\nonumber \\
 &  & +4\left(19m_{V}^{4}+41m_{h}^{2}m_{V}^{2}\right)m^{4}+\left(24m_{V}^{6}-103m_{h}^{2}m_{V}^{4}\right)m^{2}+30m_{h}^{2}m_{V}^{6}\Biggr\}\end{eqnarray}

\section{Dark matter direct detection\ \label{sec:Dark-matter-direct}}

\subsection{Fermionic DM-nucleon interaction\label{sub:Fermionic-DM-nucleon-interaction}}

When the DM is a fermion $\chi$, its coupling to quark, leading to
elastics scattering from a nucleus, can be parametrized as Eq.\ \ref{eq:FDM-nucleon-interaction},
of which the first term results in a spin-independent scattering from
a nucleus while the second term leads to a spin-dependent scattering
from a nucleus.

Consider the SI scattering first. When a WIMP interact with quarks
via a vector-like interaction, e.g. given by\begin{equation}
\mathcal{L}_{V}=b_{q}\bar{\chi}\gamma_{\mu}\chi\bar{q}\gamma^{\mu}q,\label{eq:fermion-SI-vector-lang}\end{equation}
where $b_{q}$ is a shorthand notation of the WIMP-quark vector coupling
given in the first term of Eq.\ \ref{eq:FDM-nucleon-interaction}.
In this case, the contributions of each quark in the nucleus add coherently
and large cross sections result for large nuclei. The WIMP-nucleus
cross section in this case is 
\begin{equation}
\sigma_{\chi N}^{SI}=\frac{m_{DM}^{2}m_{N}^{2}b_{N}^{2}}{\pi\left(m_{DM}+m_{N}\right)^{2}},\label{eq:fermion-xsec-SI-App}
\end{equation}
where $b_{N}=2Z\, b_{p}+\left(A-Z\right)b_{n}$ with $b_{p}=2b_{u}+b_{d}$
and $b_{n}=b_{u}+2b_{d}$. It is convenient to consider the cross
section with the single nucleon for comparing with the experiment.
Through a simple algebra calculation, we obtained the cross sections
of WIMP-proton and WIMP-neutron as given below: 
\begin{eqnarray}
\sigma_{\chi p}^{SI} & = & \frac{m_{DM}^{2}m_{p}^{2}}{\pi\left(m_{DM}+m_{p}\right)^{2}}\frac{1}{16\Lambda^{4}}\left[3\left(\alpha_{q\chi}^{R}-\alpha_{q\chi}^{L}\right)+\alpha_{\phi\chi}\frac{g}{4\cos\theta_{W}}\frac{v}{m_{Z}}\left(1-4s_{W}^{2}\right)\right]^{2}\nonumber \\
 & \approx & \left(6.98\times10^{-5}\,{\rm pb}\right)\left(\frac{{\rm TeV}}{\Lambda}\right)^{4}\left(\alpha_{q\chi}^{R}-\alpha_{q\chi}^{L}+0.013\alpha_{\phi\chi}\right)^{2},\\
\sigma_{\chi n}^{SI} & = & \frac{m_{DM}^{2}m_{n}^{2}}{\pi\left(m_{DM}+m_{p}\right)^{2}}\frac{1}{16\Lambda^{4}}\left[3\left(\alpha_{q\chi}^{R}-\alpha_{q\chi}^{L}\right)-\alpha_{\phi\chi}\frac{g}{4\cos\theta_{W}}\frac{v}{m_{Z}}\right]^{2}\nonumber \\
 & \approx & \left(6.98\times10^{-5}\,{\rm pb}\right)\left(\frac{{\rm TeV}}{\Lambda}\right)^{4}\left(\alpha_{q\chi}^{R}-\alpha_{q\chi}^{L}-0.162\alpha_{\phi\chi}\right)^{2}.
 \end{eqnarray}

Now consider the SD scattering. WIMPs could also couple to the spin
of the target nucleus through an {}``axial-vector'' interaction\begin{equation}
\mathcal{L}_{A}=d_{q}\bar{\chi}\gamma_{\mu}\gamma_{5}\chi\bar{q}\gamma^{\mu}\gamma_{5}q,\label{eq:Fermion-SD-axialvetor-lang}\end{equation}
where $d_{q}$ is the shorthand notation of the axial coupling given
in the second term of Eq.\ \ref{eq:FDM-nucleon-interaction}. The
spin-dependent WIMP-nuclei elastic scattering cross section can be
expressed as\begin{equation}
\sigma_{\chi N}^{SD}\approx\frac{32m_{DM}^{2}m_{N}^{2}}{\pi\left(m_{DM}+m_{N}\right)^{2}}\left[\Lambda_{N}^{2}J\left(J+1\right)\right],\label{eq:fermion-xsec-SD-App}\end{equation}
where $J$ is the total angular momentum of the nucleus. $\Lambda_{N}\left(\propto1/J\right)$
depends on the axial couplings of WIMPs to the quarks,\[
\Lambda_{N}\equiv\frac{a_{p}\left\langle S_{p}\right\rangle +a_{n}\left\langle S_{n}\right\rangle }{J}.\]
where \[
a_{p}=d_{u}\Delta_{u}^{p}+d_{d}\Delta_{d}^{p}+d_{s}\Delta_{s}^{p},\qquad a_{n}=d_{u}\Delta_{u}^{n}+d_{d}\Delta_{d}^{n}+d_{s}\Delta_{s}^{n}.\]
Here, $\Delta$'s are the fraction of the nucleon spin carried by
a given quark. Their values are measured to be\ \cite{Mallot:1999qb}
\[
\Delta_{u}^{p}=\Delta_{d}^{n}=0.78\pm0.02,\qquad\Delta_{d}^{p}=\Delta_{u}^{n}=-0.48\pm0.02,\qquad\Delta_{s}^{p}=\Delta_{s}^{n}=-0.15\pm0.02.\]
$\left\langle S_{p}\right\rangle $ and $\left\langle S_{n}\right\rangle $
are the expectation values of the total spin of protons and neutrons,
respectively, whose values depend on the number of protons and neutrons
in the nucleus being considered, namely, whether it is odd or even.
For odd-proton nuclei the spin-dependent WIMP-nucleus cross section
is mainly due to the WIMP-proton interactions, whereas for odd-neutron
nuclei it is dominated by WIMP-neutron scattering. For even-even nuclei
the spin-dependent cross-section is highly suppressed. For a proton
and a neutron as a target, Eq.\ \ref{eq:fermion-xsec-SD} is transformed
into the cross section form WIMP-proton (neutron) interactions with
the proton (neutron) spins $\left\langle S_{p,n}\right\rangle =1/2$
and $J=1/2$. Thus, the SD cross section of WIMP-proton and WIMP-neutron
are given by\begin{eqnarray*}
\sigma_{\chi p}^{SD} & = & \frac{24\, m_{D}^{2}m_{p}^{2}}{\pi\left(m_{D}+m_{p}\right)^{2}}\frac{1}{16\Lambda^{4}}\left[0.15\alpha_{q\chi}^{R}+0.15\alpha_{q\chi}^{L}-0.68\alpha_{\phi\chi}\right]^{2}\\
 & \approx & \left(4.183\times10^{-6}\,{\rm pb}\right)\left(\frac{{\rm TeV}}{\Lambda}\right)^{4}\left(\alpha_{q\chi}^{R}+\alpha_{q\chi}^{L}-4.53\alpha_{\phi\chi}\right)^{2},\\
\sigma_{\chi n}^{SD} & = & \frac{24\, m_{D}^{2}m_{n}^{2}}{\pi\left(m_{D}+m_{n}\right)^{2}}\frac{1}{16\Lambda^{4}}\left[0.15\alpha_{q\chi}^{R}+0.15\alpha_{q\chi}^{L}+0.54\alpha_{\phi\chi}\right]^{2}\\
 & \approx & \left(4.183\times10^{-6}\,{\rm pb}\right)\left(\frac{{\rm TeV}}{\Lambda}\right)^{4}\left(\alpha_{q\chi}^{R}+\alpha_{q\chi}^{L}+3.53\alpha_{\phi\chi}\right)^{2}.\end{eqnarray*}

\subsection{Scalar DM-nucleon interaction\label{sub:Scalar-DM-nucleon-interaction}}

When the DM is a scalar $\varphi$, it can be detected in the spin-independent
experiments via the scalar interaction with nucleus \begin{equation}
\mathcal{L}_{\varphi\varphi qq}=\frac{1}{\Lambda^{2}}\frac{\alpha_{f\phi}v}{\sqrt{2}}\varphi\varphi\bar{q}q+\frac{1}{\Lambda^{2}}\frac{\alpha_{\phi1}v^{2}}{2}\frac{m_{q}}{m_{h}^{2}}\varphi\varphi\bar{q}q,\label{eq:SDM-nucleon-interaction}\end{equation}
which leads to the scattering amplitude\[
\left\langle \mathcal{M}\right\rangle =\mathcal{C}_{\varphi q}\left\langle \bar{q}q\right\rangle ,\]
where $\left\langle \,\,\right\rangle $ denotes an average and sum
over the spins of the initial and final state quarks, respectively,
and the coefficient $\mathcal{C}_{\varphi q}$ is given by \[
\mathcal{C}_{\varphi q}=\frac{1}{\Lambda^{2}}\left(\frac{\alpha_{f\phi}v}{\sqrt{2}}+\frac{\alpha_{\phi1}v^{2}}{2}\frac{m_{q}}{m_{h^{2}}}\right).\]
The matrix element $\left\langle \bar{qq}\right\rangle $ of quarks
in a nucleon state is given in\ \cite{Jungman:1995df} by\[
\left\langle \bar{q}q\right\rangle =\frac{m_{p,n}}{m_{q}}f_{Tq}^{(p,n)}\;\left({\rm light\,\, quarks}\right);\qquad\left\langle \bar{q}q\right\rangle =\frac{2}{27}\frac{m_{p,n}}{m_{q}}f_{Tg}^{(p,n)}\:\left({\rm heavy\,\, quarks}\right).\]
Summing over quark flavors, we obtain the $\varphi$-nucleon couplings:\begin{equation}
f_{\varphi q}^{(p,n)}=\sum_{q=u,d,s}f_{T_{q}}^{(p,n)}\mathcal{C}_{\varphi q}\frac{m_{p,n}}{m_{q}}+\frac{2}{27}f_{T_{g}}^{(p,n)}\sum_{q=c,b,t}\mathcal{C}_{\varphi q}\frac{m_{p,n}}{m_{q}},\label{eq:func}\end{equation}
where\ \cite{Bottino:2001dj,Ellis:2005mb}\begin{eqnarray}
f_{T_{u}}^{(p)}\approx0.020\pm0.004,\quad & f_{T_{d}}^{(p)}\approx0.026\pm0.005,\quad & f_{T_{s}}^{(p)}\approx0.118\pm0.062,\nonumber \\
f_{T_{u}}^{(n)}\approx0.014\pm0.003,\quad & f_{T_{d}}^{(n)}\approx0.036\pm0.008,\quad & f_{T_{s}}^{(n)}\approx0.118\pm0.062\,.\label{eq:neuclon-form}\end{eqnarray}
The first term in Eq.\ \ref{eq:func} corresponds to interaction
with the quarks in the target nuclei, while the second term corresponds
to interactions with the gluons in the target through a quark loop
diagram. $f_{T_{g}}^{(p)}$ is given by $1-f_{T_{u}}^{(p)}-f_{T_{d}}^{(p)}-f_{T_{s}}^{(p)}\approx0.84$,
and analogously, $f_{T_{g}}^{(n)}\approx0.83$. Note that we have
to sum over the couplings to each nucleon before squaring because
the wavelength associated with the momentum transfer is comparable
to or larger than the size of nucleus, the so-called {}``coherence
effect''. Due to such a coherence effect with the entire nucleus,
the cross section for spin-independent interactions scales approximately
as the square of the atomic mass of the target nucleus\ \cite{Yao:2006px}.

The total spin-independent WIMP-nuclei cross section at zero momentum
transfer is given by \begin{equation}
\sigma_{\varphi N}^{SI}=\frac{m_{N}^{2}}{4\pi\left(m_{DM}+m_{N}\right)^{2}}\left[Zf_{\varphi q}^{(p)}+\left(A-Z\right)f_{\varphi q}^{(n)}\right]^{2},\label{eq:xsec-SI-scalar}\end{equation}
where $m_{N}$ is the target nuclei's mass, and $Z$ and $A$ are
the atomic number and atomic mass of the nucleus. In order to compare
with the experimental sensitivities and limits which are often described
in terms of the dark matter elastic scattering with nucleons, we derive
the $\varphi$-nucleon cross section as following:\[
\sigma_{\varphi p}^{SI}=\frac{m_{p}^{2}}{4\pi\left(m_{DM}+m_{p}\right)^{2}}\left[f_{\varphi p}^{(p)}\right]^{2},\qquad\sigma_{\varphi n}^{SI}=\frac{m_{n}^{2}}{4\pi\left(m_{DM}+m_{n}\right)^{2}}\left[f_{\varphi n}^{(n)}\right]^{2}.\]
In practices, we found $\sigma_{\varphi p}^{SI}\simeq\sigma_{\varphi n}^{SI}$.

\subsection{Vector DM-nucleon interaction\label{sub:Vector-DM-nucleon-interaction}}

Since the vector DM $\mathcal{Z}$ only talks to the Higgs boson in
our model , it can be detected in the spin-independent experiments.
The relevant Lagrangian is given by\[
\mathcal{L}_{\mathcal{ZZ}qq}=\frac{\alpha_{\phi\mathcal{Z}}}{2\Lambda^{2}}\frac{m_{q}}{m_{h}^{2}}C_{\mu\nu}C^{\mu\nu}\bar{q}q,\]
which gives rise to the following scattering amplitude\[
i\mathcal{M}_{\mathcal{ZZ}qq}=\frac{\alpha_{\phi\mathcal{Z}}}{\Lambda^{2}}\frac{m_{q}}{m_{h}^{2}}\left[\left(p_{1}\cdot p_{3}\right)g^{\mu\nu}-p_{1}^{\nu}p_{3}^{\mu}\right]\varepsilon_{\nu}^{*}\left(p_{3}\right)\varepsilon_{\mu}\left(p_{1}\right)\bar{q}\left(p_{4}\right)q\left(p_{2}\right).\]
In the extreme non-relativistic limit, $p_{1}=p_{3}=\left(m_{DM},0\right)$
and the polarization vector of the heavy vector $\mathcal{Z}$ are
purely spatial, $\varepsilon^{\mu}\left(p_{1,3}\right)=\left(0,\vec{\varepsilon}_{1,3}\right)$.
The above amplitude is thus simplified as\[
i\mathcal{M}_{\mathcal{ZZ}qq}=\mathcal{C}_{\mathcal{Z}q}\,\,\varepsilon_{\mu}^{*}\left(p_{3}\right)\varepsilon^{\mu}\left(p_{1}\right)\bar{q}\left(p_{4}\right)q\left(p_{2}\right),\]
where \[
\mathcal{C}_{\mathcal{Z}q}=\alpha_{\phi\mathcal{Z}}\frac{m_{D}^{2}}{\Lambda^{2}}\frac{m_{q}}{m_{h}^{2}}.\]
The effective WIMP-nucleon scattering cross section is thus given
by\begin{eqnarray*}
\sigma_{\mathcal{Z}p,\mathcal{Z}n}^{SI} & = & \frac{m_{p,n}^{2}}{4\pi\left(m_{D}+m_{p,n}\right)^{2}}\left[Zf_{\mathcal{Z}q}^{(p,n)}+\left(A-Z\right)f_{\mathcal{Z}q}^{(p,n)}\right]^{2},\end{eqnarray*}
where the $\mathcal{Z}$-nucleon couplings, $f_{\mathcal{Z}q}^{(p,n)}$,
are given by\[
f_{\mathcal{Z}q}^{(p,n)}=\sum_{q=u,d,s}f_{T_{q}}^{(p,n)}\mathcal{C}_{\mathcal{Z}q}\frac{m_{p,n}}{m_{q}}+\frac{2}{27}f_{T_{G}}^{(p,n)}\sum_{q=c,b,t}\mathcal{C}_{\mathcal{Z}q}\frac{m_{p,n}}{m_{q}}.\]
Again, using $m_{p}\simeq m_{n}$ and $m_{D}\gg m_{p,n}$, we obtain
the effective WIMP-nucleon cross section, $\sigma_{\mathcal{Z}p}^{SI}(=\sigma_{\mathcal{Z}n}^{SI})$
as following: \begin{eqnarray*}
\sigma_{\mathcal{Z}p}^{SI} & = & \frac{\alpha_{\phi\mathcal{Z}}^{2}}{4\pi}\left(\frac{m_{D}^{2}}{\Lambda^{4}}\right)\left(\frac{m_{p}}{m_{h}}\right)^{4}\left[f_{T_{s}}+\frac{6}{27}f_{T_{G}}\right]^{2}.\\
 & \approx & \left(2.88\times10^{-10}\,{\rm pb}\right)\alpha_{\phi\mathcal{Z}}^{2}\left(\frac{m_{D}}{100\,{\rm GeV}}\right)^{2}\left(\frac{{\rm TeV}}{\Lambda}\right)^{4}\left(\frac{100\,{\rm GeV}}{m_{h}}\right)^{4}\end{eqnarray*}

The SI elastic $\mathcal{Z}$-nucleon scattering cross sections are
plotted in Fig.\ \ref{fig:vector-SI}, along with the projected future
sensitivity of SuperCDMS, Stage-A (blue dashed line) and Stage-C (black
dashed curve). The cross sections are well below all the current bound
from CDMS collaboration, the can be probed at the future experiment
by SuperCDMS at Stage-C.

\section{Dark matter indirect detection from gamma-ray\ \label{sec:Dark-matter-indirect}}

WIMPs annihilation in the halo may lead to a flux of gamma-rays, with
both continuum and line contributions. The former can be produced
as continuum photons from final state radiation and the cascades of
other annihilation products, whereas the latter can be emission from
loop-diagrams to $\gamma\gamma$, $\gamma Z$ or $\gamma h$ final
states. Observation of monochromatic gamma rays would provide a {}``smoking-gun''
signal for the existence of WIMPs in the halo. 

The produced gamma ray flux from the annihilation of dark matter particles
can be expressed as\begin{equation}
\frac{\Phi_{\gamma}}{d\Omega\, dE}=\sum_{i}\frac{dN_{\gamma}^{i}}{dE_{\gamma}}\left\langle \sigma_{i}v\right\rangle \frac{1}{4\pi m_{DM}^{2}}\int_{l.o.s}\rho^{2}dl,\label{eq:photon-flux}\end{equation}
where $\rho$ is the dark matter density profile, $\left\langle \sigma_{i}v\right\rangle $
and $dN_{\gamma}^{i}/dE_{\gamma}$ are, respectively, the thermally
averaged annihilation cross section times the relative velocity $v$
and the differential gamma spectrum per annihilation coming from the
decay of annihilation products of final state $i$. The integration
is taken along the line of sight.

All the halo model dependence isolated in the integral which is given
in the dimensionless form\ \cite{Bergstrom:1997fj}\[
J\left(\Psi,\Delta\Omega\right)\equiv\frac{1}{8.5\,{\rm kpc}}\left(\frac{1}{0.3\,{\rm GeV}/{\rm cm}^{3}}\right)^{2}\frac{1}{\Delta\Omega}\int_{\Delta\Omega}d\Omega\int_{l.o.s}\rho^{2}dl,\]
where $\Psi$ is the angle away from the direction of the galactic
center that is observed while $\Delta\Omega$ the solid angel of the
field of view centered on $\Psi=0$. After averaging $J\left(\Psi,\Delta\Omega\right)$
over a spherical solid angle $\Delta\Omega$, one obtains the photon
flux as\ \cite{Bergstrom:1997fj}\begin{equation}
\frac{d\Phi}{dE_{\gamma}}=\left(5.5\times10^{-10}{\rm s}^{-1}{\rm cm}^{-2}\right)\frac{dN_{\gamma}^{i}}{dE_{\gamma}}\left(\frac{\left\langle \sigma_{i}v\right\rangle }{{\rm pb}}\right)\left(\frac{100\,{\rm GeV}}{m}\right)^{2}\bar{J}\left(\Psi,\Delta\Omega\right)\Delta\Omega,\label{eq:gamma-flux}\end{equation}
where the function $\bar{J}$ contains all information about the dark
matter distribution in the halo. The above equation can be factorized
into two parts: $\bar{J}\left(\Psi,\Delta\Omega\right)\Delta\Omega$
(astrophysics) and other terms describing the dark matter annihilation
and the fragmentation of its annihilation products (particle physics).
The latter can be determined from microscopic measurement, but the
former exhibits considerable uncertainties involved in the distribution
of dark matter in the Galactic center region. For example, at $\Delta\Omega=10^{-3}\,{\rm sr}$,
characteristic of ground-based Atmospheric Cerenkov Telescopes (ACTs),
typical values of $\bar{J}$ range from $10^{3}$ for the NFW profile\ \cite{Navarro:1995iw,Navarro:1996gj},
to about $10^{5}$ for the profile of Moore, \emph{et al}.\ \cite{Moore:1999nt}.
Furthermore, as the distribution of dark matter might be clumpy, this
inhomogeneity effect would enhance indirect detection rates by a {}``boost
factor'', defined as\begin{equation}
B\equiv\frac{\left\langle \rho^{2}\right\rangle }{\left\langle \rho\right\rangle ^{2}}.\label{eq:boost}\end{equation}
If the dark matter were locally distributed completely evenly, the
boost factor would be equal to one. Small-scale structure of dark
matter, however, enhance this quantity to a large value.

The particle physics model dependence enters through all the other
factor in Eq.\ \ref{eq:photon-flux}. The energy integral is roughly
$\int dE\, dN_{\gamma}^{i}/dE\sim0.5$ for all $i$, but the energy
distribution depends significantly on the annihilation channel. Note
that the resulting photon spectra depend only on the initial energies
of the annihilation products and not on the details of the WIMPs annihilation
process. The spectra have been studied using PYTHIA~\cite{Sjostrand:2006za},
and a simple analytic fit has been presented in Refs.\ \cite{Bergstrom:1997fj,Feng:2000zu}
for the most important annihilation channels as follows: \[
\frac{dN_{\gamma}}{dx}=\frac{a}{x^{1.5}}e^{-bx},\]
where $x\equiv E_{\gamma}/m_{DM}$ and $\left(a,b\right)=\left(0.73,7.76\right)$
for $WW$ and $ZZ$, $\left(1.0,10.7\right)$ for $b\bar{b}$, $\left(1.1,15.1\right)$
for $t\bar{t}$, and $\left(0.95,6.5\right)$ for $u\bar{u}$. In
most cases the spectrum produced does not vary much. The only exception
to this is the somewhat harder spectrum generated through annihilation
to $\tau^{+}\tau^{-}$. Particle physics models, however, very rarely
predict a dark matter candidate that annihilates mostly to $\tau^{+}\tau^{-}$.
In this work we focus our attention only on the annihilation channel
involving the heavy fermion, gauge boson and Higgs boson.

The integrated photon flux above some photon energy threshold $E_{th}$
is\ \cite{Feng:2000zu} \begin{eqnarray}
\Phi_{\gamma}\left(E_{th}\right) & = & \left(5.5\times10^{-10}{\rm s}^{-1}{\rm cm}^{-2}\right)\left(\frac{100\,{\rm GeV}}{m}\right)^{2}\bar{J}\left(\Psi,\Delta\Omega\right)\Delta\Omega\nonumber \\
 & \times & \sum_{i}\left(\frac{\left\langle \sigma_{i}v\right\rangle }{{\rm pb}}\right)\int_{E_{th}}^{m_{D}}dE_{\gamma}\frac{dN_{\gamma}^{i}}{dE_{\gamma}},\label{eq:integral-flux}\end{eqnarray}
where the sum is over all possible annihilation channels. Detectors
also have upper cutoffs, but these are typically irrelevant, as the
energy distribution falls steeply with energy. In this work we consider
two representative $E_{th}$: $1\,{\rm GeV}$, accessible to space-based
detectors, and $50\,{\rm GeV}$, characteristic of ground-based telescopes. 

\bibliographystyle{apsrev}

\end{document}